\documentclass[11pt,letter]{article} 
\usepackage{jheppub} %

\usepackage[ps,dvips,matrix,arrow,frame,import,curve,color]{xy}

\def\blfootnote{\xdef\@thefnmark{}\@footnotetext}

\long\def\symbolfootnote[#1]#2{\begingroup%
\def\thefootnote{\fnsymbol{footnote}}\footnote[#1]{#2}\endgroup}

\newcommand{\be}{\begin{eqnarray}}
\newcommand{\ee}{\end{eqnarray}}
\newcommand{\ben}{\begin{eqnarray*}}
\newcommand{\een}{\end{eqnarray*}}

\newcommand{\bcent}{\begin{center}}
\newcommand{\ecent}{\end{center}}
\newcommand{\benum}{\begin{enumerate}}
\newcommand{\eenum}{\end{enumerate}}
\newcommand{\bdesc}{\begin{description}}
\newcommand{\edesc}{\end{description}}
\newcommand{\bitem}{\begin{itemize}}
\newcommand{\eitem}{\end{itemize}}
\newcommand{\bquote}{\begin{quote}}
\newcommand{\equote}{\end{quote}}
\newcommand{\bhalfp}{\begin{minipage}{0.45\textwidth}}
\newcommand{\ehalfp}{\end{minipage}}
\newcommand{\bhead}{\begin{center}\bf \Large}
\newcommand{\ehead}{\end{center}\bigskip}

 \newcommand{\half}{{1\over 2}}
 \newcommand{\del}{{\partial}}
 
\def\be{\begin{equation}}
\def\ee{\end{equation}}
\def\ba{\begin{eqnarray}}
\def\ea{\end{eqnarray}}

\newcommand{\roughly}[1]{\mathrel{\raise.3ex\hbox{$#1$\kern-0.85em
\lower1ex\hbox{$\sim$}}}}

\def\2pi{\left(2\pi\right)}

\def\beq{\begin{equation}}
\def\eeq{\end{equation}}
\def\bg{\begin{eqnarray}}
\def\nd{\end{eqnarray}}
\def\bea{\begin{eqnarray}}
\def\eea{\end{eqnarray}}

\def\D3{\overline{\mbox{D3}}}

\title{F-Theory, Seiberg-Witten Curves and ${\cal N} = 2$ Dualities}

\author{Keshav Dasgupta,}
\author{Jihye Seo,}
\author{and Alisha Wissanji}
\affiliation{Ernest Rutherford Physics Building, McGill University,\\
3600 University Street, Montr{\'e}al QC, Canada H3A 2T8}
\emailAdd{keshav, jihyeseo, wissanji@hep.physics.mcgill.ca}
\abstract{F-theoretic constructions can alternatively be understood as consequences of certain ${\cal N} = 2$
Seiberg-Witten theories via type IIB $r$ D3-branes probing the quantum corrected orientifold backgrounds.
We present four models that come out from such consideration.

In Model 1, the seven-branes wrap
the flat ${\mathbb R}^4$ directions, leading to the well known $Sp(2r)$ theories. We study singularity structure of moduli space of Seiberg-Witten curve, such as maximal Argyres-Douglas loci, in order to construct 1-1 map between moduli spaces.

In Model 2, the seven-branes are wrapped on
Taub-NUT and multi Taub-NUT spaces instead of ${\mathbb R}^4$.
These configurations may succinctly explain many of the recently proposed Gaiotto-type
constructions including possible extensions to non-conformal models with interesting cascading behaviors.
In this model the UV limit would
be described by the probe D3-branes
decomposed into D5-${\overline{\rm D5}}$ pairs wrapped on vanishing 2-cycles of multi Taub-NUT space,
while the IR would remain a
4d theory. For certain arrangements of the
seven-branes and in a specific delocalization limit, this model
may be dualized to
the brane networks recently studied by Benini, Benvenuti and Tachikawa.
On the other hand, the Gaiotto dualities in Model 2
are explained by three simultaneous effects: chiral anomaly cancellations, anti-GSO projections and brane transmutations.

Model 3 is described by seven-branes wrapped on a K3 manifold and D3 and ${\overline{\rm D3}}$ probes whose number may
differ at most by 24. These constructions could lead to new
${\cal N} = 2$ models with possible dualities to
both type IIB and heterotic theories on non-K\"ahler manifolds, and involve interesting interplays between abelian
instantons and fluxes. In the limit where the number of probes becomes very large,
the physics is captured by M(atrix) theory on ${\rm K3} \times {\rm K3}$ manifold with $G$-fluxes.

Finally, Model 4
is described by
$k$ D3-branes probing intersecting seven-brane backgrounds with $Sp(2k) \times Sp(2k)$ gauge group.
These constructions could produce
new ${\cal N} =1$ heterotic dual given by $k$ small instantons wrapping 2-cycles of
a non-K\"ahler K3 manifold that is no longer
conformally Calabi-Yau.
We discuss possible constraints on these models coming
from global charge and anomaly cancellations in F-theory and study the implications of these
constraints on the global and local symmetries of
the underlying gauge theories.}
\begin{document}
\maketitle
\flushbottom
 \toccontinuoustrue

\section{Introduction}

\noindent ${\cal N} = 2$ supersymmetric Yang-Mills theory 
lies between trivial (${\cal N} = 4$) and not-fully-solvable (${\cal N} = 1$) in the sense that one can solve exactly the theory
with at most two derivatives and four fermions
and at low energies\cite{sw1, sw2}. This is mostly because the dynamics therein are
governed by holomorphic quantities. However, the Wilsonian effective action with higher derivative terms are
not governed by holomorphic quantities and therefore finding the exact solution is a challenge.

The exact solution for ${\cal N} = 2$ $SU(2)=Sp(2)$ Seiberg-Witten theory \cite{sw1}
relies on 
a few essential ingredients that
lie at the heart of the physics. First is the holomorphicity of the prepotential ${\cal F}$ which allows us to
express the $U(1)$ gauge theory completely in terms of the following action \cite{sw1, sw2}:
\bg\label{uigt}
 {1\over 16\pi} ~{\rm Im}~\int d^4x \left[\int d^2\theta {\cal F}''(\Phi) W^\alpha W_\alpha + \int d^2\theta d^2{\bar\theta}\Phi^\dagger{\cal F}(\Phi)\right],
\nd
where $\Phi$ and $W^\alpha$ are the chiral and the vector multiplets respectively, using ${\cal N} = 1$ language. In
fact, we don't have to go beyond the $U(1)$ case, as quantum corrections prevent the theory from enhancing to its full
$Sp(2)$ gauge symmetry anywhere in the moduli space. The classical enhancement point is replaced by
two points where a monopole and a dyon respectively become massless \cite{sw1, sw2}.

The second is the existence of a well-defined positive-definite metric in the moduli space that is positive definite everywhere. In other words, the holomorphic prepotential can  also be used to define the metric as:
\bg\label{hpdm}
ds^2 ~ = ~ {\rm Im}~ {\cal F}''(a) ~ da d{\bar a} ~ = ~ {\rm Im}~\tau(a) da d{\bar a} ~ = ~ \vert \partial_u a\vert^2
\tau(a) \vert du\vert^2,
\nd
with ${\rm Im}~\tau(a) > 0$ everywhere in the moduli space and $u$ (but not $a$) parametrizes the moduli space.

The third is the holomorphic prepotential ${\cal F}$ that is itself constrained by the weakly coupled limit of the underlying ${\cal N} = 2$
supersymmetry at large vev to have the following form:
\bg\label{prpon2}
{\cal F}(a) ~ = ~ {1\over 2} \tau_0 a^2 + {i\over \pi} a^2 {\rm log} {a^2\over
\Lambda^2} + {a^2\over 2\pi i} \sum_{l=1}^\infty c_l \left({\Lambda\over a}\right)^{4l},
\nd
where the first term is the classical piece with $\tau_0$ the bare coupling constant, the second is the one-loop
term, and the rest are the instanton contributions.

The last is the duality that exchanges the $U(1)$ action to its dual action. This duality is the manifestation of the underlying Montonen-Olive duality where electrically charged states are replaced by magnetically charged states. In fact, there is a whole tower of dual states governed by an underlying $SL(2, {\bf Z})$ symmetry. For our case, the dual lagrangian associated with \eqref{uigt} is given by:
\bg\label{uigtd}
 {1\over 16\pi} {\rm Im}~\int d^4x \left[\int d^2\theta {\cal F}_D''(\Phi_D) W_D^\alpha W_{D\alpha} + \int d^2\theta d^2{\bar\theta}\Phi_D^\dagger{\cal F}_D(\Phi_D)\right],
\nd
where the subscript $D$ denotes the electric-magnetic dual variables. One may also use the dual variable $a_D$ to write the metric in the moduli space \eqref{hpdm} more compactly as:
\bg\label{mims}
ds^2 ~ = ~ {\rm Im}~(da_D d{\bar a}).
\nd

The four points mentioned above are enough to tell us the exact prepotential for our case, which requires one to study the singularity structure in the moduli space. Recall that, in the Wilsonian action, the singularities in the
moduli space imply that certain massless states are integrated out. Thus finding these singular regions of the moduli
space will tell us precisely what states have been integrated out from the underlying monodromies. This information,
as shown by Seiberg-Witten \cite{sw1, sw2}, is succinctly captured by an elliptic curve that is non-trivially fibered
over the $u$-plane, i.e over the moduli space.

\begin{figure}[htb]
        \begin{center}
\includegraphics[height=4cm]{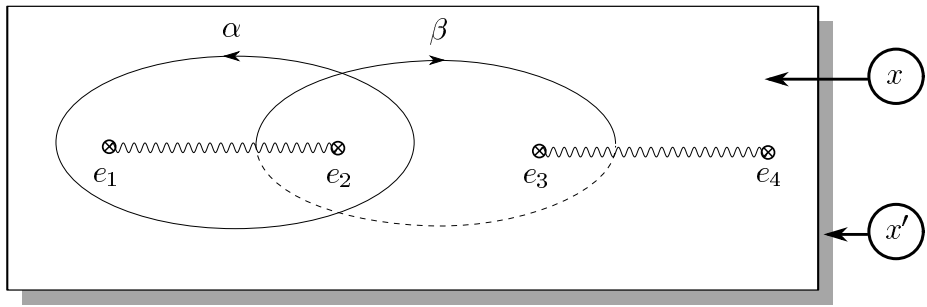}   \end{center}
        \caption{A slightly different way to see the Seiberg-Witten torus on the $u$-plane compared to what we will
present later. The Seiberg-Witten curve is given by $y^2(x, u) = (x^2-u)^2 - \Lambda^4 ={\prod}_{i=1}^{4} (x-e_i)$
The $e_i(u)$ are the
zeroes of $y^2$, and therefore branch points on $x$-plane.
The choice of the homology basis are given by the
cycles ($\alpha, \beta$), and the two sheets are shown as $x$ and $x'$. These two sheets are glued together along the cuts
that run between the branch points $e_i$.
This figure is taken from the excellent review article of Lerche \cite{LercheReview}.}
        \label{lerchefig}
        \end{figure}

The reason why elliptic curve may appear can be presented in a slightly different way, compared to the way it was originally presented in \cite{sw1, sw2}. Imagine we {\it knew} the solutions for ($a(u), a_D(u)$) to take the following
form\footnote{For example, this can be determined by solving a certain Schr\"odinger-type differential equation with a
meromorphic potential that has poles at certain
points. See \cite{BilalReview} for a more detailed treatment of this.}:
\bg\label{aad}
a(u) ~ = ~ {\sqrt{2}\over \pi}\int_{-1}^1 {dx \sqrt{x-u}\over \sqrt{x^2-1}}, \qquad a_D(u) ~ = ~ {\sqrt{2}\over \pi}\int_{1}^u {dx \sqrt{x-u}\over
\sqrt{x^2-1}}.
\nd
The above integrands have square-root branch cuts with branch points at $x=+1, -1,
u$ and $\infty$. The two branch cuts can be taken to run from $-1$ to $+1$ and  from $u$ to $\infty$. The Riemann surface of the integrand is two-sheeted with the two sheets connected through the cuts. If one adds the point at infinity to each of the two sheets, the topology of the Riemann surface of that of two
spheres connected by two tubes or a torus. A slightly different way is to construct the
torus is shown in {\bf figure \ref{lerchefig}}.
Thus the
Riemann surface of the integrands in \eqref{aad} has genus
one and is given by:
\bg\label{sebl}
y^2 ~ = ~ (x^2 -1) (x - u).
\nd
This curve is nothing but the quotient of the upper half plane $H$ modded out by $\Gamma(2)$ i.e $H/\Gamma(2)$ where $\Gamma(2)$ is a subgroup of matrices in
$SL(2, {\bf Z})$ congruent to 1 modulo 2. The ($a, a_D$) are then the integral of a one-form $\lambda$ over the 2-cycles $\gamma_1, \gamma_2$ of the torus, i.e:
\bg\label{shdies}
a_D = \oint_{\gamma_1} \lambda, \qquad a = \oint_{\gamma_2} \lambda, \qquad
\lambda =  {\sqrt{2}\over \pi} \cdot {\sqrt{x-u}\over \sqrt{x^2-1}}~dx.
\nd
This completes our short tour through Seiberg-Witten theory.
In the following subsection \ref{subsecReal}, we will warm-up with another short pedagogical analysis of the Seiberg-Witten
curves for pure $Sp(2r)$ theories, before we finally get to more technical and systematic approach to come in subsection \ref{Sp2rcurveTech}.
Our goal is to analyze the root structures for Seiberg-Witten curves and determine the effects of the quantum
corrections. This will help us find a map between moduli space of Seiberg-Witten curves and that of brane dynamics, which will be discussed in \ref{subsecMapM}.
        %

\subsection{Analysis of branch cuts along the real axis \label{subsecReal}}

\begin{figure}[htb]
        \begin{center}
\includegraphics[height=6cm]{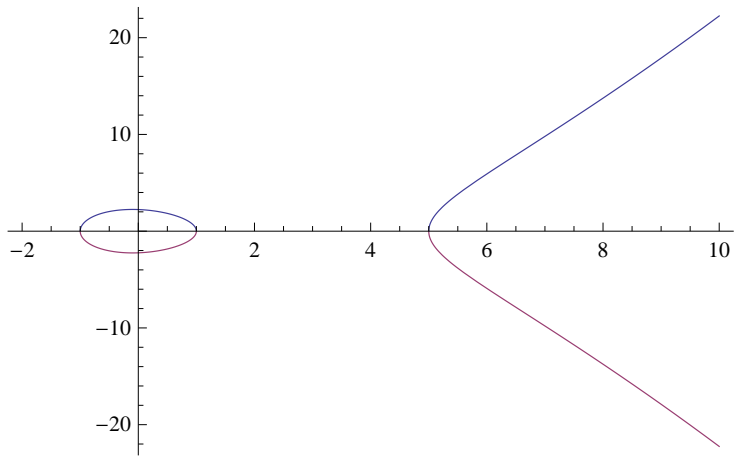}   \end{center}
        \caption{The horizontal and vertical axes stand for ${\rm Re  }[x]$ and ${\rm Re }[y]$, while $x$ and $y$ satisfy \eqref{115}. The intercepts with horizontal axis $x=\pm 1, u$ and $x=\infty$ are branch points of the Seiberg-Witten curve for the simplest $SU(2)$ case without matter \eqref{sebl}, here taken at a generic real value of moduli $u=5$.}
        \label{swfig1}
        \end{figure}

\begin{figure}[htb]
        \begin{center}
\includegraphics[height=6cm]{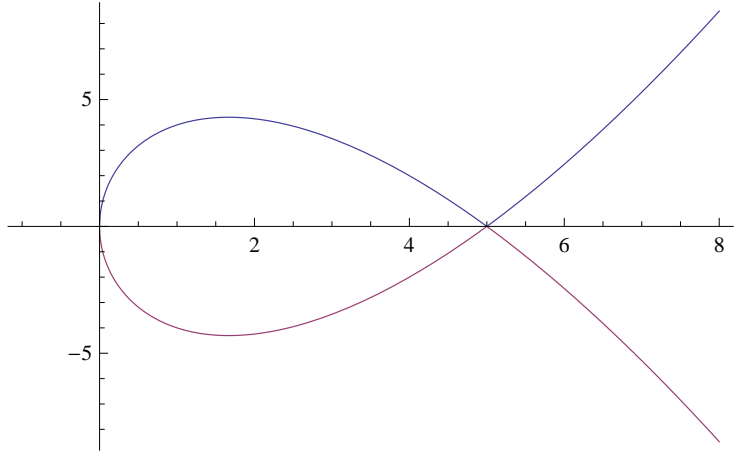}
        \end{center}
        \caption{Real loci plot for the classical $Sp(2)$ case without matter, by taking $\phi_1 = 5$, $\Lambda=0$, $r=1$ in \eqref{curve3}. The curve is singular at double roots $x=\phi_a$ for any choice of moduli.}
        \label{swfig2}
        \end{figure}

\begin{figure}[htb]
        \begin{center}
\includegraphics[height=6cm]{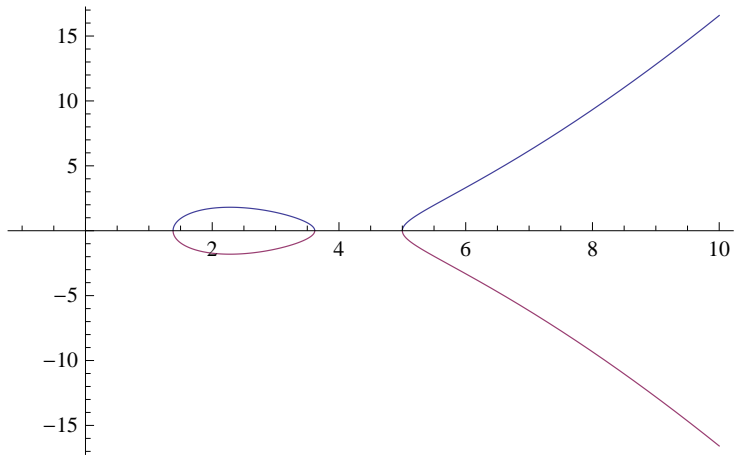}
        \end{center}
        \caption{Real loci plot for the quantum $Sp(2)$ case without matter, by taking $\phi_1 = 5$, $\Lambda=(5/16)^{1/4}$, $r=1$ in \eqref{curve3}. For such a generic choice of moduli, the curve is smooth.} \label{swfig3}
        \end{figure}

\begin{figure}[htb]
        \begin{center}
\includegraphics[height=6cm]{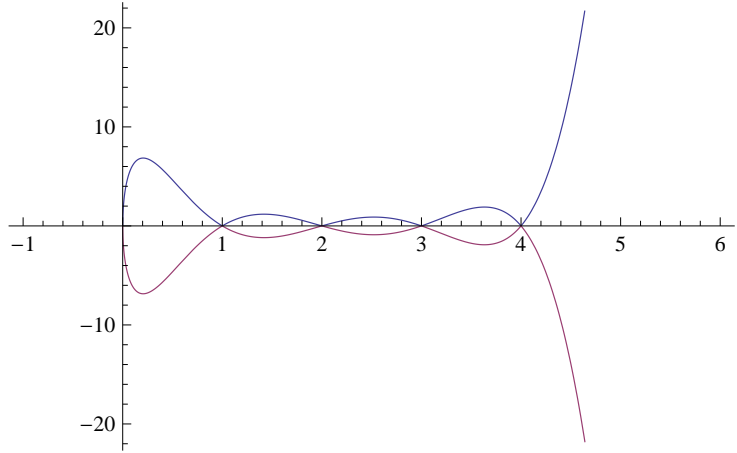}
        \end{center}
        \caption{The various singularities for the classical $Sp(8)$ case without matter. Plot of real parts of $x$ and $y$ in \eqref{curve3} with $r=4, \Lambda=0$ at a point in moduli space given by \eqref{phivalues}} \label{swfig4}
        \end{figure}

\begin{figure}[htb]
        \begin{center}
\includegraphics[height=6cm]{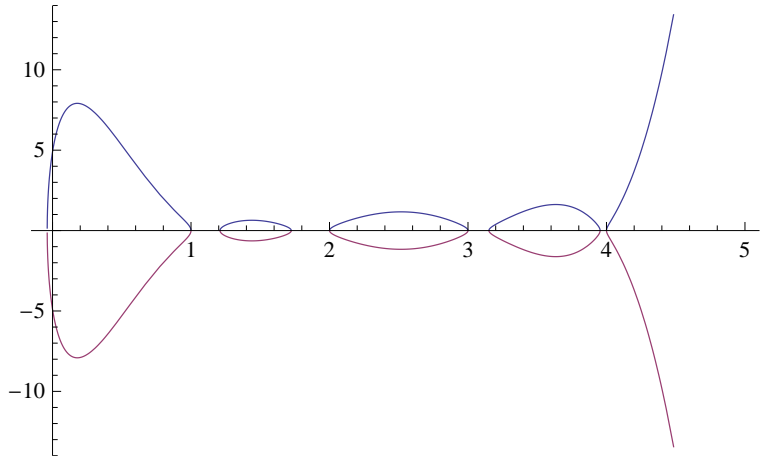}
        \end{center}
        \caption{The various branch cuts for the quantum $Sp(8)$ case without matter. Plot of real parts of $x$ and $y$ in \eqref{curve3} with $r=4, \Lambda\ne 0$ at a point in moduli space given by \eqref{phivalues}}
        \label{swfig5}
        \end{figure}

We will start off with something familiar: the well known ${\cal N} =2$ $SU(2)$ gauge theory without matter.
Consider its Seiberg-Witten curve of the form \eqref{sebl}, and for example use $u=5$:
\bg\label{curver}
y^2 = (x-1)(x+1)(x-5). \label{115}
\nd
The real cross-section (by suppressing imaginary parts of $x$ and $y$) of this curve is given as in {\bf figure \ref{swfig1}}. Once we
add back the point from infinity the curve will become a genus 1 Riemann surface, i.e a two-torus.
The above curve \eqref{curver} is in the Weierstrass form. For other groups one may also write similar kind of (hyper-)elliptic
curves. For us the gauge group that we will be mostly interested in is the non-simply laced group $Sp(2r)$ where
$r$ is the rank of the group. As derived in appendix \ref{sp2r01} using \cite{ArgyresShapere}, Seiberg-Witten curve for $Sp(2r)$ theory without matter is given by:
\bg\label{curve3}
y^2 = \prod_{a=1}^r (x-\phi_a^2) \left[x\prod_{b=1}^r (x-\phi_b^2) + 16\Lambda^{2r+2}\right],
\nd
whose quantum effects come from demanding $\Lambda \ne 0$\footnote{In \eqref{sebl} and \eqref{115}, non-zero value of $\Lambda=1$ is already plugged in.}. In a classical limit of $\Lambda \rightarrow 0$, all the $\phi_a$ becomes double roots, creating singularity there. For simplicity we choose $Sp(2)$ case with $\phi_1 = 5$ in some appropriate units. The
real cross-section for the classical case $\Lambda=0$ is depicted in {\bf figure \ref{swfig2}}: note that the curve is singular at double roots $x=\phi_a$.
To see how the curve changes under quantum corrections, let us take $\Lambda\ne0$. The real
cross-section for the quantum curve then is as shown in {\bf figure \ref{swfig3}} for $\Lambda \equiv (5/16)^{1/4}$.
It is reassuring to see how the quantum curve smoothing out the singularity coming from double roots of classical curve. This behavior is
generic, and will occur for all higher rank cases. To see this, let us consider a little more complicated example of $r=4$
and take:
\bg\label{phivalues}
\phi_1 = 1, \qquad \phi_2 = 2, \qquad \phi_3 = 3, \qquad \phi_4 = 4.
\nd
For this, the classical curve is given by {\bf figure \ref{swfig4}}: This has the expected singular points from the
vanishing of the discriminant.
As one would have expected, the quantum curve splits the classical singularities of double roots into a pair of branch points on $x$-plane, as
depicted in {\bf figure \ref{swfig5}}. Again, considering a point at infinity will give us a correct genus of $r=4$ as expected.

The classical curve with $\Lambda=0$ is singular everywhere in the moduli space, in that it contains double roots. The quantum correction term $\Lambda$ lifts this singularity, so that the curve will be smooth at a generic point in a moduli space. By observing where in the moduli space the quantum curve becomes singular - in other words, the singularity loci of the moduli space of the Seiberg-Witten curve - we will learn about physics, such as appearance of massless BPS dyons and Argyres-Douglas theories in subsections \ref{subsecRank1} and \ref{Sp2rcurveTech}.

\subsection{First look at F-theory embeddings \label{subsecF1D3}}

\begin{figure}[htb]
        \begin{center}
\includegraphics[height=6cm]{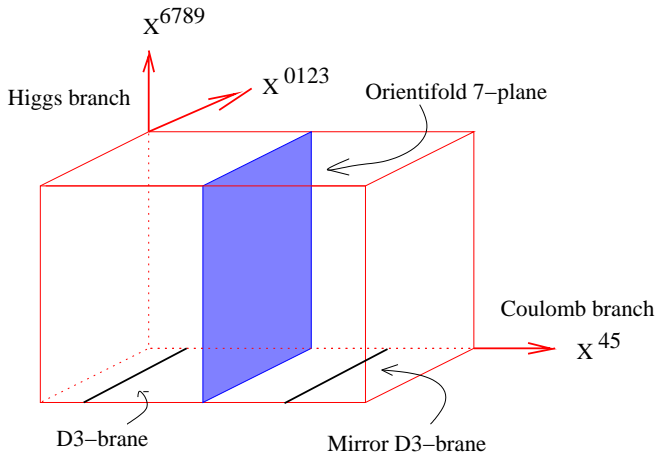}
        \end{center}
        \caption{In F theory
        a single D3-brane probing an orientifold seven-plane background maps to the
classical picture of $Sp(2)$ Seiberg-Witten theory.} \label{SWF1}
        \end{figure}

Let us
discuss possible F-theory embeddings of pure $Sp(2r)$ theories using
constructions similar to the one discussed in Douglas-Lowe-Schwarz \cite{dougschwa}.
Our starting point would be the quantum curve for the well known $Sp(2)$ case without matter \cite{KLYTsimpleADE}:
\bg\label{sp2case}
y^2 = (x^2 - u )^2 -\Lambda^4.
\nd
F-theory uplift of this is surprisingly simple, thanks to insightful works of \cite{senF, bds}. Classically a single probe D3-branes is kept in the background
of an orientifold
seven plane i.e and O7-plane. The D3-brane spans the spacetime directions $x^{0,1,2,3}$ and the O7-plane is
parallel to the D3-brane and spans directions $x^{0,1,2,3}$ as well as $x^{6,7,8,9}$. Therefore both the
D3 probe and
the O7-plane are pointlike on the $x^{4,5}$ directions which, in turn, will be our $u$-plane. This is depicted by the
familiar figure given as {\bf figure \ref{SWF1}}.
Classically the full $Sp(2)$ gauge
symmetry is 
enhanced only at the
point where the D3-brane coincide with the O7-plane: the {\it mirror} D3-brane will coincide with the original
D3-brane and will enhance the gauge symmetry by adding in the massless $W_\pm$ bosons. At all other points in the $u$-plane, the
gauge-symmetry will be broken to $U(1)$.

\begin{figure}[htb]
        \begin{center}
\includegraphics[height=6cm]{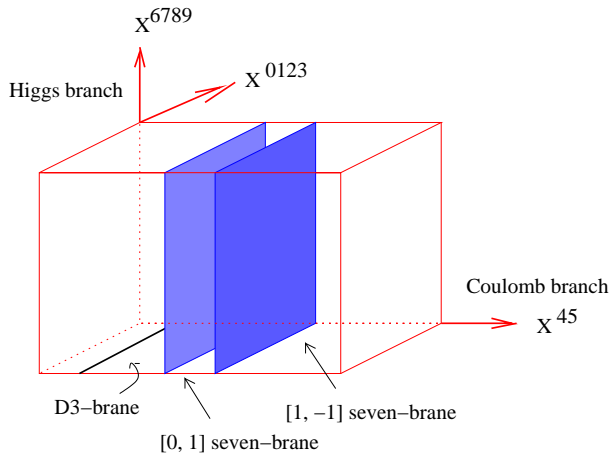}
        \end{center}
        \caption{The quantum corrections in the $Sp(2)$ theory maps to the splitting of the orientifold plane
into two ($p, q$) seven-branes in F-theory.} \label{SW2}
        \end{figure}

Under quantum corrections the situation is different. In the Seiberg-Witten theory, the quantum corrections come
from the one-loop contributions and from the infinite tower of instanton corrections. These corrections, as viewed
from the D3-brane point of view, will split the O7-plane into two ($p, q$) seven-branes of charges ($0, 1$) and
($1, -1$) shown in {\bf figure \ref{SW2}}.
 These two seven-branes are the monopole and the dyon points respectively in the original Seiberg-Witten theory.

Thus classically we have a D3-brane probing a ${\mathbb R}^2/{\mathbb Z}_2$ singularity with
${\mathbb Z}_2 \equiv \Omega (-1)^{F_L} {\cal I}_{45}$ where ${\cal I}_{45}: x^{4, 5} \to -x^{4, 5}$ and $\Omega$ is the
orientifold operation as in {\bf figure \ref{SWF1}}. Quantum effects split O7-plane into ``monopole'' and ``dyon'' seven-branes as in {\bf figure \ref{SW2}}. When the probe D3-brane come close to the ``monopole'' seven-brane the (0, 1) string between them
become arbitrarily light implying that a fundamental hypermultiplet becoming arbitrarily light. Similarly when the probe D3-brane
approaches the other ``dyonic'' seven-brane, the ($1, -1$)
string between them become arbitrarily light, implying, in turn,
that another hypermultiplet is becoming arbitrarily light. Thus we see that the probe D3-brane never gains any massless $W_\pm$
bosons, so never has any enhanced gauge symmetry. Instead, there are two-points in the $u = x^4 + i x^5$ plane where a
monopole hypermultiplet
and a dyon hypermultiplet become massless respectively. Finally, compactifying the $u$-plane using extra
seven-branes (demanding zero global monodromy) will lead us to the F-theory model of Vafa \cite{vafaF}
\footnote{Adding $N_f=4$ D7-branes parallel to O7-plane, we obtain $SU(2)$ and $N_f=4$ superconformal theory, which is studied in detail in \cite{Billo:2010mg, Billo:2011uc, Billo':2011pr}. In this simplest superconformal F-theory configuration, they compute non-perturbative effects induced by D-instantons, and make a connection with Nekrasov partition function \cite{Nekrasov} and AGT conjecture \cite{AGT}. The 8d physics
on the 7-branes can teach us about the 4d physics on the 3-brane \cite{Billo:2010mg, Billo:2011uc}. Axion-dilaton field $\tau$ knows about 8d chiral ring correlators, which are computed in  \cite{Billo':2011pr}  using localization technique.
It will be interesting to study this for the case with multiple D3-branes.}.
 
The above consideration then tells us that F-theory physics probed by D3-branes is Seiberg-Witten theory in disguise, at least locally.
This is of course
the scenario depicted in \cite{senF, bds}. The full global completion would be, for example, to {\it join} four
copies of ${\cal N} = 2, N_f = 4$ Seiberg-Witten theory \cite{sw2} so that global monodromies vanish. Then a
deformation from the orientifold point will allow us to span the full F-theory moduli space.

\subsection{Organization of the paper}

The paper is organized in the following way. The models that we want to study here
all belong to the same class of multiple D3-branes probing the F-theory backgrounds of
seven-branes and orientifold seven-planes. Following sections addressing each models can be read independently. The simplest
configuration is then $r$ D3-branes probing the usual parallel D7/O7 background along flat directions, 
giving $Sp(2r)$ Seiberg-Witten theory with antisymmetric matter. This is studied in section \ref{model1section}. We compare the brane picture and SW geometry. As a first step toward proving F-theory/SW-geometry duality for higher ranks, we start by studying the pure $Sp(2r)$ SW curve.
We analyze the maximal Argyres-Douglas points in pure $Sp(2r)$ and discuss a potential 1-1 map between the moduli space of the brane dynamics and that of the Seiberg-Witten curve for higher ranks. At rank 1, antisymmetric matter is empty, so the mapping is rather simple.
However for higher ranks we cannot ignore the anti-symmetric matter and therefore
it does make a difference\footnote{We thank Philip Argyres for discussion on this topic.}. 
The 1-1 mapping now is rather
subtle, and we point out various issues related to this. We speculate on some ways to resolve the puzzles and point
out an ideal setting where such 1-1 mapping could in-principle be realized.

In section \ref{model2section}, we generalize the simplest model of section \ref{model1section} to a slightly more non-trivial scenario by wrapping the
seven-branes and orientifold planes on multi Taub-NUT spaces, which is still ${\cal N}=2$ theory despite the new complication in the geometry. 
With D3-branes probing this background,
we argue that these configurations may succinctly explain many of the recently proposed
Gaiotto-type models \cite{gaiotto} on the
world-volumes of the D3-branes. In subsection \ref{subsecDDbarTN}, we show how to generate it by decomposing the
D3-branes as {\it fractional} D3-branes. These fractional D3-branes are D5-brane anti D5-brane (${\rm D5}$-$\overline{\rm D5}$) pairs, and they
wrap 2-cycles of the multi Taub-NUT (TN) space\footnote{To be precise, a {\it single} D3-brane is decomposed as
two fractional D3-branes, i.e two {\it half} D3-branes. A fractional D3-brane can be viewed as a D5-brane wrapped
on a ${\mathbb P}^1$ with fluxes \cite{DDG}.
Similarly a fractional D3-brane can also be viewed as a $\overline{\rm D5}$ wrapped
on a ${\mathbb P}^1$ with a slightly different choices of fluxes whose details are explained in subsection
\ref{subsecDDbarTN}.
Therefore when we say that the fractional D3-branes are ${\rm D5}$-$\overline{\rm D5}$ pairs we mean each D5 and
$\overline{\rm D5}$ carry half a unit of D3-brane charge respectively. Another alternative way to express this would be
to arrange the world-volume fluxes in such a way that the wrapped D5-brane carry an {\it integer} D3-brane charge and the
${\overline{\rm D5}}$-brane
carry no world-volume fluxes but cancels the RR charge of the D5-brane. Both the viewpoints lead
to the same results and will be discussed in subsection \ref{subsecDDbarTN}.
See also \cite{mukdas, poln2} for details.}.
Various ways of wrapping the TN 2-cycles will lead to
various different sets of product gauge groups.

The wrapped seven-branes on the TN lead to interesting physics when we switch on {\it time-dependent} gauge
fields on the 2-cycles, in subsection \ref{subsecTimeVary}.  These time-varying gauge fields may lead to chiral anomalies along the TN circle (at
infinity). Additionally, the wrapped ${\rm D5}$-$\overline{\rm D5}$ pairs undergo certain interesting transmutations from the
varying vector fields. These transmutations also effect the tachyons between the ${\rm D5}$-$\overline{\rm D5}$ brane pairs.
We argue that
 the Gaiotto dualities in Model 2
are explained by three simultaneous effects: chiral anomaly cancellations, anti-GSO projections and brane transmutations.
Various conformal cases of the Gaiotto models 
are discussed in subsubsection \ref{ConfGMap}, using the above set of ideas. Interestingly,
under some limiting situation, that we discuss in subsubsection \ref{subsecNetwork}, our model may be dualized to the
brane networks of \cite{bbt}. These mappings serve as consistency checks of our scenario.

In subsubsection \ref{subsecUVIRgrav}, we discuss the gravity duals of these models using our brane anti-brane picture.
Since the UV limit are
described by the probe D3-branes
decomposed into ${\rm D5}$-$\overline{\rm D5}$ pairs wrapped on vanishing 2-cycles of multi Taub-NUT space in our models, the
corresponding gravity duals should capture the
six-dimensional gauge theories. On the other hand, since
the IR remains a
four-dimensional theory, the gravity duals should again capture the underlying physics. 
In subsubsection \ref{subsecUVIRgrav} we give brief derivations in both of
these limits and argue how to see two different scenarios at the UV and IR scales.

Since our constructions are in F-theory, we can probe both the conformal as well as non-conformal scenarios. In
subsubsection \ref{subsecNonConf}, we give various new non-conformal scenarios. In one of the scenario we show how {\it cascading} models
appear naturally from our constructions. It was suspected for sometime that the non-conformal deformations
of the Gaiotto models should show cascading behaviors. 
Here we argue why this is most natural.

In section \ref{model3section} we discuss Model 3.
This is described by seven-branes wrapped on a K3 manifold and probed by D3 and ${\overline {\rm D3}}$ probes whose number may
differ at most by 24. This is an interesting model because the internal space is compact. 
In F-theory,
this is the ${\rm K3} \times {\rm K3}$ four-fold, and so should follow the constraints of \cite{becker, SVW, DRS}. In subsections \ref{FhetSugra}
and \ref{subsecModel3D3IIB}, we argue that
these constructions 
lead to new
${\cal N} = 2$ models with possible dualities to
both type IIB and heterotic theories on non-K\"ahler manifolds, and involving interesting interplays between abelian
instantons and fluxes.

Interestingly, in the limit where the number of brane anti-brane probes becomes very large, we show in subsection \ref{MatrixTheory}, how
M(atrix) theory on ${\rm K3} \times {\rm K3}$ manifold could enter the story. However the M(atrix) description is rather
non-trivial because of the underlying anomaly cancellation condition 
inserts extra $G$-fluxes on the
four-fold.

Under certain choices of fluxes, the supersymmetry could be broken to ${\cal N} =1$. In subsection \ref{subsecModel3N12geomDual}, we speculate on
certain new effects 
 of this. For example we discuss the puzzling question of
how branes on one side of the duality {\it disappear} and are replaced by fluxes. We refer to few 
known
examples in both type IIB as well as $SO(32)$ heterotic theory where this phenomena has been explained. Using these
examples,
we argue therein how to understand the underlying
${\cal N} =1 $ gauge/gravity dualities from the considerations of subsection \ref{FhetSugra}.

In the section \ref{model3section}, we also discuss how certain ${\cal N} = 1$ non-conformal cascading models could naturally appear
from our set-up. The contents of this section should be considered as a continuation of the ideas presented
in subsubsection \ref{subsecNonConf} wherein the cascading ${\cal N} = 2$ models are presented. We show how the ${\cal N} =1$ picture
appears from certain intersections of ALE and ${\mathbb P}^1$ curves in the limit where we impose non-compactness.
A more elaborate set of connections to other kind of cascading models could also appear if more non-trivial
fibrations of the TN space over the compactified $u$-plane are allowed. We speculate on various manifestations of
these ideas.

In section \ref{model4section} we discuss Model 4, which is dynamics of $k$ D3-branes probing intersecting seven-brane backgrounds with $Sp(2k) \times Sp(2k)$ gauge group. In the
literature this class of models \cite{gimpol} have been studied, mostly {\it without} the probe D3-branes.
In subsection \ref{subsecModel4HetFlux5}, we show that once the probes are taken, these constructions could produce
new ${\cal N} =1$ heterotic dual given by $k$ small instantons wrapping 2-cycles of
a non-K\"ahler K3 manifold that is no longer
conformally Calabi-Yau. As far as we know, the non-K\"ahler K3 manifold is a new construction as all previous
models have only considered the conformally K3 case. Also these ${\cal N} = 1$ examples are different from the
examples that we study in \ref{model3section}.
In subsection \ref{model4subsecIIBO7}, we discuss how the F-theory picture looks
like with the non-K\"ahler K3 base. We also analyze the model from M-theory point of view where the
heterotic small instantons become M5-branes.
We then discuss possible constraints on these models coming
from global charge and anomaly cancellations in F-theory and study the implications of these
constraints on the global and local symmetries of
the underlying gauge theories. We end our paper with discussions on how to extend all the four models
to incorporate more interesting gauge theory phenomena.

The appendices are arranged in the following way. In appendix \ref{sp2r01} we derive Seiberg-Witten curve and 1-form for pure $Sp(2r)$ theory, starting from those with matter. In appendix \ref{fantasy} we fantasize on how to save the naive 1-1 mapping proposed
in subsubsection \ref{naiveMap}, circumventing the objection given in subsubsection \ref{towardMap}.
In appendix \ref{higgsing01} we study various higgsings of $Sp(2r)$.
In appendix
\ref{modular01}, we explicitly compute the one-instanton contributions to $\tau$ from the expressions for prepotential given in \cite{dhoker}, for
$Sp(2r)$ curve with flavors where at least two of them are massless.
In appendix \ref{tachyonappendix} we discuss how tachyons can be made massless by choosing appropriate background
fluxes. We do a simple analysis for flat backgrounds. Finally, in appendix \ref{integrab}, we discuss the integrability
of various ${\cal N} =2 $ models studied here, and speculate issues related to ${\cal N} = 1$ integrability.


\subsection{What are the new results in this paper?}
In this paper we will be investigating a large class of models (shown in {\bf table \ref{4ModelTable}}),
so for the benefit of the reader, let us point out what we
consider new results that have come out from our work.
\begin{table}[htb]
 \begin{center}
\begin{tabular}{|c|c|c|c|}\hline
Model & Type IIB on & F-theory on & SUSY \\ \hline
1 &  $ \left( {{\bf T}^2\over \Omega\cdot (-1)^{F_L}\cdot {\cal I}_{45}}\right)
\times {{\mathbb R}^4} \times {\mathbb R}^{0123}$ & $ \left({\rm K3} \right)\times {{\mathbb R}^4} \times {\mathbb R}^{0123}$ & ${\cal N}=2$ \\ \hline
2 & $  \left({{\bf T}^2\over \Omega\cdot (-1)^{F_L}\cdot {\cal I}_{45}}\right)
\times {{{\mathbb R}^4}\over {\mathbb Z}_2} \times {\mathbb R}^{0123}$ & $ \left({\rm K3} \right)\times {{{\mathbb R}^4}\over {\mathbb Z}_2} \times {\mathbb R}^{0123}$ & ${\cal N}=2$  \\ \hline
3 &$ \left({{\bf T}^2\over \Omega\cdot (-1)^{F_L}\cdot {\cal I}_{45}} \right)\times {\rm K3} \times {\mathbb R}^{0123}$   &
$ \left({\rm K3}\right) \times {\rm K3} \times {\mathbb R}^{0123}$  & ${\cal N}= 2, 1$   \\ \hline
4 & $ \left({{\bf T}^2\over \Omega\cdot (-1)^{F_L}\cdot {\cal I}_{45}} \times
 {{\bf T}^2\over \Omega\cdot (-1)^{F_L}\cdot {\cal I}_{89}} \right) \times {\mathbb R}^2 \times {\mathbb R}^{0123}$ &
$\left({\rm CY}_3\right) \times {\mathbb R}^2 \times {\mathbb R}^{0123}$ & ${\cal N}=1$   \\ \hline
  \end{tabular}
\end{center}
  \caption{The four different models, probed in each cases by multiple D3-branes, that we will be studying in this paper.
All the four models have new interesting physics hitherto unexplored that will be the subject of this paper. The
${\rm CY}_3$ defined above in the last row is typically of the form ${\bf T}^2 \ltimes ({\mathbb P}^1 \times {\mathbb P}^1)$. Similarly for each row, a first factor in $\left( ~ \right)$ on the right column is an elliptic fibration over the first factor in $\left( ~ \right)$ on the left column. For Models 1,2, and 3, the directions for each space is $x^{4,5}, x^{6,7,8,9},x^{0,1,2,3}$, while for Model 4 it is given as  $x^{4,5}, x^{8,9},x^{6,7},x^{0,1,2,3}$.}
  \label{4ModelTable}
\end{table}

\noindent We start with Model 1 which is
multiple D3-branes probing F-theory on K3. For this model we present:
\begin{itemize}
\item Detailed study of curves and singularity structure, and including maximal Argyres-Douglas points for pure $Sp(2r)$ gauge group.
\item An attempt to construct a mapping (homeomorphism, if possible) between moduli space of brane dynamics and that of pure Seiberg-Witten curve for the multi-probe case. The arguments based on combinatorics and dimensionality narrow down the possible choices, but perhaps too much. We discuss several puzzles related to this mapping and speculate ways to resolve them.
\end{itemize}

\noindent Model 2 is physics of multiple D3-branes probing F-theory on ${\rm K3} \times {\rm TN}_n$ with the
seven-branes wrapped on the multi Taub-NUT spaces i.e ${\rm TN}_n$ spaces. For this model we present:
\begin{itemize}
\item An alternative way to study Gaiotto-type models using fractional branes to probe the F-theory
geometry with multi Taub-NUT spaces.
\item A study of a class of Gaiotto dualities via chiral anomaly cancellations, anti-GSO projections and
brane transmutations in a ${\rm TN}_n$ background with wrapped seven-branes.
\item A study of the ultra-violet and infra-red geometries using branes anti-branes system, including an
analysis of the holographic duals.
\item A study of many conformal examples in Gaiotto models and a map to the brane network picture. Explicit
examples of 
dualities in these models.
\item A study of many new non-conformal models directly from F-theory, including an interesting
example of non-conformal cascading ${\cal N} = 2$ theories. Examples of the mapping to cascading ${\cal N} = 1$ theories.
\end{itemize}

\noindent Model 3 is dynamics of D3 anti-D3 brane pairs probing F-theory on ${\rm K3} \times {\rm K3}$ with
appropriate background fluxes to cancel both the anomalies as well as the tachyons. For this model we present:
\begin{itemize}
\item Supergravity solution for brane anti-brane configurations probing ${\rm K3} \times {\rm K3}$ in F-theory
with $G$-fluxes.
Examples of anomaly-free consistent backgrounds.
\item The idea of the duality between abelian instantons and $G$-fluxes in M-theory. Examples of models
that manifest this duality.
\item A connection to a non-trivial M(atrix) theory on ${\rm K3} \times {\rm K3}$ with fluxes.
\end{itemize}

\noindent Model 4 is physics of multiple D3-branes probing intersecting seven-branes and seven-planes
background or F-theory on a certain Calabi-Yau three-fold. For models in this category, we present:
\begin{itemize}
\item A dual map to the heterotic theory on a non-K\"ahler K3 manifold that is not a
conformally Calabi-Yau manifold.
\item New examples of type IIB and M-theory compactifications on non-K\"ahler manifolds.
\end{itemize}
Plus a few short results, unlisted here, are scattered throughout the text. Note that the paper is arranged in such a way that 
 a reader interested in a particular model can skip ahead and go to the relevant section.

\newpage

\addtocontents{toc}{\protect\setcounter{tocdepth}{3}}
\section[Model 1: Multiple D3's probing orientifold background]{Model 1: Multiple D3-branes probing orientifold background \label{model1section}}

After having constructed all the necessary background details in subsection \ref{subsecF1D3}, 
the simplest generalization is to incorporate multiple D3-branes as probes for the same orientifold geometry. This picture
has been discussed first by \cite{dougschwa} where they used the spectral curve construction to analyze the geometry. 
When we have $r$ D3 branes
probing the O7 geometry, we obtain $Sp(2r)$ gauge theory \cite{dougschwa}, which is captured by $Sp(2r)$ SW theory with an antisymmetric matter \cite{AMP}. 
Here
we will instead study the Seiberg-Witten curve and singularity structure for pure $Sp(2r)$ theory to discuss the
implications for the F-theoretic geometry, {\it in the limit when the antisymmetric matter is very heavy}. This way the subtleties due to the 
addition of anti-symmetric matter will not influence our analysis. The full analysis with anti-symmetric matter will be dealt elsewhere.

Here we start with the main idea that massless BPS states are captured by vanishing 1-cycles of SW curve. 
As reviewed in \cite{LercheReview}, BPS mass is given by the magnitude of the central charge, and 
the central charge of a BPS state is given by integrating 1-form $\lambda_{\rm SW}$ over 1-cycle $\nu$, i.e: 
\begin{equation} Z= \oint_\nu \lambda_{\rm SW}. \label{Zlambda} \end{equation}
For pure $SU(r+1)$ and $Sp(2r)$ cases, $\lambda_{\rm SW}$ does not blow up near vanishing cycles \cite{SeoUPenn}, and 
vanishing 1-cycle gives massless BPS states. For $SU(r+1)$ theories with flavors,  \cite{GST} shows that this is true only up to a subtlety related to scaling dimensions of various moduli and SW 1-form. See \cite{EHIY} and \cite{EH} for earlier works on scaling behaviour at Argyres-Douglas loci.   %

But before we move ahead,
first
we would like to revisit the Seiberg-Witten curve of the familiar rank 1 $Sp(2) \equiv SU(2)$ case where subtleties like the existence of 
anti-symmetric matter are absent.

\subsection{The rank 1 case revisited: $SU(2)$ versus $Sp(2)$ \label{subsecRank1}}

\begin{figure}[htb]
        \begin{center}
\includegraphics[height=3cm]{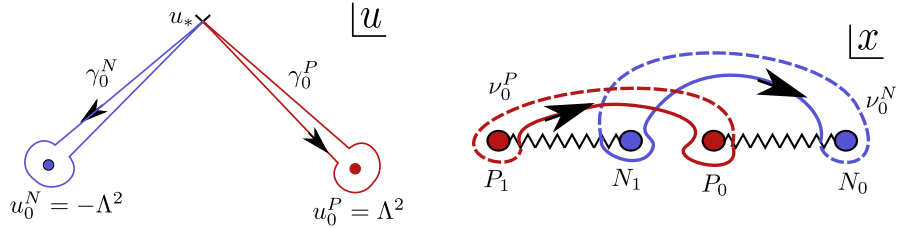}
        \end{center}
        \caption{Vanishing 1-cycles of $SU(2)$ SW curve given in the form of \eqref{su2Lerche}. As we move around on the $u$-plane (in the moduli space), branch points on the $x$-plane move around. If we move around a singular locus in $u$-plane (on a non-contractible loop), the branch points collide with each other and then change with each other.}\label{su2jihye}
        \end{figure}

Our approach for studying this would be as discussed in {\bf figure \ref{su2jihye}}: we move around in a moduli space surrounding a singular locus so that we are on a
non-contractible loop. Once there, we can then observe how the branch points move around.
This can give us all the vanishing cycles, which correspond to massless BPS dyons.

In the literature, Seiberg Witten curves for SU(2) are written in multiple
forms.
In \cite{sw2} the curve is expressed in the following way:
\begin{equation}
y^{2}=(x-u)\left( x(x-u)+\frac{1}{4}\Lambda ^{4}\right)   \label{sp2SW}
\end{equation}%
whose generalization includes $Sp(2r)$ curves given in \cite{ArgyresShapere}.
On the other hand, the alternative form
\begin{equation}
y^{2}=(x^{2}-u)^{2}-\Lambda ^{4}=(x^{2}-{u+\Lambda ^{2}})(x^{2}-{u-\Lambda
^{2}})  \label{su2Lerche}
\end{equation}%
given in \cite{KLYTsimpleADE} has an advantage of immediate generalization into Seiberg-Witten curves with gauge groups in A-D-E series.

We will consider Seiberg-Witten curves for pure $Sp(2)=SU(2)$ theory written in two different forms given in \eqref{sp2SW} and \eqref{su2Lerche}. Let us name those curves $Sp(2)$ and $SU(2)$ curves respectively, given their generalizations into $Sp(2r)$ and $SU(r+1)$ curves. We will discuss their vanishing cycles in detail to demonstrate that
these two curves indeed capture the same physics. More elaborate exposition and higher rank cases for both $Sp$ and $SU$ will be given in \cite{SeoDasgupta}.

\begin{figure}[htb]
        \begin{center}
\includegraphics[height=4cm]{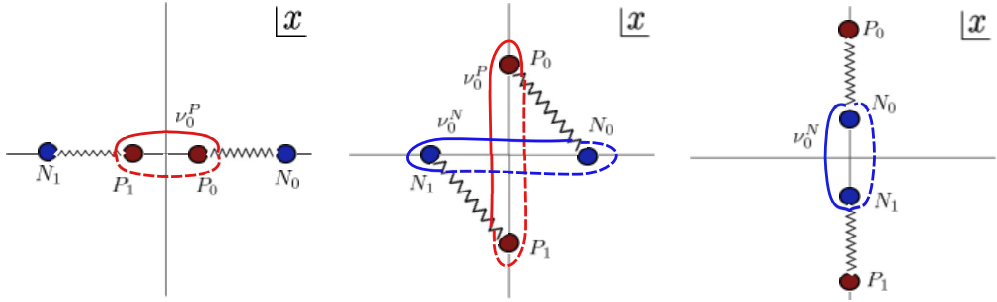}
        \end{center}
        \caption{A clearer view of how the branch points collide in the $x$-plane when
we change the moduli in the $u$-plane for the $SU(2)$ case.}\label{rank1sufig}
        \end{figure}

As explained in \cite{KLYTsimpleADE}, the monodromy of SU(2) curve captures
both massless dyon and monopole. The curve \eqref{su2Lerche} has four branch
points
\begin{equation}
 N_{0,1} = \pm \sqrt{u+\Lambda^2}, \qquad P_{0,1}=\pm \sqrt{u-\Lambda^2}, 
\end{equation} which are all distinct at a generic value of modulus $u$. 
 As we vary $u$, a different pair of branch points will collide at different
singular point in the moduli space: $N_0$ and $N_1$ collide as $u\rightarrow -\Lambda^2$ and $P_0$ and $P_1$ collide as $u\rightarrow  \Lambda^2$  as denoted with blue and red in {\bf figure
\ref{su2jihye}}. Explicitly how this happens is described in {\bf figure
\ref{rank1sufig}}. As we change the moduli of $u$ along the real axis, we see
how branch points move on $x$-plane. From the left, each figure happens at
different values of $u$ as:
\begin{equation}
u\sim -\Lambda ^{2}, \qquad -\Lambda^{2}< u<  \Lambda^{2}, \qquad u\sim  \Lambda ^{2}
\end{equation} respectively.
In the middle of {\bf figure
\ref{rank1sufig}}, all
the branch points are separated, and two cycles are drawn which correspond
to a monopole and a dyon. As we vary the value of $u$ (moving to left and right figures), each cycle vanishes
at a point in the moduli space (all drawn in consistent colors),
corresponding to either a monopole or a dyon becoming massless.

\begin{figure}[htb]
        \begin{center}
\includegraphics[height=6cm]{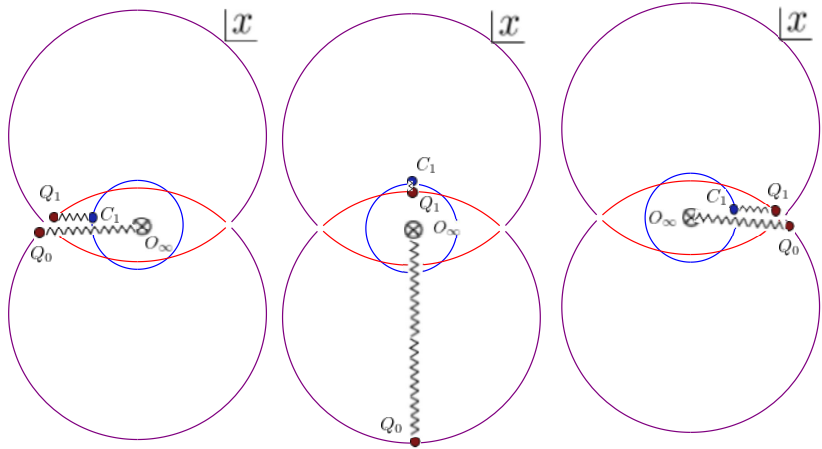}
        \end{center}
        \caption{For the $Sp(2)$ case the branch points moves around once we
change the phase of the moduli $u$. One may see that the behavior is slightly
different from the $SU(2)$ cases plotted earlier. Blue, purple, and red tracks are trajectories of branch points $C_1, \ Q_0$, and $Q_1$ respectively of \eqref{sp1fourpoints}, as we vary phase of $u$ while fixing its magnitude $|u|={\Lambda ^{2}}$. From the left, the values of $u$ are $\Lambda^{2}, \  i\Lambda^{2},  \ -\Lambda^{2}$ respectively. In other words, the phase of $u$ are $0,\ \pi/2, \ \pi$.}\label{rank1spnoaxes}
         \end{figure}

Now, let us examine the monodromy of the Sp(2) curve of \eqref{sp2SW}. We
have total four branch points: three take finite values of $x$ and
one is at infinity. Shift $x$ by $x\rightarrow x+u$ in \eqref{sp2SW} to obtain
\begin{equation}
y^{2}=(x+u)\left( x(x+u)+\frac{1}{4}\Lambda ^{4}\right).   \label{sp2SWshift}
\end{equation}%
Let us bring a branch
point at infinity to the origin, so that it is easier to keep track of how
cycles change. Perform $x\rightarrow 1/2x, y\rightarrow x^2 y$ transformation to obtain
\begin{equation}
y^{2}=x(1+2 u x)\left( 1 +2 ux+ \Lambda ^{4} x^2 \right),   \label{sp2SWshift2}
\end{equation}%
whose four branch points are
\begin{equation}
O_{\infty}=0, \qquad  C_1=-\frac{1}{2u}, \qquad Q_{0,1}=-\frac{1}{\Lambda^2}\left(u\pm\sqrt{-\Lambda^2+u^2}\right).\label{sp1fourpoints}
\end{equation}
At generic value of $u$, all four branch points are separated, but as we vary $u$, they can collide with each other.  {\bf Figure \ref{rank1spnoaxes}} shows how branch points move on $x$-plane under
changing the phase of moduli $u$ while its magnitude is fixed at $|u|={\Lambda ^{2}}$.
Each of three non-zero branch points follows the track with the
corresponding color. Blue, purple, and red tracks are trajectories of branch points $C_1, \ Q_0$, and $Q_1$.

\begin{figure}[htb]
        \begin{center}
\includegraphics[height=6cm]{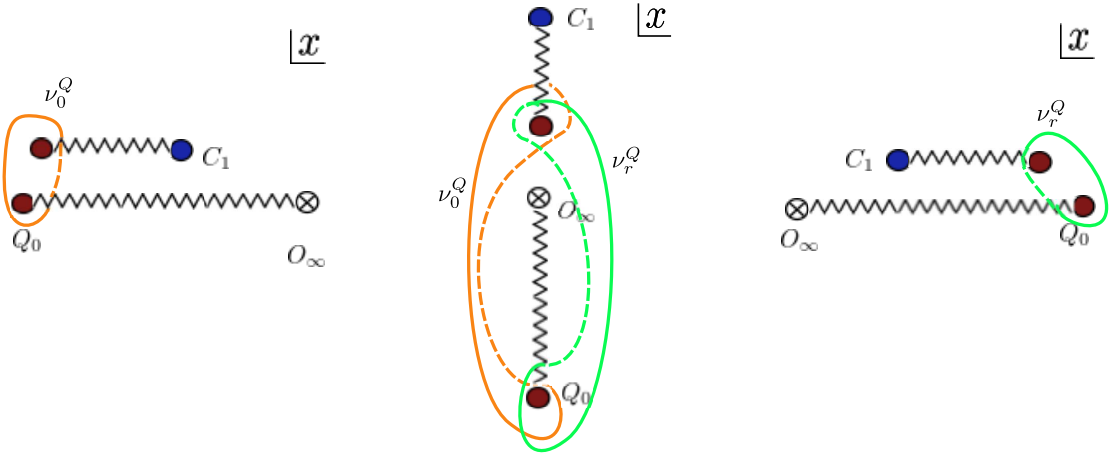}
        \end{center}
        \caption{This is similar to the earlier figure for the $Sp(2)$ case except that we magnify the
various branch points and the branch cuts. One may now easily compare the $Sp(2)$ story with the corresponding
$SU(2)$ case.} \label{rank1sp}
        \end{figure}

{\bf Figure \ref{rank1sp}} shows the magnified view of the
branch points and branch cuts. In the middle figure, we have two 1-cycles in
orange and green, which vanish in left and right figures respectively. As in
$SU(2)$ case in {\bf figure \ref{rank1sufig}}, they again correspond to massless dyon and monopole at appropriate locations in moduli space. Note that
each cycle connects the same pair of branch points, but through different
trajectories.

The simple picture above can be extended to higher rank gauge groups. In the
following we will elaborate the $Sp(2r)$ case and discuss how various non-trivial singularities in the moduli
space could be approached from various different angles. More details on monodromies
around branch points etc. for pure $Sp(2r)$ and $SU(r+1)$ will appear soon in \cite{SeoDasgupta}.

\subsection{Singularity structure of pure $Sp(2r)$ curve \label{Sp2rcurveTech}}

Here we will study the singularity structure of the moduli space of the Seiberg-Witten curve for pure $Sp(2r)$ theory. What one is most interested in would be the singularity of the Seiberg-Witten theory itself, which knows about Seiberg-Witten curve and also Seiberg-Witten 1-form. 
 Here we make an argument that the singularity of pure $Sp(2r)$ Seiberg-Witten curve is a subset of singularity of Seiberg-Witten theory. In other words, we can learn much about the singularity of Seiberg-Witten theory, just by looking at Seiberg-Witten curve.

As in \eqref{Zlambda}, BPS mass is given by integrating 1-form $\lambda_{\rm SW}$ over 1-cycle.
 In this section, we will look at various vanishing 1-cycles: as long as 1-form is not a delta function near the vanishing 1-cycles, 1-form will integrates to zero. We observe that SW 1-form vanishes near the vanishing 1-cycles of $Sp(2r)$ Seiberg-Witten curve. Therefore, all the singularity we observe from the Seiberg-Witten curve will indeed be true singularity of Seiberg-Witten theory.

On the other hand, it is not forbidden to integrate 1-form over non-vanishing 1-cycle only to get zero value for the integration.  In that case, Seiberg-Witten theory will contain massless BPS state, which is not captured by vanishing 1-cycle of Seiberg-Witten curve. This will be highly non-generic, but currently we do not know whether this is prohibited either. So we will keep this as an option.

Starting from the Seiberg-Witten curve in a hyper-elliptic form, we will analyze the singularity structure, whose simplest kind is captured by vanishing discriminant.
For a given hyperelliptic curve $y^2=f_n(x)$ with
$f_n(x)$ a degree-$n$ polynomial in $x$
\begin{equation}
f_n(x)=\sum_{i=0}^n a_i x^i = a_n \prod_{i=1}^n (x-e_i), \label{polynomialF}
\end{equation}
the singularity is captured by the discriminant, which captures degeneracy of roots and is defined as
\begin{equation}
\Delta_x \left( f_n(x) \right) =  a_n^{2n-2} \prod_{i<j} (e_i-e_j)^2 . \label{DiscDef}
\end{equation}
We will use this to analyze singularity structure of the Seiberg-Witten curves.

The Seiberg-Witten curve
for pure $Sp(2r)$ gauge theories in \eqref{curve3} factorizes into
\bg\label{sprhs12}
y^{2} =f_{Sp(2r)} (x) = f_C f_Q
\nd
with
\bg\label{rhsdef}
{f}_{C}=  \prod_{a=1}^{r} (x-C_a), \quad C_a = \phi_a^2, \quad
{f}_{Q} =  x f_C +16\Lambda ^{2r+2} =\prod_{i=0}^{r} (x-Q_i).
\nd
Note that the $f_C = f_Q=0$ occurs only if $\Lambda=0$. In quantum theories, we demand
$\Lambda \neq 0$ and we learn that ${f}_{C}$ and ${f}_{Q}$ never share roots. The discriminant on the curve effectively factorizes as
\begin{equation}
\Delta_x f_{Sp(2r)} =  (16 \Lambda^{2r+2})^{2r} (\Delta_x f_C )( \Delta_x f_Q ).
\end{equation}

So far, we only considered the Seiberg-Witten curve, ignoring how the Seiberg-Witten 1-form changes in the moduli space of the curve. Considering Seiberg-Witten 1-form might change singularity structure of the moduli space. For example, there can be some delta function behaviour, where we integrate infinitely large 1-form over a vanishing cycle, only to get finite amount. In this case, the singularity will be removed. On the other hand, we might see some new singularity, if integrating 1-form over a non-vanishing cycle gives zero.
It is possible that the SW curve alone might not capture everything, and SW form might bring some drastic changes to the singularity structure. 

As derived in appendix \ref{sp2r01} using \cite{ArgyresShapere}, Seiberg-Witten 1-form for $Sp(2r)$ theory without matter is given by:
\begin{eqnarray}
\lambda  & =& a \frac{dx}{2\sqrt{x}}  \log \left( \frac{   x\prod \left(  x-\phi_{a}^{2}\right) + 8\Lambda^{2r+2} +\sqrt{x} y   }{  x\prod \left(  x-\phi_{a}^{2}\right) + 8\Lambda^{2r+2} -\sqrt{x} y  } \right), \label{OneForm}
\end{eqnarray}
for further convenience, which can be written as,
\begin{eqnarray}
\lambda  & =& a \frac{dx}{2\sqrt{x}}  \log \left( \frac{   x f_C + 8\Lambda^{2r+2} +\sqrt{x} y   }{  x f_C + 8\Lambda^{2r+2} -\sqrt{x} y  } \right)  \label{OneForm1} \\
& =& a \frac{dx}{2\sqrt{x}}  \log \left( \frac{  f_Q - 8\Lambda^{2r+2} +\sqrt{x} y   }{ f_Q -8\Lambda^{2r+2} -\sqrt{x} y  } \right) \nonumber.
\end{eqnarray}
Let us observe how 1-form behaves near a vanishing cycle. First we will restrict in moduli space of the curve, so that we are near vanishing discriminant locus of the curve. Secondly, we restrict further along the Riemann surface, meaning that along the Riemann surface, we go to a neighborhood near a vanishing cycle. More explicitly we take $\Delta=0$ and $y=0$. For given hyperelliptic curve, we will require the RHS to have a double root of $x$, and by going near that region, we require $y=0$, which means $f_C=0$ or $f_Q=0$. Plugging in $y=0,\ f_C=0$ and $y=0, \ f_C=0$ will give $1$ inside the $\log( \ ) $ of the first and the second line of \eqref{OneForm1} respectively. Therefore, we confirm that 1-form vanishes near a vanishing 1-cycle. Therefore, singularity of Seiberg-Witten curve will survive as singularity of Seiberg-Witten theory, with a possibility still remaining that Seiberg-Witten 1-form might add, but not subtract singularity. Singularity of SW curve is a subset of SW theory, and in this paper, we study the former.

At a generic point in $r_{\mathbb{C}}$-dimensional moduli space, $\Delta_x f \ne 0$ holds and all the branch points are separated. As we bring branch points together, we will be reduced to a subspace with lower dimension. It will eat up $1_{\mathbb{C}}$ degree of freedom to bring two branch points together on $x$-plane. Each time we demand a branch point to collide with another, we lose $1_{\mathbb{C}}$ degree of freedom. For example, in a rank-2 case, if we demand two $Q_i$'s to collide with each other, it will happen in a codimension $1_{\mathbb{C}}$ locus in a $2_{\mathbb{C}}$-dimensional moduli space. This loci can be parametrized by $1_{\mathbb{C}}$ number. For example, $f_Q$ will take a following form:
\begin{equation}
f_Q  =x\prod_{a=1}^{2}\left(  x-\phi_{a}^{2}\right)  +16\Lambda^{6}=(x-a)^{2} (x-b)
\end{equation} where $a$ is the repeated root with multiplicity 2.
Demanding all the coefficients match order by order for $x$, we get \begin{equation} b=-\frac{16\Lambda^{6}}{a^{2}}, \qquad
\left\{  \phi_{1}^{2},\phi_{2}^{2}\right\}  =\left\{  \left(  a-\frac
{8\Lambda^{6}}{a^{2}}\right)  \pm \frac{4 \Lambda^3 \sqrt{ (a^3 +4 \Lambda^6)}}{a^2} \right\}.
\end{equation}
The Weyl-invariant moduli $u =- \sum   \phi_a^2$ and $v=\prod
\phi_a^2$ also are expressed in terms of  $1_{\mathbb{C}}$ parameter
\begin{equation}
u=- 2a + \frac{16\Lambda^{6}}{a^{2}}, \qquad
v= a^2 - \frac{32 \Lambda^6}{a}.
\label{SigmaQofSp4}
\end{equation}

\subsubsection{Maximal Argyres-Douglas theories occur at $r+1$ points in ${\cal M}_{f_{Sp(2r)}}$}
 Argyres-Douglas (AD) \cite{ArgyresDouglas}
loci occur where we bring $n \ge 3$ branch points together on $x$-plane, such that the curve takes a cusp form
\bg\label{cuspformj}
y^2 = (x-a)^n \times \cdots.
\nd

Particularly,
maximal AD points ($n=r+1$) occur when we bring all $Q_i$ points together
$Q_i=Q$. This happens at $r+1$ points in moduli space, whose location we compute here.
If we demand all the $r+1$ roots of $f_Q$ to collide ($Q_i=Q$), we will use up all the $r_{\mathbb{C}}$ degrees of freedom in the moduli space, and we will end up at countable number of points in the moduli space. These are maximal Argyres-Douglas points, where the curve develops $A_{r}$ singularity.

\begin{figure}[htb]
   \begin{center}
        \includegraphics[width=3 in]{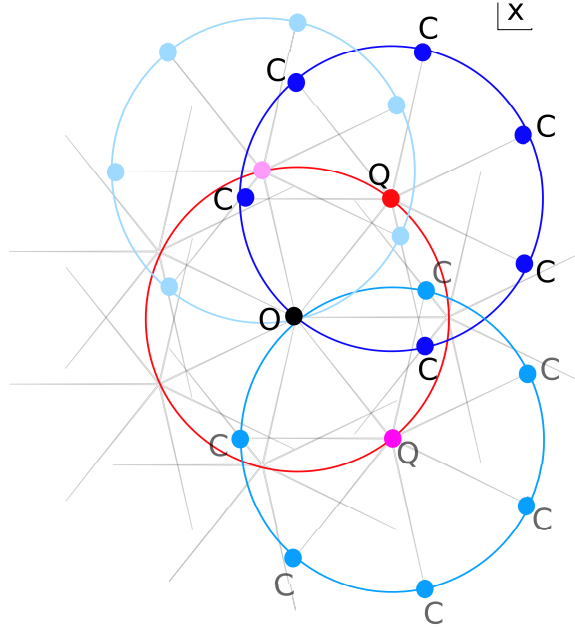}
    \end{center}
    \caption{Location of roots on $x$-plane at maximal Argyres-Douglas points for pure $C_6=Sp(12)$ theory. Note that there are $r+1$ choices of roots. As seen from \eqref{Qvalue}, the magnitude of $Q$ is fixed however, its phase takes one of $r+1$ values (a few of which are drawn in red color, in varying shades). For each choice of $Q$, $C_i$'s will surround the given $Q$ avoiding the origin as in \eqref{Cvalues}. Though physical connection is unclear for the time being, note the resemblance to the {\bf figure 3} in \cite{ADJMwall}.}
    \label{spMaxAD}
   \end{figure}

We obtain solutions with no free parameter satisfying the relation below:
\begin{equation}
f_Q=x\prod_{a=1}^{r}\left(  x-\phi_{a}^{2}\right)  +16\Lambda
^{2r+2}=(x-Q)^{r+1} . \label{fQQ}
\end{equation}
First, demand all the coefficients match order by order for $x$. From the constant piece,
$16\Lambda^{2r+2}=(-Q)^{r+1}$, we get
\begin{equation}
Q=- \exp\left(  \frac{2\pi i}{r+1}k\right)  \left(  16\right)  ^\frac{1}{r+1} \Lambda^{2}, \qquad k\in \mathbb{Z}. \label{Qvalue}
\end{equation}
Note that the phase part gives $(r+1)$ different choices for $Q$, which are all $Z_{r+1}$-symmetric among themselves, by the phase rotation on the complex plane. The physical interpretation of this phase rotation symmetry of the SW curve is unclear to us, it would be interesting to see whether this will affect the allowed orders of instantons just as in $SU(2)$ Seiberg-Witten case. Recall that $Sp(2)$ SW theory had instanton correction in every 4th order, and do we expect that $Sp(2r)$ SW theory will have instanton correction in every 2(r+1)th order due to the $Z_{r+1}$ phase rotation symmetry of the SW curve? 
See {\bf figure \ref{spMaxAD}} for arrangements of $Q$'s.

Now let us solve for gauge invariant moduli $u_{{i}}$'s in terms of $Q$. From \eqref{fQQ}, we demand that the coefficients of the following
\begin{align}
f_Q & =  x\prod_{a=1}^{r}\left(  x-\phi_{a}^{2}\right)  +16\Lambda^{2r+2} \nonumber \\
&  =x\prod_{a=1}^{r}\left(  x-C_{a}\right)  +(-Q)^{r+1}\nonumber \\
&  =x^{r+1}-\left(  \sum_{a=1}^{r}\phi_{a}^{2}\right)  \ x^{r}+\left(
\sum_{a=1}^{r}\phi_{a}^{2}\phi_{b}^{2}\right)  x^{r-1}+\cdots+x\prod_{a=1}%
^{r}\left(  -\phi_{a}^{2}\right)  +(-Q)^{r+1}%
\end{align}
must match with those of
\begin{equation}
(x-Q)^{r+1}=x^{r+1}-(r+1)Qx^{r}+\cdots=\sum_{k=0}^{r+1}\left(
\begin{array}
[c]{c}%
r+1\\
k
\end{array}
\right)  x^{r+1-k}(-Q)^{k}.
\end{equation}
Therefore, we obtain following solutions for $u_i$'s:%
\begin{align}
u_1 =&\left( - \sum_{a=1}^{r}\phi_{a}^{2}\right)    =-(r+1)Q= \left(
\begin{array}
[c]{c}%
r+1\\
1
\end{array}
\right) (-Q)^{1},\\  u_2=&
\left(  \sum_{a=1}^{r}\phi_{a}^{2}\phi_{b}^{2}\right)     =\left(
\begin{array}
[c]{c}%
r+1\\
2
\end{array}
\right)  Q^{2},\\
&\cdots \nonumber \\
u_i=&   \left(
\begin{array}
[c]{c}%
r+1\\
i
\end{array}
\right)  (-Q)^{i}, \\
&\cdots \nonumber \\
u_r=&\prod_{a=1}^{r}\left(  -\phi_{a}^{2}\right)     =\left(
\begin{array}
[c]{c}%
r+1\\
r
\end{array}
\right)  (-Q)^{r}.
\end{align}
With the value of $Q$ inserted from \eqref{Qvalue}, these relations above locate maximal Argyres-Douglas points in the moduli space. Given that \eqref{Qvalue} allowed for $r+1$ different choices of phase for $Q$, again we have $Z_{r+1}$ symmetric $r+1$ points on the moduli space, where maximal Argyres-Douglas singularity occurs. Again the symmetry refers to the phase rotation, and the physical interpretation of this symmetry is unclear to us. 

Let us now solve for $C_i$'s.
As a simple exercise, recall that if we know symmetric combinations of $a$ and $b$, namely $a+b=-u_{1}, \ ab=u_{2}$ we can solve for $a$ and $b$ in terms of $u_i$'s
from the fact that
$a$ and $b$ are solutions to equation
\begin{equation}
x^{2}-(a+b)x+ab=x^{2}+u_{1}x+u_{2}=0.
\end{equation}
We can do a similar trick, given that all the expressions above are symmetric in
$C_i$'s: \{$C_i$\}'s are all $r$ solutions to the equation
\begin{equation}
\frac{(x-Q)^{r+1}-(-Q)^{r+1}}{x}=\sum_{k=0}^{r}\left(
\begin{array}
[c]{c}%
r+1\\
k
\end{array}
\right)  x^{r-k}(-Q)^{k}=0. \label{symeqn}
\end{equation}
Note that the left hand side is a $r$-degree polynomial in $x$. It is easier to solve a similar looking $(r+1)$-degree polynomial instead, which is \eqref{symeqn} times $x$ on both left and right hand sides, given as:
\begin{equation}
(x-Q)^{r+1}-(-Q)^{r+1}=\sum_{k=0}^{r}\left(
\begin{array}
[c]{c}%
r+1\\
k
\end{array}
\right)  x^{r+1-k}(-Q)^{k}=0, \label{symeqnx}
\end{equation}
which has $(r+1)$ solutions by solving $(x-Q)^{r+1}%
-(-Q)^{r+1}=0$ or equivalently
\begin{equation}(x-Q)^{r+1}=(-Q)^{r+1}.
\end{equation}
Taking $(r+1)$'th root on both hand sides, we can easily solve for $x$ in the following way:
\begin{align}
(x-Q)  &  =\mathrm{exp} \left(  \frac{2\pi i}{r+1}k\right)  (-Q),\\
x  &  =Q\left(  1-\mathrm{exp} \left(  \frac{2\pi i}{r+1}k\right)  \right), \qquad k \in \mathbb{Z}. \label{r1roots}
\end{align}
Expressed in \eqref{r1roots} are $r+1$ roots of \eqref{symeqnx}, which also includes a trivial root $x=0$ when $ k \in (r+1) \mathbb{Z}$. Recalling the relation
\begin{equation}
x \eqref{symeqn}  \rightarrow \eqref{symeqnx},
\end{equation}
the $r$ solutions of \eqref{symeqn} are all the $(r+1)-1=r$ solutions of \eqref{symeqnx} excluding $x=0$. Therefore, solutions of \eqref{symeqn} are
\begin{equation} x    =Q\left(  1-\mathrm{exp} \left(  \frac{2\pi i}{r+1}k\right)  \right), \qquad k \in \mathbb{Z}, \quad  k \notin (r+1) \mathbb{Z}, \label{rroots}
\end{equation} and these are the values of \{$C_i$\}'s.
 The above formula \eqref{r1roots} shows that $r+1$ solutions are
surrounding on a circle with radius $|Q|$ around $x=Q$ point as drawn in {\bf figure \ref{spMaxAD}}. Among all $(r+1)$
solutions $x=0$ solution is excluded for \{$C_i$\}. Therefore, we obtain $r$ different values for $C_i$'s for a given $Q$ as below:
\begin{equation}
\{ C_i \}=
\{ \phi_{a}^{2} \}= \left\{  Q\left(  1-\exp\left(  \frac{2\pi i}{r+1}k\right)  \right)  \right\} , \qquad k \in \mathbb{Z}, \qquad k \notin (r+1) \mathbb{Z}. \label{Cvalues}
\end{equation}
Note that all the $C_i$'s take different values. One can interpret that bringing $Q_i$'s together makes $C_i$'s to repel each other by the quantum distance determined by $\Lambda$: the similar phenomena also happen for two maximal Argyres-Douglas points for $SU(r+1)$ \cite{LercheReview}.

Here we showed that for pure $Sp(2r)$ theory maximal Argyres-Douglas theories occur at $r+1$ points which are all symmetric to one another. For pure $SO(2r+1)$ theory, \cite{Seo} finds $2r-1$ maximal Argyres-Douglas points. Scaling behaviour at maximal Argyres-Douglas points for pure ABCDE SW theory were studied in \cite{EHIY} and \cite{EH}, and there are two such points in moduli space for ADE groups. For B and C, number of maximal Argyres-Douglas points seem to be determined by dual Coxeter number \cite{Seo}, and scaling behaviour is also being studied there.

\subsubsection{Singularity structure of $Sp(4)=C_2$}
\begin{figure}[htb]
\begin{center}
\includegraphics[width=0.8\textwidth]{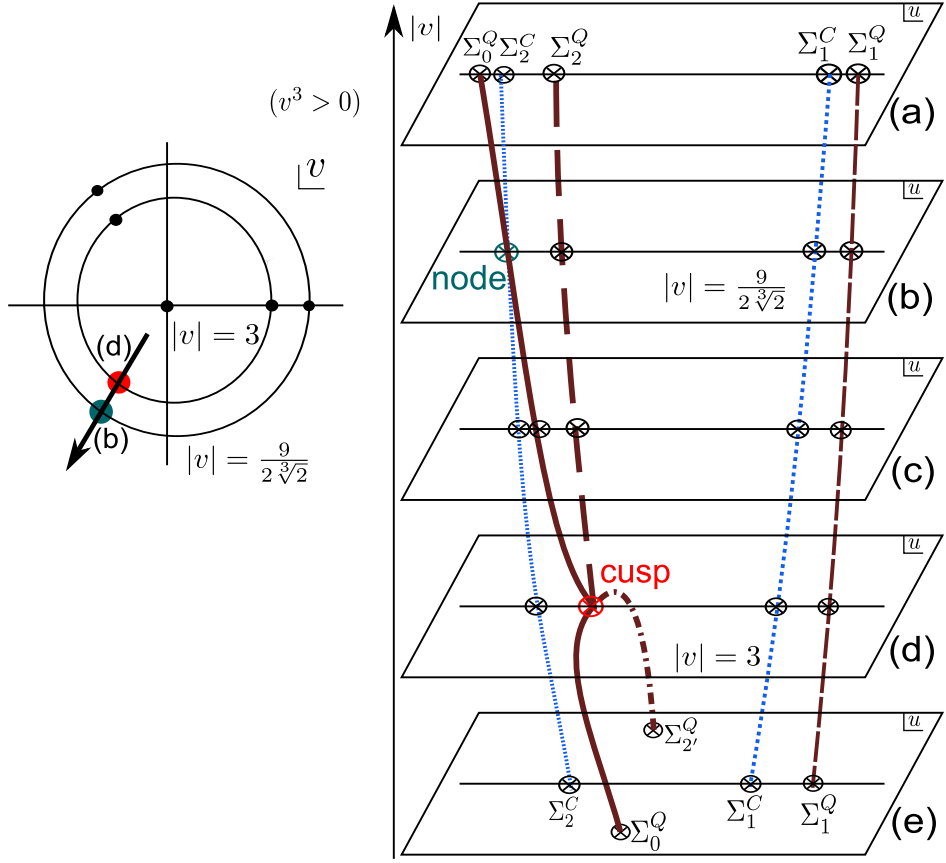}
\end{center}
\caption{A slice of moduli space for pure $Sp(4)$ Seiberg-Witten theory.
At $v=0$, $\Sigma^C_1$ and $\Sigma^C_2$ intersect. More details for the singularity loci of $Sp(4)=C_2=B_2=SO(5)$ theory will be discussed in \cite{SeoDasgupta}
along with with monodromies and singularity structures of pure $SU(r+1)$ and $Sp(2r)$ theories.}
\label{sp4jihye1}
\end{figure}

Moduli space of pure $Sp(4)=C_2$ theory is $2_{\mathbb C} = 4_{\mathbb R}$
 dimensional. Inspired by \cite{KLYTsimpleADE}, let us take a $3_{\mathbb R}$ dimensional slice of the
moduli space for viewing convenience. For example, to reduce one real degree of freedom, we will fix the
phase of $v$, as in the left figure of
{\bf figure \ref{sp4jihye1}}.
Each of the five $u$-planes, marked by (a) through (e) are slices at
different magnitude of $v$. The blue and brown curves denoted by $\Sigma$'s are where we have at
least one massless dyons. These codimension $1_{\mathbb C}$ singular loci are captured by vanishing
discriminant of the curve. When $\Sigma$'s intersect, we have a worse singularity:
massless dyons coexist at these codimension $2_{\mathbb C}$ loci. The SW curve degenerates into either cusp or node form. The shape of intersection loci of $\Sigma$'s also take cusp or node form respectively, each leading to different
kind of singularity (mutually non-local and local). Similar phenomena occur for pure $SU(3)$ theory \cite{KLYTsimpleADE}.

\begin{figure}[htb]
        \begin{center}
\includegraphics[height=16cm]{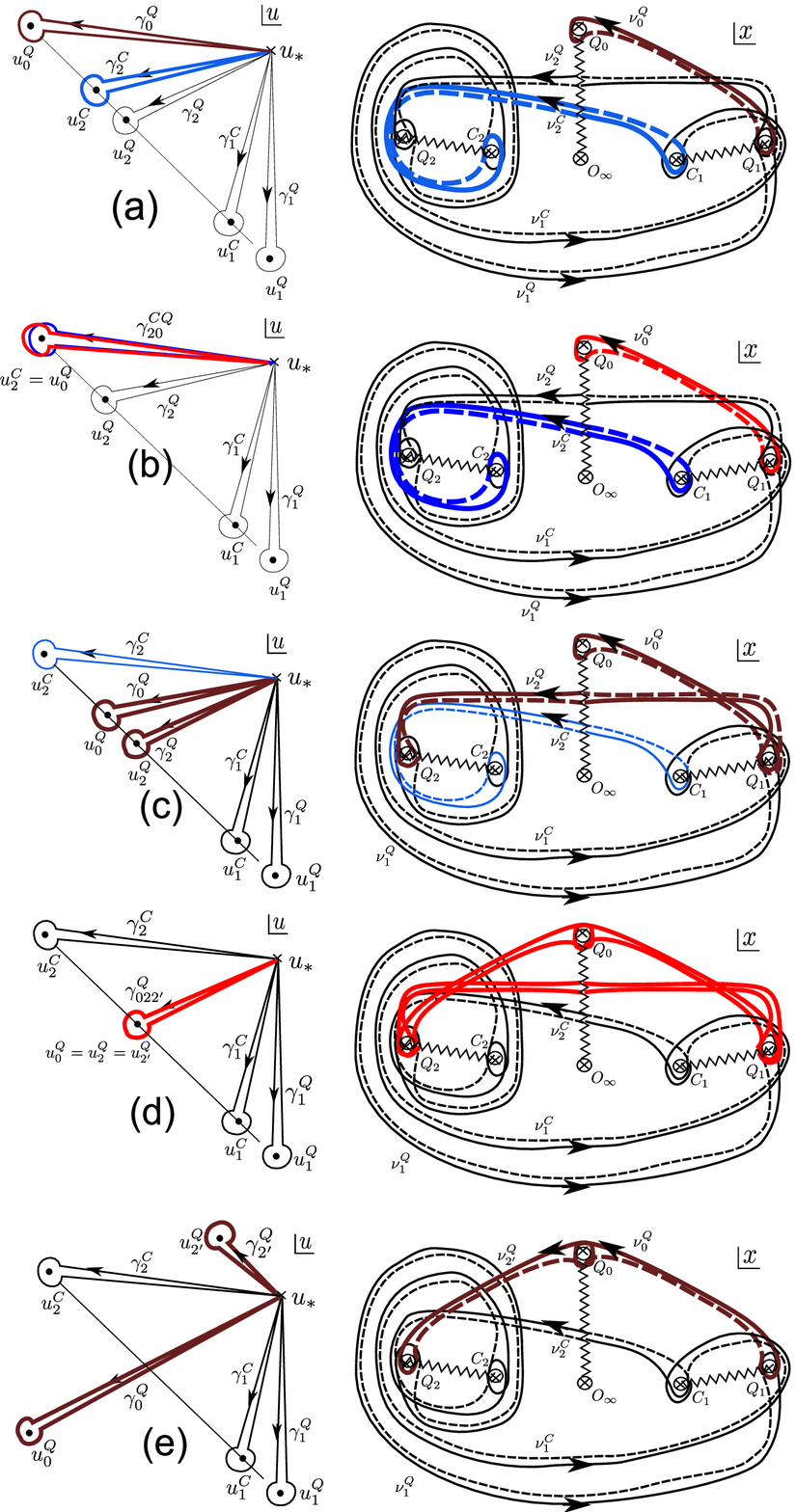}
        \end{center}
        \caption{Singularity structure at slices of moduli space for pure $Sp(4)$ Seiberg-Witten theory. Vanishing cycles will be analyzed in \cite{SeoDasgupta}.}\label{sp4jihye2}
        \end{figure}

For each $u$-plane marked by (a) to (e) of {\bf figure \ref{sp4jihye1}}, we
have drawn corresponding $x$-planes in {\bf figure \ref{sp4jihye2}} displaying vanishing cycles
on $x$-plane, for each slice.  Let us have a closer look. staring from the
top slice marked as (a).

\begin{description}
\item[(a)] At first, all the five singularity are separated on $u$-plane, which translate into 5 vanishing cycles on $x$-plane. Especially, two vanishing cycles $\nu_{0}^{Q}$ (brown) and $\nu_{2}^{C}$ (blue) vanish at two different moduli loci 
 $u_{0}^{Q}$ and $u_{2}^{C}$ respectively.
\item[(b)] Singularity loci $\Sigma_{0}^{Q}$ and $\Sigma_{2}^{C}$ intersect at $|v|=\frac{9}{2\sqrt[3]{3}}$, with node-like
crossing. On $u$-plane, two points $u_{0}^{Q}$ and $u_{2}^{C}$
coincide at this slice of moduli given by $|v|=\frac{9}{2\sqrt[3]{3}}$. The SW curve degenerates into a node-like singularity $
y^{2}=(x-a)^{2}(x-b)^{2}\times \cdots $. Two pairs of points collide with
each other pairwise: it is not Argyres-Douglas form. Instead, the corresponding singularity is such that we have two massless dyons which are mutually local.
\item[(c)] Change the moduli, now that $u_{0}^{Q}$ and $u_{2}^{C}$ are separated. Dyon charges of vanishing cycles di not change as we go through node-like singularity of (b).
Note that two reds $u_{0}^{Q}$ and $u_{2^{\prime }}^{Q}$
are separated but they are running toward each other.  
\item[(d)] As we change the magnitude of moduli $v$, $\Sigma _{0}^{Q}$ and $%
\Sigma _{2}^{Q}$ intersect tangentially at $|v|=3$. Two points $u_{0}^{Q}$ and $%
u_{2}^{Q}$ on $u$-plane collide, and the vanishing cycles $\nu
_{0}^{Q}$ and $\nu _{2}^{Q}$ merge. In other
words, three points collide at the same time on $x$-plane. The curve
degenerates into cusp form $y^{2}\sim (x-a)^{3}\times \cdots $, giving
Argyres-Douglas theory with $SU(2)$ singularity, with two mutually non-local massless BPS dyons.
\item[(e)] Now $u_{0}^{Q}$ and $u_{2}^{C}$ are separated. However note that the BPS dyon charges of vanishing cycles changed as we go through cusp-like (or Argyres-Douglas) singularity of (d). Instead of $\nu_2^Q$, now $\nu_2^{Q^\prime}$ is the new vanishing 1-cycle.
\end{description}

\subsection{Multiple D3-branes probing orientifold geometry \label{subsecMapM}}

From the singular behavior of the Seiberg-Witten curves of $Sp(2r)$ theories, we can ask how they could
be realized from our F-theory set-up. This is where we face the following puzzle. In the $Sp(2)$ case we know that
the gauge theory monopole and dyon points map to the two ($p, q$) seven-branes that appear from the splitting of the
orientifold plane. In the $Sp(2r)$ case, as we saw from the above subsection \ref{Sp2rcurveTech},
there are multiple singularity loci with various dimensions, including $r+1$ maximal Argyres-Douglas points. How are these seen from the F-theory viewpoint? Furthermore in F-theory
we only have an access to the $x^4 + ix^5 \equiv u$ plane, whereas the $Sp(2r)$ case has a $2r_{\mathbb R}= r_{\mathbb C}$ dimensional Coulomb
branch. How are the informations on the multi-dimensional Coulomb branch projected to two-dimensional $u$-plane
in F-theory probed by $r$ D3-branes?

All the above points need to be first answered before we go about analyzing the dualities. The type IIB (or the
F-theory) background probed by multiple D3-branes is generically given by:
\bg\label{joba}
&&{\rm Type~ IIB ~on} ~~{{\bf T}^2\over \Omega\cdot (-1)^{F_L}\cdot {\cal I}_{45}}
\times {{\mathbb R}^4} \times {\mathbb R}^{0123}
~ = ~  \nonumber \\
&&{\rm F~ Theory ~on}~~ {\rm K3} \times {{\mathbb R}^4} \times {\mathbb R}^{0123}
\nd
where the type IIB seven-branes wrap ${{\mathbb R}^4} \times {\mathbb R}^{0123}$ and the probe D3-branes are oriented
along
the Minkowski ${\mathbb R}^{0123}$ directions. In the following we will however only consider the local version of the
model \eqref{joba} i.e consider ${\mathbb R}^2$ instead of ${\bf T}^2$. This means that all the extra massive
charged states associated with ${\mathbb R}^2 \to {\bf T}^2$ are integrated out.

Our conjecture now is that even for the multiple D3-brane probes, on the
$u = x^4+ix^5$ plane the orientifold seven-plane again decompose into {\it two} monopole/dyon points. Thus although
from the Seiberg-Witten curve we expect various singularity loci in a higher dimensional Coulomb branch, in F-theory all these
dynamics are captured by the orientifold plane splitting into {\it two} distinct ($p, q$) seven-branes.

This conjecture of the orientifold plane splitting into two distinct ($p, q$) seven-branes can easily be argued
intuitively. When there is a single probe D3-brane the splitting is well registered by many authors \cite{senF,bds}.
Once we increase the number of D3-branes probes and move the probes away from the O7-plane, why would the dynamics
of the O7-plane be affected by the presence of probe branes? A more formal proof, by projecting the various singularity loci from a higher dimensional Coulomb branch to the two-dimensional $u$-plane, will be discussed elsewhere. Here
we will suffice with our intuitive understanding of the issue.

We would like to find a mapping between the moduli space of the Seiberg- Witten curve for pure
$Sp(2r)$ theory $ {\cal M}_{f_{Sp(2r)}}$ and that of the D3-brane dynamics ${\cal M}_{r{\mathrm D3}}$ given
 by location of $r$ D3-branes on $u$-plane. Both moduli spaces are $r_{\mathbb C}$-dimensional, and at this point we do not have reasons to believe that one theory contains more information that the other. Therefore we hope to find a 1-1 mapping between these two moduli spaces, which are also homeomorphism, so that a neighborhood will be mapped into a neighborhood on the other side, i.e:
 \begin{equation}
{\cal B} : {\cal M}_{f_{Sp(2r)}} \rightarrow {\cal M}_{r {\mathrm D3}},\qquad {\cal C} :  {\cal M}_{r {\mathrm D3}} \rightarrow    {\cal M}_{f_{Sp(2r)}}.
\end{equation}
In other words, we want both ${\cal B}, {\cal C}$ to be homeomorphism, which are inverses of each other. (Here the naming is such that ${\cal B}$ stands for brane dynamics, and ${\cal C}$ stands for the curve).

Most rigorous way to build this mapping would be computing the BPS masses on either side and finding a map at a generic point in moduli space. This is considerably harder to evaluate,
therefore, as a first step, we will study various singular loci with massless states, and see how the singular loci map with each other, as a group (rather than as a point).

Two helpful guiding principles are consistency is dimension (number of degrees of freedom) and combinatorics. We prefer ${\cal B}, {\cal C}$ to preserve the correct combinatorics and dimensionality.

   Using the fact that we have O7-plane splitting into two and that branch points are grouped into two, one might naively suggest to have monopole point to be responsible for the degeneration of $f_Q$ and the dyon for the $f_C$ for example. However, this doesn't work very well, because between $f_C$ and $f_Q$ there is not a clean $Z_2$ symmetry while there is a clear $Z_2$ symmetry between monopole and dyon; more over, all the maximal Argyres-Douglas points come from having $f_Q$ to be maximally degenerate.

  Alternatively, we could get a hint from the fact that locations of monopole and dyon points and $f_Q$ depend on $\Lambda$, while $f_C$ has no dependence on $\Lambda$. We propose the following scenario:
  \begin{itemize}
  \item D3-branes at monopole and dyon location are responsible for the degeneracy of $f_Q$, while the degeneracy of $f_C$ can be captured by how we are distributing $(r-i-j)$ D3-branes elsewhere and $u=\infty$
  \item If we have D3-branes on top of each other, then we will have singularity enhancement for $f_C$.
  \end{itemize}
This preserves homeomorphicity, and works very well for various intersection loci.

\begin{figure}[htb]
        \begin{center}
\includegraphics[height=4cm]{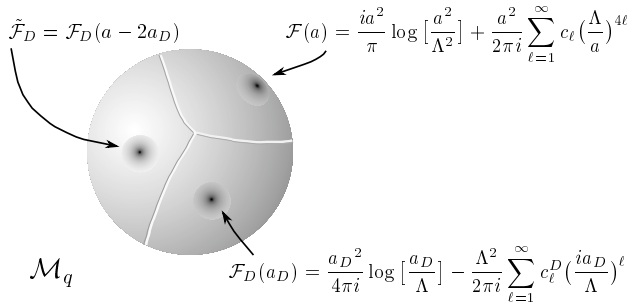}        \end{center}
        \caption{Three special points on the moduli space of pure $SU(2)$ Seiberg-Witten theory. The top right point corresponds to singularity at $u \rightarrow \infty$, and the given expression for prepotential ${\cal F}$ is valid for large $u$. The bottom right point corresponds to a magnetic monopole point at $u=\Lambda^2$, and the left point corresponds to a dyon point at $u=-\Lambda^2$.
This figure is again taken from Lerche's review article \cite{LercheReview}.}
\label{lercheMq}\end{figure}

\begin{table}[htb]
 \begin{center}
\begin{tabular}{|c|c|c|c|c|}\hline \multicolumn{4}{|c|}{${\cal M}_{\rm D3}$} & {${\cal M}_{f_{Sp(2)}}$} \\ \cline{1-4}
monopole & dyon & $\infty$ & elsewhere & \\ \hline
1&0&0&0& $\Sigma^Q_i$ for $i=0,1$ \\
0&1&0&0& \\ \hline
0&0&1 &0& $M_{\infty}$ (not related to a vanishing cycle) \\ \hline
0&0&0 &1& generic location in moduli space with $\Delta_x f \ne0$ \\ \hline
  \end{tabular}
\end{center}
  \caption{One D3-brane probing the quantum corrected orientifold background. First four columns denote location of D3-branes (different locus in $ {\cal M}_{\mathrm D3}$). Its image by ${\cal C}$ map is given in the last column as a locus in $  {\cal M}_{f_{Sp(2)}}$.}
  \label{sp2table}
\end{table}

Rank 1 case maps smoothly: dynamics of a single D3-brane versus Seiberg-Witten curve with pure $Sp(2)$ gauge group. Recall from {\bf figure \ref{rank1sp}} that two vanishing cycles for rank 1 case come from colliding two $Q_i$ points along two different trajectories. In some sense, monopole and dyon point only cared about location of $Q_i$ points. When D3-brane is infinitely far away from the origin, there is also a monodromy associated with infinity given as $M_\infty$
(see for example the review article of Lerche \cite{LercheReview}). Mappings between moduli spaces are summarized in {\bf table \ref{sp2table}} below for the rank 1 case.
Note that $f_C$ does not know about $\Lambda$, so origin or infinity on $u$-plane are the only special location for D3-branes to enhance singularity of $f_C$. Getting some hints from {\bf figure \ref{lercheMq}}, we distinguish three points on the $u$-plane. We claim that singularity coming from $f_Q$ is related to locating D3-branes at $u=\pm \Lambda^2$, while the singularity of $f_C$ is partially captured by D3-branes at $u=\infty$.

Rank 1 case presented in {\bf table \ref{sp2table}} suggests that D3-brane probe treats some locations on $u$-plane distinctively. We assume that multiple D3-brane probes will also single out these locations, and introduce the following notation
\begin{equation}
[ r_m, r_d ; r_{\infty}, \{ r_1, r_2, \cdots \}  ], \label{notationM}
\end{equation}
where $r_m, r_d, r_{\infty}$ denote number of D3-branes located at monopole, dyon, and
at $u = \infty$ points on the $u$-plane, and $r_i$'s denote groupings of leftover D3-branes put elsewhere. We put `;' inside \eqref{notationM} in order to divide $f_Q$ and $f_C$ effects: numbers on the left of `;' affects singularity of $f_Q$, and those on the right of `;' affects singularity of $f_C$. We may omit some of the elements when it is unambiguous (for a given rank, when all the rest is zero).

The success of the 1-1 mapping for the rank 1 case might wrongly suggest that this could be straightforwardly
extended to higher rank cases. The situation at hand is more subtle, part of the reason being the existence of anti-symmetric matter which was absent for 
the rank 1 case. 
To see this, the reader may look up our discussion of a  
 {\it naive} way of 1-1 mapping between moduli space of multiple D3-branes and that of Seiberg-Witten curve in appendix \ref{naiveMap} 
taking the anti-symmetric matter to be so heavy that it can be integrated out from the dynamics at low energies. 
The argument will be  based on dimensionality of various singular loci and combinatorics of point-like loci. However, we will see that it conflicts with known properties of D3-brane dynamics - namely it will not give correct Argyres-Douglas loci. Therefore, in \ref{towardMap} below, we will give its modified version and discuss a partial mapping between these moduli spaces (again keeping the anti-symmetric matter to be very heavy)
which does not seem to be 1-1. We will present a few puzzles with this new mapping - in terms of dimensionality, combinatorics, and division between $f_Q$ and $f_C$. We will speculate various ways to resolve these puzzles, and end with a 
fantasy about under what condition the naive picture given in \ref{naiveMap} could have worked.


\subsubsection{Towards (1-1) mapping between moduli spaces \label{towardMap}}

Let us start by testing the naive mapping given in appendix \ref{naiveMap} by considering correct D3-brane dynamics and
in the process finding some mismatch arising from non-locality of
various states\footnote{The discussion in this subsection was essentially
clarified to us by Ashoke Sen. We thank him for numerous useful correspondences.
We also thank Olivier DeWolfe and Jeff Harvey for useful comments.}. We will again begin by assuming that the anti-symmetric matter is heavy so that 
it does not influence the dynamics of the system. The effect of the light anti-symmetric matter on the 1-1 mapping will be dealt elsewhere. 

We make necessary modifications in {\bf tables \ref{sp4table}} and {\bf \ref{sp2rtable}}, however they suffer from new puzzles. We speculate on various ways to resolve the puzzles and also fantasize over a possibility where a naive map could survive physical tests.

When the open string ends on D3-branes, it is charged with respect to the $U(1)$'s carried by the D3-branes. We use a vector notation for charges carried by the open string, putting charges with respect to different D3's in different elements in the charge vector.
\begin{equation}
 \big( (p_1, p_2, p_3, \cdots, p_k), (q_1, q_2, q_3, \cdots, q_k) \big)
\end{equation}
where $p_i, q_i$ are electric and magnetic U(1) charges with respect to $i$'th D3-brane in the system
and $k$ is the number of D3-branes in the system. 

The intersection number of charge vectors is given as
\begin{eqnarray}
 &&\big( (p_1, p_2, p_3, \cdots, p_k), (q_1, q_2, q_3, \cdots, q_k) \big) \cap  \big( (p_1^\prime, p_2^\prime, p_3^\prime, \cdots, p_k^\prime), (q_1^\prime, q_2^\prime, q_3^\prime, \cdots, q_k^\prime) \big) \nonumber \\
 &&~~~~~~= q_i p_i^\prime - p_i q_i^\prime.
\end{eqnarray}
When this number is (non)zero, two states are said to be mutually (non)local.

Let us consider a few cases in rank 2 case, and analyze open strings states in the brane picture.
Massive states might play some role, but let us restrict our attention to massless states in the paper.
There might be 3-junction strings as well, with three ends on 3- or 7-branes. However, if we only consider massless states, 3-junction string will have two or more ends on D3-branes and the rest on 7-brane - and all these D3-branes and possible 7-brane must be on top of each other, in order to keep the 3-junction string massless. Because 7-branes are separated by a non-zero distance ($2\Lambda^2$) on $u$-plane, we expect that 3-junction string with two or more ends on 7-brane to be massive\footnote{We will ignore states like 3-3-7 with fluxes on the 7-brane, and assume them to be heavy. These exotic states might 
contribute to the anti-symmetric matter, and therefore speculations about their contributions to the 1-1 mapping will be delegated for future work.\label{footnote337}}. 

When we have multiple D3's on top of $[p,q]$-seven brane, (excluding 3-junction string for now), we will have massless strings in 3-3 and 3-7 sectors.
On $[p,q]$-seven brane, only $(p,q)$ string may end\footnote{By $[p,q]7$ brane, we mean a 7-brane where $(p,q)$ string can end. This is a different notation from \cite{bbt}, where $[p,q]7$ brane means a 7-brane where $(p,q)5$ brane can end.
For example, (0,1) string or D1-brane may end on (1,0)5-brane or NS5-brane and also on $[0,1]7$-brane, while (1,0) string or F1 may end on (0,1)5-brane or D5-brane and also on $[1,0]7$-brane or D7-brane.}. Therefore, in 3-7 sector, the open string can be only $(p,q)$ type, and with one end on $[p,q]$-seven brane and another end on one of the multiple D3-branes.

For simplicity, let us consider a system with only 3 D3-branes, which are on top of $[p,q]$-7 brane.
Massless open strings are presented in {\bf table \ref{massless}}.

\begin{table}[h!]
 \begin{center}
\begin{tabular}{|c|cccccc|c|}\hline
Sector & $p_1$ & $p_2$ & $p_3$ &  $q_1$ & $q_2$ & $q_3$   & non-local wrt some states in \\ \hline
$3_1$-$3_2$ & $\big( (p_{12}\big.$, & $-p_{12}$, & $0)$, & $(q_{12}$ , & $ -q_{12}$ , & $\big.0) \big)$ & $3_i$-$3_j$'s, $3_1$-7, $3_2$-7\\
$3_1$-$3_3$ & $\big( (p_{13}\big.$, & $0$,& $-p_{13})$ ,& $(q_{13}$, & $0$, & $  \big.-q_{13}) \big)$ &  $3_i$-$3_j$'s, $3_1$-7, $3_3$-7\\
$3_2$-$3_3$ & $\big( (0\big.$, & $ p_{23}$, & $ -p_{23})$, & $ (0$, & $ q_{23}$, & $  \big.-q_{23}) \big)$ & $3_i$-$3_j$'s, $3_2$-7, $3_3$-7\\  \hline
$3_1$-$7_{[p,q]}$ & $\big( (p\big.$, & $ 0$, & $ 0)$, & $ (q$, & $ 0$, & $  \big.0) \big)$ & $3_1$-$3_2$, $3_1$-$3_3$\\
$3_2$-$7_{[p,q]}$ & $\big( (0\big.$, & $ p$, & $ 0)$, & $ (0$, & $ q$, & $  \big.0) \big)$ & $3_1$-$3_2$, $3_2$-$3_3$ \\
$3_3$-$7_{[p,q]}$ & $\big( (0\big.$, & $ 0$, & $ p)$, & $ (0$, & $ 0$, & $  \big.q) \big)$ & $3_1$-$3_3$, $3_2$-$3_3$\\ \hline
  \end{tabular}
\end{center}
\caption{While $p, q$ are determined by the type of 7-brane, $p_{ij}, q_{ij} \in {\mathbb Z} $ are not fixed - between a pair of D3-branes, any type of string may end at the same time. Note the sign choices in front of $p_{ij}, q_{ij}$, which denotes the orientation of the open string. In the second column, we denote the electric and magnetic charges with respect to $U(1)$'s of D3-branes.}
\label{massless}
\end{table}

Note that $3_i$-$3_j$ sector allows various $p_{ij}, q_{ij}$ choices, and they are mutually non-local. Between 2 D3-branes, any types of $(p,q)$ strings may end at the same time.
 Within one sector, we already have infinite number of massless states which are mutually non-local to each
other\footnote{At this point, the concept of {\it state} breaks down.}.
Such a phenomenon does not occur in 3-7 sector, because given type of 7-brane restricts values of $p, q$ charges: namely only $(p,q)$ string may end on $[p,q]$ seven brane.

Since there easily occur infinite number of massless states which are mutually non-local, we will be counting number of sectors (which are finite) instead of number of massless states (which are infinite but countable).
When we put $n$ D3-branes near 7-brane,
mutually non-local states occur \begin{itemize}
\item within a $3_i$-$3_j$ sector: $ \left(
\begin{array}
[c]{c}%
n\\
2
\end{array}
\right)= \frac{n(n-1)}{2}$,
\item between $3_i$-$3_j$ and $3_i$-$3_k$ sectors: $ 3 \left(
\begin{array}
[c]{c}%
n\\
3
\end{array}
\right)= \frac{n(n-1)(n-2)}{2}$,
\item and between $3_i$-$3_j$ and $3_i$-7 sectors: $ 2 \left(
\begin{array}
[c]{c}%
n\\
2
\end{array}
\right)= {n(n-1)}$,
\end{itemize}
where after `:', we note number of ways to choose such (pairs of) sectors\footnote{Since infinite number of states become massless, it will make more sense to count open string sectors, rather than counting states in each sector.}.

The counting above explains that putting a single brane near 7-brane will not give mutually non-local massless states (the combination vanishes for $n=1$). We need at least 2 D3-branes on top of each other in order to have mutually non-local massless states, and this can happen even far away from 7-brane. Further moving them near 7-brane provides extra mutual non-locality.

Having some D3's at monopole and some others at dyon will give us some Argyres-Douglas theory, if there are at least two D3's on top of each other. However, we will get the maximal numbers of massless sectors which are mutually non-local, if we put all the D3's at the same location and on top of monopole or dyon. That is clear from counting pairs of mutually non-local massless sectors as given above. Putting all near a 7-brane will maximize number of pairs of massless sectors which are mutually nonlocal.

From the curve analysis in the rank 2 case, Argyres-Douglas loci is made up of three points, and all of them are maximal Argyres-Douglas points. However, 2 D3-brane picture tells us that just putting D3's on top of each other {\it anywhere} already may provide Argyres-Douglas theory. We obtain more singular Argyres-Douglas theory by demanding D3's to coincide at special locations where 7-brane lies. In some sense, D3-brane picture is telling us that we have two types of Argyres-Douglas loci in the moduli space of brane dynamics - one with less singularity but with one degree of freedom (since the D3 pair can be anywhere on $u$-plane) and the other with more singularity with no free parameter left (since D3 pair is now pinned down to be at a 7-brane location).

\begin{table}[h!]
 \begin{center}
\begin{tabular}{|c|c|c|c|c|}\hline \multicolumn{4}{|c|}{${\cal M}_{2 \rm D3}$} & {${\cal M}_{f_{Sp(4)}}$} \\ \cline{1-4}
monopole & dyon & $\infty$ & elsewhere & \\ \hline
2&0&0 &0& 3 maximal Argyres Douglas points of $\Sigma^Q_i \cap \Sigma^Q_j$ \\
0&2&0&0& (Puzzle: combinatorics) \\ \hline
1&1&0&0& (Puzzle: com, Q/C) 2 mutually local massless BPS? \\ \hline
1&0&1&0& One massless dyon and $M_{\infty}$ (?) \\
0&1&1&0&  or 2 mutually local massless BPS (?) \\ \hline
1&0&0&1& $\Delta_x f_C \ne 0$ but $\Delta_x f_Q = 0$ \\
0&1&0&1& one massless dyon \\ \hline
0&0&2 &0& Two $\Sigma^C$ vanish (related to $v=0$ loci) - some AD? \\ \hline
0&0&1 &1& Some sort of $M_\infty$ ? \\  \hline
0&0&0 &$\{1,1\}$ & generic location with $\Delta_x f \ne 0$ \\ \hline
0&0&0 &$\{2\}$&  $\Delta_x f_C = 0$ but $\Delta_x f_Q \ne 0$ - some AD? (Puzzle: dim) \\ \hline
  \end{tabular}
\end{center}
\caption{Two D3-branes probing the quantum corrected orientifold background. The number of {\emph {leftover}} D3-branes (put elsewhere) corresponds to the number of degrees of freedom.}
\label{sp4table}
\end{table}

When we consider the higher rank case, the mapping does not seem to improve.
From the rank $r$ curve analysis, we know that we have $r+1$ maximal Argyres Douglas points.
The corresponding picture on the brane dynamics is when we have $r$ D3's probing the O7 geometry.
The most singular configuration by D3-branes would be when we put all of them at 7-brane. If we argue that O7 still splits into two 7 brane, then we have only 2 configurations for maximal Argyres-Douglas points. It appears that $r+1$ points on the curve moduli space will somehow map to 2 points on the moduli space of D3-brane dynamics. But we expected 1-1 mapping between the moduli spaces. 
Namely, Argyres-Douglas loci on the curve happen at least for codimension-two or higher.
However, now it occurs at codimension-one loci already, just by putting two D3's on top of each other.

\begin{table}[h!]
 \begin{center}
\begin{tabular}{|c|c|c|c|c|}\hline \multicolumn{4}{|c|}{${\cal M}_{r {\rm D3}}$} & {${\cal M}_{f_{Sp(2r)}}$} \\ \cline{1-4}
monopole & dyon & $\infty$ & elsewhere & \\ \hline
$r$&0&0 &0& $r+1$ maximal Argyres Douglas points \\
0&$r$&0&0&  (Puzzle: combinatorics) \\ \hline
$r-1$&1&0&0&  some Argyres-Douglas loci (partial, not maximal)\\
$\cdots$ & $\cdots$ &  0 & 0 & (Puzzle: dimension, combinatorics) \\
1&$r-1$&0&0& \\ \hline
 1&0&0&$\{ 1^{r-1} \}$& $\Delta_x f_C \ne 0$ but $\Delta_x f_Q = 0$ \\
0&1&0&$\{ 1^{r-1} \}$& $\Sigma^Q_i$ loci with one massless dyon\\ \hline
0&0&$r_\infty$ &$\{r_1,r_2,\cdots\}$& AD if any $r_i\ge2$ (Puzzle: dimension)  \\ \hline  \end{tabular}
\end{center}
  \caption{Multiple D3-branes probing the quantum corrected
orientifold background. (A partial table) The number groupings of D3-branes put elsewhere corresponds to the number of degrees of freedom.}
  \label{sp2rtable}
\end{table}

We observe that three new types of puzzles occur in the modified (partial) mapping between moduli spaces given in {\bf tables \ref{sp4table} and \ref{sp2rtable}}.
\begin{itemize}
 \item Combinatorics (com): We get $m$-$n$ mapping instead of $1$-$1$ mapping with $m, n \in Z$. For example, for the maximal Argyres-Douglas points, it seems we need to map $(r+1)$-2.
 \item Dimensionality (dim): Putting two D3 on top of each other (codim-1 in ${\cal M}_{{\rm D3}}$) seems to give us Argyres-Douglas loci, which is codim-2 in $  {\cal M}_{f_{Sp(4)}}$.
 \item Distinction between $f_Q$ and $f_C$ singularity enhancement (Q/C): As listed above {\bf figure \ref{lercheMq}}, we distinguished the sources of singularities of $f_Q$ and $f_C$ to be D3-branes put at monopole/dyon and elsewhere/$\infty$ respectively. However, with the modification, this distinction between Q/C breaks down for certain cases.
\end{itemize}



\paragraph{Speculations toward 1-1 mapping:} We naturally expect 1-1 mapping between D3-brane dynamics and Seiberg-Witten geometry. We experience the difficulty in building 1-1 mapping, and some evidence point towards that it might be even 1-$\infty$ mapping where this $\infty$ is not even countable, in that dimensionality does not match correctly. How do we resolve these problems in the scenario where the anti-symmetric matter is very heavy? 
Here we point out several ways we could obtain 1-1 mapping.
\begin{itemize}
 \item One could argue that our F-theory picture with D3-brane probing O7 is incomplete: it might be that not all maximal Argyres-Douglas points would appear on the brane-side.
 F theory configuration 
 might be hiding some information still, for example hiding $(r-1)$ other configurations for maximal Argyres-Douglas theories.
 \item One possible way is to modify our old conjecture about O7-splitting.
 When $r$ D3-brane probes an O7 geometry, it may be that O7 splittings into $r+1$ different $(p,q)$7-branes. Putting all $r$ D3-branes together at one of these $r+1$ locations on $u$-plane will give us maximal Argyres-Douglas points, just as we get $r+1$ maximal Argyres-Douglas points in the moduli space of the $f_{Sp(2)}$. However it does not seem physical because it is unnatural to expect that probes with smaller dimensions would make such a drastic change to the dynamics of much higher dimensional objects.
 \item Can it be not 1-1? It is not forbidden to think that the mapping between moduli spaces is not 1-1 in the limit where we have isolated the 
contribution from anti-symmetric matter.  
However it does not seem very plausible because in all limits the mapping should work consistently.
\item When D3's are on top of each other all $(p,q)$ strings between them become massless. However, we can consider deforming the theory away from such a singular locus, by taking D3's apart from each other. In such a case, all the $(p,q)$ strings will become massive, however only a few of them will become the lightest, and that will be determined by the coupling constant, which is again affected by relative location with respect to various 7-branes. For generic choice of coupling constant only a certain $(p,q)$ and $(-p,-q)$ will become the lightest. However for a less generic choice of coupling constant, we might have a few states with the equal tension. In this sense, it is possible that putting D3's on top of each other at a generic point in $u$-plane might not actually give us Argyres-Douglas theory, resolving the dimensionality puzzle mentioned above.
  \item It might be that massive states play some role here. For example the anti-symmetric matter that we assumed heavy, affects the dynamics.  
However, note that there is no open string connecting monopole and dyon 7-brane. They have different $(p,q)$ charge, so no open string is allowed to end on them at the same time. However one could introduce 3-string junction as mentioned earlier in footnote \ref{footnote337}. For example, the antisymmetric matter appears only when we have multiple D3-branes. Having multiple D3-branes may allow us 3-string junction among two D3-branes and one 7-brane, which may be responsible for the antisymmetric matter the kind that we speculated earlier. 
It will be interesting to see how this could
change the picture.  
\end{itemize}

Also, this might call for a more rigorous way of mapping the states. Namely, we are only mapping the massless states (therefore various singular loci in the moduli space). A brute-force computation for BPS masses in both moduli spaces should be a more solid way of finding a 1-1 mapping.

Before ending this section we should point out that in appendix \ref{fantasy} we fantasize on how to save the naive 1-1 mapping proposed
in apendix \ref{naiveMap}, circumventing the objection given in subsection \ref{towardMap}.
\subsection{Supergravity duals for the conformal cases}

So far we considered pure cases, i.e without seven-branes in our orientifold picture.
However once we add D7-branes the theory can be made conformal
if we can cancel the RR charge of the O7-plane locally by inserting four D7-branes on top of the O7-plane.

As before, this system could be probed by multiple D3-branes. In the special case of zero axio-dilaton background, the
world-volume theory on the probe D3-branes would be an ${\cal N} = 2, N_f = 4$  conformal theory with $SO(8)$
global symmetry \cite{sw2}. In the limit when the number of probes becomes very large, the
strongly coupled CFT will be captured by
an AdS supergravity background, i.e type IIB on $AdS_5 \times S^5/{\mathbb Z}_2$ background \cite{FS}.

The above is the simplest case. There could be other arrangements of the seven-branes that allow non-zero but
constant background axio-dilaton, for examples the ones studied in \cite{DM1}. All these correspond to CFTs.
The supergravity duals for the conformal cases may now appear when we have large number of
D3-branes to probe the constant coupling backgrounds. These have already been studied
in detail in \cite{FS} (see also the sequel \cite{AFM}) so we will be very brief. Therefore in the following we
will simply summarize the story, and no new informations will be added.

The supergravity duals of \cite{FS} are based on the constant coupling limits studied in \cite{DM1} as mentioned above.
These
constant coupling limits appear as the exceptional global symmetries of the underlying gauge theory on the
probe D3-branes. If we parametrize the $u$-plane as $u = x^4 + ix^5$, then the metric near the vicinity of the
coincident seven-branes with $E_6, E_7$ and $E_8$ global symmetries are given as:
\bg\label{engs}
\big\vert u^{-2/3} du \big\vert^2, \qquad \big\vert u^{-3/4} du \big\vert^2, \qquad \big\vert u^{-5/6} du \big\vert^2,
\nd
respectively. The rest of the discussions are straightforward. The near horizon geometries are now given by
$AdS_5 \times S^5/{\mathbb Z}_n$ with $n = 3, 4, 6$ respectively \cite{FS}. The ${\cal N} = 1$ extension to the story
was done in \cite{AFM} where the near horizon geometries associated to various constant coupling cases were discussed.
These near horizon geometries are relevant to Model 4 in section \ref{model4section}.

This concludes our disussion about Model 1. In the next section we will describe Model 2 wherein we will make our first non-trivial change in the 
geometry.

\newpage

\section[Model 2: Multiple D3's probing seven-branes on a Taub-NUT]{Model 2: Multiple D3-branes probing seven-branes on a Taub-NUT background \label{model2section}}

\begin{figure}[htb]
        \begin{center}
\includegraphics[height=6cm]{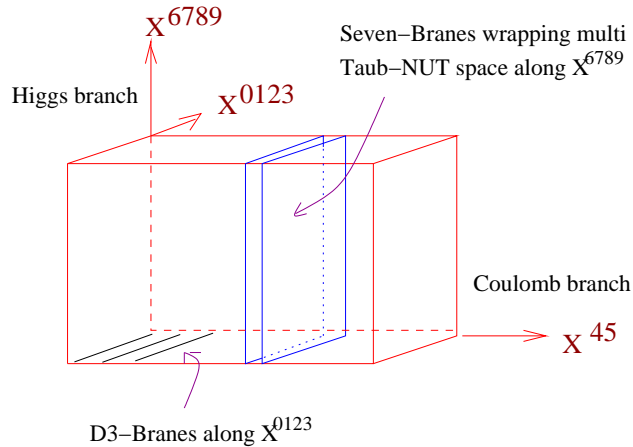}
         \end{center}
          \caption{Multiple D3-branes probing seven-branes on a multi Taub-NUT geometry.}          \label{taubNUTbranefig}
        \end{figure}

Our next construction would be to generalize the above to a type IIB D3/D7 system. One of the simplest generalization
is to replace the ${\mathbb R}^4$, on which we have the seven-branes and planes wrapping, to a more non-trivial
four-dimensional space. The simplest non-compact example is an ALE space ${{{\mathbb R}^4}\over {\mathbb Z}_2} $,
or more locally, a Taub-NUT space. The supersymmetry of this configuration still remains ${\cal N} = 2$ as one
can incorporate this without breaking further supersymmetries in the system. Thus our second set of examples
falls into:
\bg\label{joba2}
&&{\rm Type~ IIB ~on} ~~{{\bf T}^2\over \Omega\cdot (-1)^{F_L}\cdot {\cal I}_{45}}
\times {{{\mathbb R}^4}\over {\mathbb Z}_2} \times {\mathbb R}^{0123}
~ = ~ \nonumber \\
&&{\rm F~ Theory ~on}~~ {\rm K3} \times {{{\mathbb R}^4}\over {\mathbb Z}_2} \times {\mathbb R}^{0123}
\nd
probed by a single D3-brane\footnote{Although we will write the background as ${{{\mathbb R}^4}/{\mathbb Z}_2}$
or more generally as
${{{\mathbb R}^4}/{\mathbb Z}_k}$, we will {\it always} take the asymptotically locally flat case i.e
introduce constant terms for the radius of the asymptotic circles in the harmonic function for
${{{\mathbb R}^4}/{\mathbb Z}_k}$. See \cite{wittenTN} for a more recent exposition.}.
As we will discuss below, once we increase the number of D3-branes, we can also
make the Taub-NUT space multi-centered otherwise without breaking further supersymmetries. The brane configuration is
given by {\bf figure \ref{taubNUTbranefig}} and {\bf table \ref{directionstable}}.
\noindent In our configuration the D3-branes are oriented along the spacetime $x^{0,1,2,3}$ directions. The seven-branes
(and seven-planes) are parallel to the D3-branes and also wrap multi-centered Taub-NUT space oriented along
$x^{6,7,8,9}$. Therefore as before, the Coulomb branch will be the complex $u \equiv x^4 + ix^5$ plane, whereas the
Higgs branch will be along the Taub-NUT space.

\subsection{Brane anti-brane on a Taub-NUT background \label{subsecDDbarTN}}

Let us start with a singular ${\mathbb Z}_2$ ALE space
along directions
$x^{6,7,8,9}$. The node is really a 5-plane filling the remaining
directions. Close to the singular point $x^{6,7,8,9}=0$, the space can be replaced by
a 2-centre (separated in $x^6$) Taub-NUT metric with coincident (in $x^{0,1,2,3}$) centres. This
is equivalent to saying that we have two coincident Kaluza-Klein
monopoles. We also know \cite{aspinwall} that the ${\mathbb Z}_2$ orbifold hides
half a unit of $B_{\rm NS}$ flux through the shrunk 2-cycle
$\Sigma$. The four moduli associated to this ALE space are three
geometrical parameters, which can be thought of as the blowup of the
ALE to form a smooth Eguchi-Hanson metric, and the $B_{\rm NS}$ flux \cite{aspinwall}.

Take a D3-brane transverse to the ALE space, filling the
directions $x^{0,1,2,3}$. (More generally we start with $r$ such
D3-branes.) When the ALE space is singular, the
world-volume theory of the 3-brane has two branches: a Higgs branch,
when the brane is separated from the singularity along
$x^{6,7,8,9}$, and a Coulomb branch when the brane hits the
singularity and dissociates into a pair of fractional branes which
can move around only in the $x^{4,5}$ directions. However, if the ALE
space is blown up, then the Coulomb branch gets disconnected from the Higgs branch because the 3-brane cannot
dissociate supersymmetrically into pair of fractional branes.

The fractional D3-brane is interpreted as D5-brane wrapped on a ${\bf P}^1$ with fluxes or
$\overline{\rm D5}$-brane wrapped on a ${\bf P}^1$ with different choice of fluxes (see details
below). Therefore an integer D3-brane would be a pair of D5-branes whose
D5-brane charges cancel, hence they are really a ${\rm D5}$-$\overline{\rm D5}$ (D5-brane --
anti-D5-brane) pair (see also \cite{mukdas}, \cite{poln2, poln3, aharonyn2}
where a somewhat similar model has been discussed).
However, they carry D3-brane charge by virtue of
the Chern-Simons coupling on D5-branes. Denoting the world-volume gauge field
strength on the D5-brane by $F_1$, we have the coupling
\bg\label{cscoupling}
\int (B_{\rm NS} - F_1)\wedge C_4
\nd
where $C_4$ is the self-dual 4-form potential in the type IIB
string. At the orbifold point we have $\int_\Sigma B_{\rm NS} = {1\over 2}$ and
hence half a unit of D3-brane charge. The $\overline{\rm D5}$ (anti-D5-brane) (whose world-volume gauge field strength is
denoted by $F_2$) will have a coupling
\bg\label{antics}
-\int (B_{\rm NS} - F_2)\wedge C_4.
\nd
Now let us also turn on a
world-volume gauge field strength $F_2$ on the anti-D5-brane and give it
a flux of $+1$ unit through the vanishing 2-cycle $\Sigma$ (more
generally, we assign unit flux to the relative gauge field $F_- = F_2
- F_1$). In this configuration, the ${\rm D5}$-$\overline{\rm D5}$ pair has
in total D3-brane charge equal to 1, or more generally:
\bg\label{charge}
\int (F_2 - F_1) \wedge C_4 \equiv \int_{\Sigma \times {\mathbb R}^{0123}} F_- \wedge C_4.
\nd

\begin{figure}[htb]
        \begin{center}
\includegraphics[height=6cm]{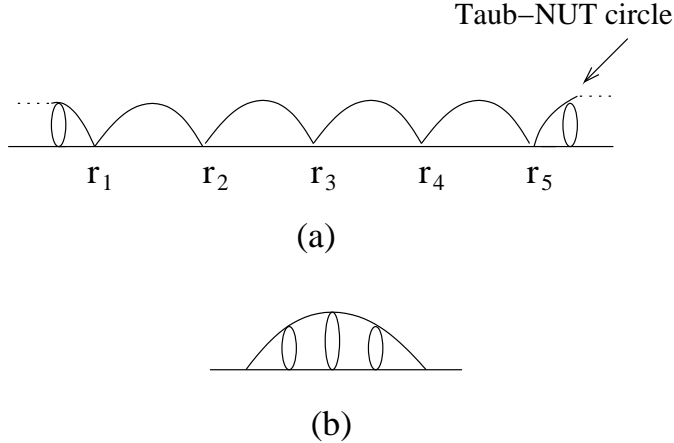}        \end{center}
        \caption{In fig (a) the singularities of Taub-NUT space are shown. The compact direction is the
Taub-NUT circle that is fibered over the base. The various points at which the circle (which is along $x^6$ direction
in the text) degenerate are the singular points. 
Between two singular points form a
two-cycle, as shown in fig (b), once we assume that the $x^6$ circle is degenerating along a line parametrized by
the $x^7$ coordinate. Thus the ${\bf P}^1$'s are labelled by $x^{6, 7}$ coordinates in the text.}
\label{tnfrac}\end{figure}

In a slightly more generalized setting with multi Taub-NUT space the situation is somewhat similar. To see this, let
us consider a Taub-NUT space with $m$ singularities as shown in {\bf figure \ref{tnfrac}}. Once we bring the D3-branes
near the Taub-NUT singularities, they decompose as $m$ copies of D5-${\overline{\rm D5}}$ wrapping the various two-cycles
of the Taub-NUT space. Each of the wrapped $k$'th
D5's can be assumed to create a fractional D3-brane on its world-volume via
the world-volume $F_{1,k}$ fluxes by normalising the total integral of $F_1$ over all the two-cycles to equal the number
of integer D3-branes.
The ${\overline{\rm D5}}$ branes, on the other hand, are used only to cancel the
D5 charges as their world-volume fluxes are taken to be zero. These fractional D3-branes can now move along the
Coulomb branch as expected. T-dualising this configuration gives us D4-branes between the NS5-branes which may be
broken and moved along the Coulomb branch. The above way of understanding the fractional branes has two immediate
advantages:

\begin{itemize}
\item Since every ${\bf P}^1$ of the multi Taub-NUT space is wrapped by D5-${\overline{\rm D5}}$,
and the system is symmetrical, one is restricted to
switching on same gauge fluxes on each of the ${\bf P}^1$'s. However for non-compact ${\bf P}^1$'s this restriction
doesn't hold as the wrapped branes give rise to flavors and not colors, and one may switch on different fluxes.
This will be used to understand the Hanany-Witten
brane creation process later in the text.
\item Although the number of D5-${\overline{\rm D5}}$ on each of the ${\bf P}^1$'s are the same, we can
wrap {\it additional} D5-branes on each of the ${\bf P}^1$. The number of these additional D5-branes could be different
for every ${\bf P}^1$'s. Such a configuration will break conformal invariance leading to
cascading theories\footnote{Instead, if we wrap
additional ${\overline{\rm D5}}$'s on the ${\bf P}^1$'s, they will break supersymmetry. Also the number of
D5-${\overline{\rm D5}}$'s on each ${\bf P}^1$ should remain same so as to cancel the tachyons across each
wrapped ${\bf P}^1$'s.}.
\end{itemize}

In the above analysis of fluxes, we have been ignoring one subtlety related to the
orientifold action $\Omega$ discussed at the beginning of this section. Due to the orientifold
action we expect all $B_{\rm NS}$ and $B_{\rm RR}$ fluxes that do not have one component along the
$x^{4,5}$ directions to be projected out. Therefore it would seem that due to the orientifold action
the D3-branes cannot apparently dissociate into pairs of ${\rm D5}$-$\overline{\rm D5}$ branes. However as is well known,
quantum corrections in F-theory can take us away from the orientifold point, and we can study the theory
completely in terms of {\it local} and {\it non-local} seven-branes without resorting to orientifold planes.
Therefore we will henceforth assume that the F-theory background probed by the $r$ D3-branes is described
completely in terms of the seven-branes wrapped on multi Taub-NUT space which, alternatively, would also mean
that we have a ${\bf P}^1$ with 24 seven-branes, i.e:
 \bg\label{t2def}
{{\bf T}^2\over \Omega\cdot (-1)^{F_L}\cdot {\cal I}_{45}} ~ \longrightarrow ~ {\bf P}^1.
\nd
Thus going to the Coulomb branch, by tuning all of $x^{6,7,8,9}$ to 0,
the picture is somewhat different. At this point a D3-brane splits
into a pair of fractional branes which can move independently along
$x^{4,5}$. The geometric orbifold singularity now cannot be blown up
any more. This is easy to see on the T-dual type IIA side, where the
D4-brane splits into two pieces that stretch along the two intervals
between the two NS 5-branes (one from each side of the $x^6$ circle).
These partially wrapped 4-branes can move independently along the
NS 5-branes, namely in the $x^{4,5}$ directions. 

To summarize, the type IIB picture on the Coulomb branch is that the
relative world-volume gauge field strength $F_-$ on the ${\rm D5}$-$\overline{\rm D5}$ pair must be turned on over the 2-cycle $\Sigma$ and
gives rise to a 3-brane in the space transverse to that cycle. The
spacetime $B_{\rm NS}$ flux over $\Sigma$ changes the relative tensions of
the wrapped D5-brane and anti-D5-brane keeping the total constant. The directions of
various branes and fluxes in out set-up therefore is given in {\bf table \ref{directionstable}}. Supersymmetry is preserved because the
${\rm D5}$-$\overline{\rm D5}$ pairs wrap vanishing 2-cycles of the multi Taub-NUT space, in addition to the conditions mentioned
earlier. There are also additional 
background fluxes like axio-dilaton, two-forms and four-forms. The metric of the
Taub-NUT space will be deformed due to the backreactions of the branes and fluxes, that we will discuss later.
All these effects conspire together to preserve ${\cal N} = 2$ supersymmetry on the fractional probe D3-branes. Note that
if the Taub-NUT cycles are blown-up then, in the presence of the seven-branes, supersymmetry will be broken.

\begin{table}[h!]
 \begin{center}
\begin{tabular}{|c||c|c|c|c|c|c|c|c|c|c|}\hline Directions & 0 & 1 & 2
& 3 & 4 & 5 & 6 & 7 & 8 & 9 \\ \hline
D5 & --  & --   & --   & --  & $\ast$  & $\ast$ & -- & -- & $\ast$
& $\ast$\\  \hline
$\overline{\rm D5}$ & --  & --   & --   & --  & $\ast$  & $\ast$  & --  & -- &
$\ast$  & $\ast$\\  \hline
D7 & --  & --   & --   & --  & $\ast$  & $\ast$ & -- & -- & --  & --
\\  \hline
Taub-NUT & $\ast$  & $\ast$  & $\ast$  & $\ast$ & $\ast$  &
$\ast$  & --  & -- & --  & -- \\  \hline
Fluxes & $\ast$  & $\ast$  & $\ast$  & $\ast$ & $\ast$  & $\ast$
& -- & -- & $\ast$ & $\ast$ \\  \hline
  \end{tabular}
\end{center}
  \caption{The orientations of various branes and fluxes in out set-up. The dashed lines for the
branes are the directions parallel
to the world-volume of the branes; and for the fluxes and Taub-NUT space are the directions along which we have
non-trivial fluxes and metric respectively. The stars denote orthogonal spaces. Supersymmetry is preserved in the
presence of vanishing cycles and background fluxes.}
  \label{directionstable}
\end{table}

\subsection{Anomaly inflow, anti-GSO projection and brane transmutation \label{subsecTimeVary}}

There is an interesting subtle phenomena that
happens to our system when we switch on a time-varying vector potential $A_\mu(t)$ along the degenerating
one-cycle of the Taub-NUT space. 
However before we go about discussing this in detail, we want to point out
an important property of the
underlying Taub-NUT space, namely, the existence of a normalizable harmonic two form $\Omega$. For $m$-centered
Taub-NUT there would be equivalently $m$ normalizable harmonic forms $\Omega_i, i = 1, 2, \cdots , m$. The existence
of these harmonic forms are crucial in analyzing the phenomena that we want to discuss. 

To see what happens when we switch on time-varying Wilson line, note first that
the seven-brane wrapping the Taub-NUT space will give rise to a D3-brane
bound to it. The charge of the D3-brane is given by the non-trivial
$B_{\rm NS}$ background on the Taub-NUT. To see this, consider some of
the couplings on the world volume of the D7-brane (we are neglecting
constant factors in front of each terms):
\bg\label{coupdseven}
\int *C_0 ~+~ \int C_4 \wedge F \wedge B_{\rm NS} ~+~
\int  C_4 \wedge F\wedge F ~+~ \cdots.
\nd
These couplings are derived from the Wess-Zumino coupling $\int C\wedge
e^{B-F}$, where $C$ is the formal sum of the RR potentials.  The first
term $\int *C_0$ gives the charge of the D7-brane.

Coming back to our phenomena, notice that we cannot turn on a flat
connection on this space. Instead, a self-dual connection can be
turned on. This self-dual connection is of the form:
\bg\label{selfdual}
F= dA = \Omega,
\nd
where $\Omega$ is the unique normalizable harmonic two-form on the
Taub-NUT space. This harmonic two form, being normalizable, goes to
zero at infinity, hence we have a flat connection there. At infinity
there is an $S^1$ and therefore the flat connection corresponds to
a Wilson line\footnote{Multi-center Taub-NUT also known as asymptotically locally flat (ALF) space,
locally asymptotes at infinity to ${\mathbb R}^3 \times {\bf S}^1$. When the size of the ${\bf S}^1$ goes to
infinity, we get the asymptotically locally Euclidean (ALE) space.}.

The above choice of background \eqref{selfdual} however doesn't take the fluctuations of gauge fields into account.
A more appropriate choice for our case is to decompose the field strength $F$ as
\bg\label{fdecom}
F = \Omega + F_1.
\nd
instead of just \eqref{selfdual}.
Now $F_1$ will appear as a gauge field on the D7 (or Taub-NUT plane).
Inserting \eqref{fdecom} in \eqref{coupdseven} and integrating out $\Omega$,
we get the required D3-brane charge (see also \cite{mukdas} for more details).
This confirms that a bound state of a D3 with the D7-brane appears once we
switch on a self-dual connection (which is of course the Wilson line for our case).

However the situation at hand demands a {\it time-varying} gauge field on the
world volume of the D7-brane. A typical time-varying gauge field $A_\mu$ can be
constructed from $F_1$ in \eqref{fdecom} by making it time-dependent. Such a time-varying gauge
field creates a chiral anomaly along the $S^1$ at the
asymptotic region of the Taub-NUT space. This $1+1$ dimensional anomaly is of the form \cite{BDGreen}
\bg\label{anomaly}
\int d^2x ~\omega \epsilon^{ab} \partial_a A_b,
\nd
 where $\omega$ is the gauge transformation parameter. Another way to see this anomaly is
to dualize the D7-brane and the Taub-NUT space into a D6/D4 system oriented along
$x^{0,1,2,3,4,5,6}$ and $x^{0,6,7,8,9}$ respectively\footnote{Use the
following set of dualities to go from one picture to another:
T-dualities along $x^{6, 1, 2, 3}$ then a S-duality followed by another
T-duality along $x^6$.}. The chiral anomaly is along the
$x^6$ direction.

It is now time to figure out the term that could cancel the anomaly. The term that we need here is given by:
\bg\label{anomterm}
S = \int\, G_5\wedge A\wedge F
\nd
on the world volume of the seven-brane. Here $G_5 = dC_4$, the pullback of the
background four-form, in the absence of any source. The cancellation
takes place via anomaly inflow. We have a coupling, \eqref{anomterm}, in
$(7+1)d$ spacetime. Along a $(1+1)d$ subspace of this, chiral fermions
propagate and give rise to the anomaly \eqref{anomaly}.
Since D3-brane is
the source for $G_5$, we find that changing the Wilson line
produces a change of flux of $G_5$. In other words, a gauge
transformation $\delta A = d\omega$ on the world-volume will vary
\eqref{anomterm} by:
\bg\label{vary}
-\int dG_5 \wedge (\omega F)
\nd
Since $dG_5 \ne 0$ in the presence of a source of $G_5$ flux\footnote{Since $G_5$ is self-dual, this switches on
a D3-brane with orientations along $x^{0,1,2,3}$ directions.},
we end
up with:
\bg\label{anomcancel}
\delta S = - \int d^2 x~ \omega\, \epsilon^{ab}\del_a A_b
\nd
resulting in the inflow which cancels the anomaly \eqref{anomaly} by creating a D3-brane\footnote{Another way to
see this is the following. Switching on a D3-brane amounts to switching on an instanton action of the form
$${\cal L}_{\rm eff} = \int d^4y~\theta ~\epsilon_{abcd} F^{ab}F^{cd}$$ along the Taub-NUT world volume. Here
$\theta$ is the remnant of the D3-brane term in the CS coupling of the D7-brane or the more popular, $\theta$-term
of gauge theory.
Now as shown by \cite{callan} a gauge transformation will effectively give us
$$\delta_\omega {\cal L}_{\rm eff} = -\int d^2y ~\omega \epsilon^{ab} F_{ab}$$ with the correct minus sign to
cancel the anomaly. Incidentally if the above action becomes non-abelian, exactly similar computation will again
cancel the underlying anomaly. Therefore the upshot is that, switching on a D3-brane
oriented along $x^{0,1,2,3}$ directions will cancel the gauge anomaly in the system.}.

The story is however {not} complete. There is an {\it additional} phenomena that happens
simultaneously that actually reduces the number of D3-branes instead of increasing it (as we might
have expected from the above discussion). This additional phenomena relies on the dissociation
of the D3-branes into ${\rm D5}$-$\overline{\rm D5}$ pairs discussed in the previous subsection. Recall that the
tachyon between the D5 and the $\overline{\rm D5}$ is cancelled for $F_- \equiv F_2-F_1=\pm 1$. Here we set
\bg\label{f2f1}
F_2 = 1, \qquad F_1 = 0,
\nd
which also implies that $F_- = 1$ in \eqref{charge}, giving rise to a unit D3-brane charge.

\begin{figure}[htb]
        \begin{center}
\includegraphics[height=4cm]{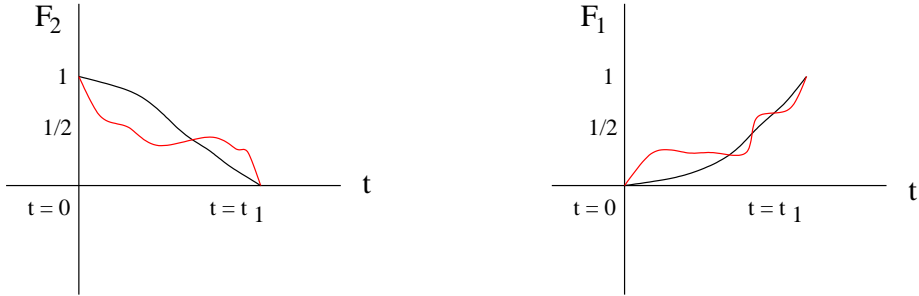}        \end{center}
        \caption{The gauge fluxes on the D5 and the $\overline{\rm D5}$ branes are changed with time. The red and the
black lines denote two different changes with the same initial and final values.} \label{fluxchange}\end{figure}

Now imagine that we change the world volume fluxes as described in {\bf figure \ref{fluxchange}}: In the beginning, $(F_1, F_2)=(0,1)$ at $t=0$, and then at some time they both become $\frac{1}{2}$, finally at $t=t_1$ $(F_1, F_2)=(1,0)$.
As seen in the figure, when
both the $F_i$ fluxes approach the mid value $F_i = {1\over 2}$ the D3-brane charge in \eqref{charge} vanishes.
However at $t = t_1$ when
\bg\label{fcha}
F_2 ~ = ~ 1~ \to ~ {1\over 2} ~ \to~ 0, \qquad F_1~ = ~ 0~ \to ~ {1\over 2} ~ \to~ 1
\nd
then one may easily check that the ${\rm D5}$-$\overline{\rm D5}$ pair give rise to an anti-D3-brane ($\overline{\rm D3}$). However since the {\it difference}
between the fluxes have still remained 1, the tachyon between the D5 and the $\overline{\rm D5}$ continue to remain massless
and the supersymmetry doesn't get broken by this process. Therefore this process transmutes a D3-brane into an
$\overline{\rm D3}$! Thus switching on a time-varying Wilson line has the following two effects:
\begin{itemize}
\item Chiral anomaly cancellation via anomaly inflow and creation of a new D3-brane.
\item D3-brane transmutation to an $\overline{\rm D3}$ brane via flux change.
\end{itemize}
\noindent Together these two effects would remove one of the existing D3-brane in the system. Therefore the color
degree of freedom would change via this process. If we do this multiple times, we can reduce the number of D3-branes in the
model. Once we go to the brane network model, we will show that the above phenomena are related to the brane annihilation and the Hanany-Witten effect.

\noindent Of course
in the {\it absence} of the Taub-NUT space, none of the above arguments would work, and
so there would be no brane creation. This is perfectly consistent with our
expectation.

Now imagine that we switch on gauge fluxes $F = n \Omega$. This would imply that we have
two new sources of the form:
\bg\label{nsour}
{n^2\over 2} \int C_4, \qquad {\rm and} \qquad n\int C_4 \wedge (B_{\rm NS} - F_1).
\nd
The latter doesn't break supersymmetry as was explained in \cite{mukdas}. In fact overall the
supersymmetry will never be broken if we take $r$ D3-branes and we simultaneously consider
$m$-centered Taub-NUT. Therefore the $r$ D3-branes wrap $m$ different vanishing 2-cycles.

From the above discussions we see that we have two models that are {\it dual} to each other while preserving
supersymmetry. The duality criteria for our case can be presented in the following way:
\begin{itemize}
\item $r$ D3-branes probing seven-branes wrapping a $m$-centered Taub-NUT space. The D3-branes dissociate
as $m$ copies of ${\rm D5}$-$\overline{\rm D5}$
pairs that move along the Coulomb branch as depicted in
{\bf figure \ref{wrappedTN}}.
The seven-branes could
be arranged to allow for any global symmetries and axio-dilaton moduli, including the conformal cases.
\item Time-varying
self-dual connections on the seven-branes along the Taub-NUT direction that create {\it and} transmute
${\rm D5}$-$\overline{\rm D5}$ sources by changing the $F_i$ in \eqref{cscoupling} and \eqref{antics}. Due to this
some D3-branes may annihilate,
thereby changing the local and possibly the global symmetries
of the model.
\end{itemize}
Our claim therefore is the following. The above two dual descriptions, coming from chiral anomaly cancellation,
D3-brane creation and D3-brane transmutation, are related by the recently proposed Gaiotto dualities. In the
next subsection we will supply more evidences for this conjecture.

Note that the total moduli in both the models are exactly similar, although both color and flavor degrees of freedom
may apparently differ. The seven-branes could be arranged such that we could either have F-theory at constant
couplings {\it a la} \cite{DM1}, or non-constant couplings. However due to the underlying F-theory constraints, the
flavor degrees of freedom remain below 24 although the color degrees of freedom could be anything arbitrary.
In addition to that there is also a M-theory uplift of our model that is quite different from the M-theory brane
constructions studied by Witten \cite{wittenM} and Gaiotto \cite{gaiotto}. We will discuss this soon.

\subsection{Mapping to Gaiotto theories and beyond}

After having constructed our model, let us try to map to some of the Gaiotto's constructions. Our first map will be
to the brane network model studied by \cite{bbt} recently. We will then argue how gravitational duals for
our models, at least in the conformal limit, may be derived. These gravity solutions should be compared to the
recently proposed gravity duals given in \cite{GM}. Our model can also be extended to the non-conformal cases
just by moving the seven-branes around. In fact we will see an interesting class of {\it cascading} ${\cal N} = 2$
models appearing naturally out of our constructions.
Additionally, new states in the theory could appear in the generic cases when the
D3-brane probes are connected by string junctions or string networks. In the later part of this section we will give
some details on these issues, extending the scenario further.

\subsubsection{Mapping to the type IIB brane network models \label{subsecNetwork}}

Recently the authors of \cite{bbt} have given a set of interesting brane network models that may
explain certain conformal constructions of the Gaiotto models, including ways to see how the Gaiotto
dualities occur from the networks. The obvious question now is whether there exist some regime of
parameters in our set-up that could capture the brane network models of \cite{bbt}.

\begin{figure}[htb]
        \begin{center}
\includegraphics[height=6.5cm]{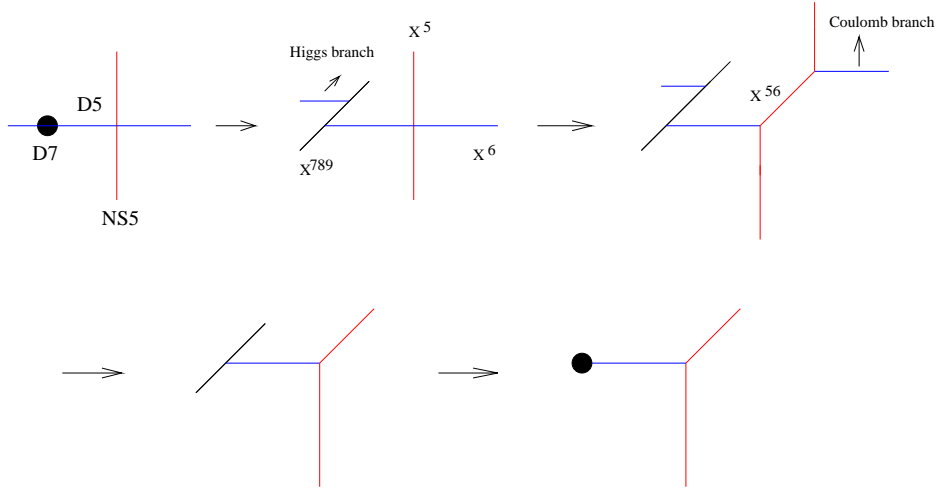}
        \end{center}
        \caption{The simplest brane junction from our Taub-NUT configuration.} \label{TN2I}
        \end{figure}
It turns out the mapping to \cite{bbt} is not straightforward. The orientations of various branes in our
set-up are given in {\bf table \ref{direction2}}. A naive T-duality along $x^4$ and $x^6$
direction will convert the Taub-NUT
space to a NS5-brane oriented along $x^{0,1,2,3,4,5}$ and the D7-brane into another D7-brane oriented along
$x^{0,1,2,3,4,7,8,9}$. However the ${\rm D5}$-$\overline{\rm D5}$ pairs will continue as ${\rm D5}$-$\overline{\rm D5}$ pairs although with a slightly different
orientation.


\begin{table}[h!]
 \begin{center}
\begin{tabular}{|c||c|c|c|c|c|c|c|c|c|c|}\hline Directions & 0 & 1 & 2
& 3 & 4 & 5 & 6 & 7 & 8 & 9 \\ \hline
D5 & --  & --   & --   & --  & $\ast$  & $\ast$ & -- & -- & $\ast$
& $\ast$\\  \hline
$\overline{\rm D5}$ & --  & --   & --   & --  & $\ast$  & $\ast$  & --  & -- &
$\ast$  & $\ast$\\  \hline
D7 & --  & --   & --   & --  & $\ast$  & $\ast$ & -- & -- & --  & --
\\  \hline
Taub-NUT & $\ast$  & $\ast$  & $\ast$  & $\ast$ & $\ast$  &
$\ast$  & --  & -- & --  & -- \\  \hline
  \end{tabular}
\end{center}
  \caption{The orientations of various branes in out set-up. Same as
the earlier table but now the flux informations are not shown.}
  \label{direction2}
  \end{table}

\begin{figure}[htb]
        \begin{center}
\includegraphics[height=5cm]{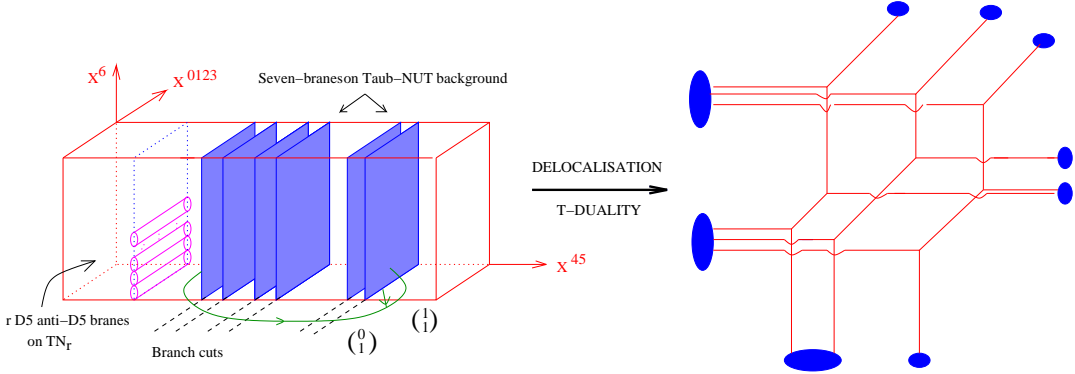}
        \end{center}
        \caption{Under special arrangement of the seven-branes, delocalization and T-dualities map our model
to the brane network studied in \cite{bbt}. For other configurations there are no simple
map to the brane networks. The blue patches on both sides represent the seven-branes. The $r$ ${\rm D5}$-$\overline{\rm D5}$ pairs
are wrapped on vanishing 2-cycles of a multi Taub-NUT space with $m$ centers. 
} \label{bbt}
        \end{figure}

This is not what we would have expected for the model of \cite{bbt}. Furthermore, because of the fluxes as well
as other fields that have non-trivial dependences along the $x^{4,6}$ directions T-dualities along these directions
are not possible. Therefore there is no simple map to the brane network model of \cite{bbt}. However we
can go to a corner of the moduli space of solutions where:
\begin{itemize}
\item[(a)] We are at low energies i.e at far IR, so the ${\rm D5}$-$\overline{\rm D5}$ pairs behave as fractional D3-branes,
\item[(b)] We have delocalized completely along the two T-duality directions $x^{6,4}$.
\end{itemize}
Under these two special cases together, we can T-dualize along $x^{6,4}$ directions to convert our
configuration to the brane network model of \cite{bbt} as depicted in {\bf figures \ref{TN2I}} and {\bf \ref{bbt}}.
In {\bf figure \ref{TN2I}} the configuration in {\bf table \ref{direction2}} is T-dualized following the above criteria
to get to the brane intersection model in the top left of the figure. Motion in the Coulomb branch is precisely the
decomposition of the D3-brane into ${\rm D5}$-$\overline{\rm D5}$ pair, such that each of them support a fractional D3-brane.
Once we have the fractional D3-branes we can move one of them along the $x^{4,5}$ direction\footnote{Of course this is
the generic case. But for wrapped ${\rm D5}$-$\overline{\rm D5}$-branes there could be
situations where in the T-dual set-up the D5-brane may
terminate on NS5-brane (much like the one in \cite{mukdas}).}.
On the other hand we also
need to break the D5-brane on the seven-brane and move this
along the Higgs branch, as depicted in {\bf figure \ref{TN2I}}. This is achieved by expressing the fractional
D3-brane (on the D5-brane) as an {\it instanton} on the seven-brane and then further decomposing the instanton
as fractional instantons on the seven-brane. Moving one set of fractional branes along the Higgs branch will
eventually give us the brane junction studied by \cite{bbt} as shown in {\bf figure \ref{TN2I}}.

Clearly this T-dual mapping works most efficiently with fractional D3-branes and ignoring their
${\rm D5}$-$\overline{\rm D5}$ origins. As we saw before, this dissociation is crucial in the
presence of multi Taub-NUT space and therefore the mapping to \cite{bbt} only works under special
circumstances. It also means that once we map our model to \cite{bbt} we may lose many informations
of our model. In particular all the high energy informations, like the presence of ${\rm D5}$-$\overline{\rm D5}$ pairs, fluxes
and massless tachyons are completely lost on the other side. But certain low energy informations do map
from our model to \cite{bbt}. For example a crucial ingredient of \cite{bbt} is the Hanany-Witten
brane creation process that occurs when we move the D7-brane across the NS5-brane. The D7-brane is located at
$x^6_{(1)}$ and the NS5-brane is located at $x^6_{(2)}$. The relative motion of the D7-brane will induce
following T-duality map:
\bg\label{jogini}
\int d^2y~\left[{\partial x^6_{(1)} \over \partial t} - {\partial x^6_{(2)} \over \partial t}\right]
~\longrightarrow~
\int d^2y~\epsilon^{06}\partial_0 A_6,
\nd
which is of course one term of the chiral anomaly $\int \omega \epsilon^{ab}\partial_a A_b$ as we saw before. A
cancellation of the chiral anomaly therefore maps to the brane creation picture of \cite{bbt}, although the
brane transmutation in our model (that relies on the dissociation of D3-brane into ${\rm D5}$-$\overline{\rm D5}$ pair) cannot be
seen directly from the T-dual model (although there may exist some equivalent picture).

Another interesting ingredient of \cite{bbt} is the so-called s-rule that preserves supersymmetry. In this
configuration the D5-branes ending on same D7-branes must end on {\it different} NS5-branes, i.e not more than one D5-brane may end on a
given pair of NS5-brane and D7-brane, otherwise supersymmetry will be broken. At low energy we saw that T-duality can map our model to
\cite{bbt}. The $m$-centered Taub-NUT space can map to the multiple configuration of the NS5-branes. Similarly, ($p, q$) five branes can be understood as explained above. The ${\rm D5}$-$\overline{\rm D5}$ pairs wrap the vanishing cycles of the
multi Taub-NUT geometry and we may keep $r$ pairs of ${\rm D5}$-$\overline{\rm D5}$ with $m$-centered Taub-NUT space. This means that there
may not be a simple map of the s-rule of \cite{bbt} to our set-up. This is understandable
because making a single T-duality to type IIA
along $x^6$, and removing the seven-branes, give us the NS5/D4 configuration where multiple D4-branes can end on
NS5-branes. However in this mapping all information of the {\it non-local} seven-branes are completely lost including
informations about exceptional global symmetries etc. Thus our F-theory model including number of seven-branes captures additional information.

\subsubsection{The UV/IR picture and gravity duals \label{subsecUVIRgrav}}

In the limit when we take the number of ${\rm D5}$-$\overline{\rm D5}$ pairs to be very large, we expect the near horizon
geometry to give us the gravity duals of the associated theories. One may arrange the seven-branes in such
a way that the axio-dilaton coupling doesn't run. In that case the corresponding theories should be conformal
at least both at UV and IR. Recently Gaiotto and Maldacena \cite{GM} have studied the gravity duals of
some of the Gaiotto models and have provided explicit expressions for the IR pictures. In this subsection we will
provide some discussions on this using our set-up. More detailed derivations will be provided in the sequel to this
paper.

\begin{figure}[htb]
        \begin{center}
\includegraphics[height=6cm]{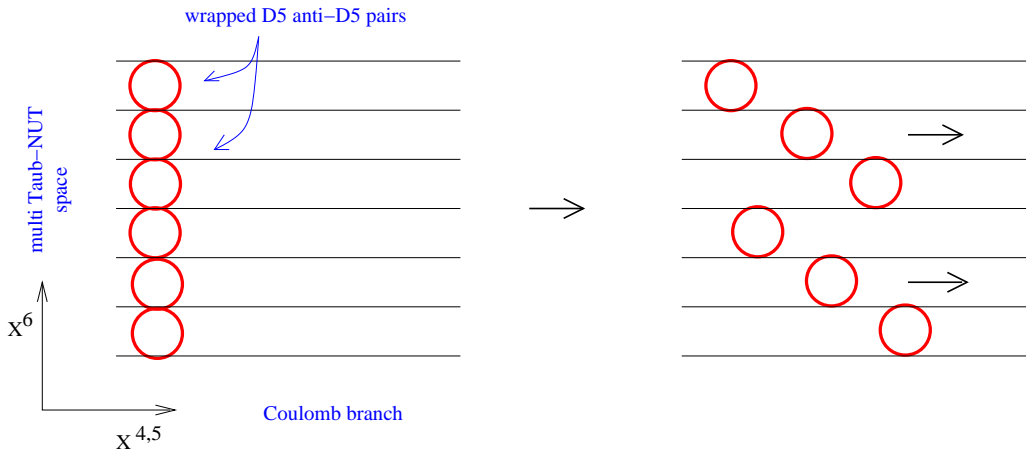}
         \end{center}
       \caption{${\rm D5}$-$\overline{\rm D5}$ branes wrapped on 2-cycles of a multi Taub-NUT space. When each of these
$N$ set of
${\rm D5}$-$\overline{\rm D5}$ pairs wrap vanishing cycles of the multi Taub-NUT space, they can give rise to
$[SU(N)]^m$ gauge groups where $m$ denote the number of vanishing Taub-NUT 2-cycles. In the next figure
various such pairs are broken and moved along the Coulomb branch.}  \label{wrappedTN}
        \end{figure}

One aspect of the gravity dual should be clear from the F-theory model that we present here: the UV of the theory
should be different from the IR. In fact we expect the UV to be a six-dimensional theory whereas the IR should be
four-dimensional. This is bourne out from the following observations. In the UV, when the system is probed using
high energy wavelengths, the complete large $m$ Taub-NUT singularities should be visible. Therefore UV description
should be given by a large $r$ ${\rm D5}$-$\overline{\rm D5}$ pairs wrapped on the vanishing 2-cycles of the muti Taub-NUT space.
Because of the presence of ${\rm D5}$-$\overline{\rm D5}$ pairs we expect the theory should become six-dimensional, and by arranging
the seven-branes appropriately the UV should be a 6d SCFT.

On the other hand the IR is simple. Since at IR we are probing the geometry with large wavelengths, the subtleties
of the geometry will be completely washed out and we will see only a simple Taub-NUT space with no non-trivial cycles.
The ${\rm D5}$-$\overline{\rm D5}$ pairs wrapping this geometry will effectively behave as four-dimensional,
and therefore the IR geometry
should be four-dimensional.

We can make this a bit more precise. The supergravity solution for pairs of ${\rm D5}$-$\overline{\rm D5}$ branes
on a {\it flat}
background can be given in the following way\footnote{It would be interesting to compare the analysis below with the one
done in \cite{poln3} where supergravity solution related to D7 and fractional D3-branes is studied. Our analysis is
very different from the one in \cite{poln3} as we will be studying the system from its D5-${\overline{\rm D5}}$
perspective, and not from its D3 perspective,
so as to capture the UV and IR behaviors.} \cite{townsend1, townsend2, bk}:
\bg\label{bakkarch}
ds^2 =&&-V_1^{-1} V_2^{-1/2}(dx_0 - kdx_7)^2 + V_2^{-1/2}dx^2_2 + V_2^{-1/2} dx_1^2
+ V_2^{-1/2} dx_3^2 \nonumber\\
&& \qquad + V_2^{1/2} (V_1^{-1}dx_6^2 + dx_7^2) + V_2^{1/2} (dx_4^2 + dx_5^2 + dx_8^2 + dx_9^2),
\nd
where stability and supersymmetry requires us to switch on an electric field $F_{0i}$
with $\vert F_{0i}\vert^2 < 1$ and a magnetic field $F_{67}$
with opposite signs on the ${\rm D5}$-$\overline{\rm D5}$ pairs. This is slightly different choice of the
world-volume fluxes compared
to the ones that we took in the previous subsections and in appendix \ref{tachyonappendix}. One may however easily
verify that both the
choices result in identical physics\footnote{In this framework
one might worry about the fundamental string oriented parallel to the
seven-branes i.e along $x^6$ direction. This can be dissolved in one of the D7-brane and then moved away in the
$u$-plane, so that local ${\cal N} = 2$ supersymmetry remains unaffected. This is equivalent to the statement that
we can go to a frame where only world-volume magnetic field is turned on and the electric field is zero.
In the framework studied in the earlier
sub-section and in appendix \ref{tachyonappendix} there are only fractional D3-branes and no fundamental strings.}.

Due to the existence of an electric field (with say $i = 6$) there would be bound fundamental strings, and due to the
magnetic fields $F_{67}$ there would be bound D3-branes. The D3-brane charge is then typically given by
$F^{(2)}_{67} - F^{(1)}_{67}$ as we saw before. If we keep the five-branes at the same point in the $u = x^4 + i x^5$
plane but separate the anti five-branes very slightly along the $r = \sqrt{(x^8)^2 + (x^9)^2}$ directions, then
\bg\label{dista}
V_i = 1 + \alpha_i\left[{1\over \vert u\vert^2 + r^2}+  {1\over \vert u\vert^2 + (r- \epsilon)^2}\right], ~
k = \beta\left[{1\over \vert u\vert^2 + r^2} - {1\over \vert u\vert^2 + (r- \epsilon)^2}\right], \nonumber\\
\nd
where $\alpha_{1,2}, \beta$ are functions of $g_s, l_s$; and the number of
fundamental strings, D3-branes and five-branes respectively.
Note that in $k$ the two terms come with a relative minus sign, so that when $\epsilon \to 0$, $k$
vanishes.

The above picture is not complete as we haven't yet accounted for the multi Taub-NUT space and seven-branes. Let us
first consider the multi Taub-NUT space oriented along $x^{6,7,8,9}$ directions. The multi Taub-NUT
space modifies the $x^{6,7,8,9}$ directions in the following way:
\bg\label{dhosta}
ds^2_{\rm TN} = \left(1 + \sum_{\sigma}{1\over \vert \vec{w}-\vec{w_\sigma}\vert}\right) d\vec{w}^2 +
\left(1 + \sum_{\sigma}{1\over \vert \vec{w}-\vec{w_\sigma}\vert}\right)^{-1} \left(dx^6 + \sum_\sigma F^\sigma_i
dx^i\right)^2, \nonumber\\
\nd
where $\sigma$ denotes 
Taub-NUT singularities and $\vert \vec{w}\vert = \sqrt{\vert r\vert^2 + (x^7)^2}$
denotes the distance along the Taub-NUT space.

In addition to the Taub-NUT space, we also have the seven-branes distributed in some way to give rise to the
global symmetries in the theory. For generic distribution of the seven-branes the resulting gauge theory is not
conformal. The metric orthogonal to the seven-branes along the $u$-plane is given by the following
expression (see for example \cite{gsvy}):
\bg\label{ortho}
ds^2_u = \tau_2(u) \Bigg\vert \eta^2(\tau(u))\prod_{i=1}^{24} {du\over (u - u_i)^{1/12}}\Bigg\vert^2,
\nd
where $\tau_2(u)$ is the imaginary part of $\tau(u)$ on the $u$-plane and $\eta(\tau)$ is the $\eta$-function (we are
using the notations of \cite{gsvy}).

Now combining \eqref{ortho}, \eqref{dhosta} and \eqref{bakkarch} we obtain the total
background metric. The $dx^6$ component of \eqref{bakkarch} should be replaced by the
$U(1)$ fibration metric of \eqref{dhosta} and the ($dx^4, dx^5$) part of \eqref{bakkarch} should be replaced by the
backreaction from the seven-branes, i.e \eqref{ortho}. Together, the final picture would be pretty involved, and
will take the following form:
\bg\label{totla}
ds^2 =&&-f_1V_1^{-1} V_2^{-1/2}(dx_0 - kdx_7)^2 + f_2V_2^{-1/2}dx^2_2 + f_3V_2^{-1/2} dx_1^2
+ f_4V_2^{-1/2} dx_3^2 \nonumber\\
&& + f_5V_1^{-1}V_2^{1/2} \left(dx^6 + \sum_\sigma F^\sigma_i dx^i\right)^2
+ f_6V_2^{1/2} \tau_2(u) \Bigg\vert \eta^2(\tau(u))\prod_{i=1}^{24} {du\over (u - u_i)^{1/12}}\Bigg\vert^2 \nonumber\\
&& + f_7 V_2^{1/2} dx_7^2 + f_8V_2^{1/2}(dx_8^2 + dx_9^2),
\nd
where we expect ($f_5, f_7, f_8$) to be functions of ($\vert u\vert, \vec{w}$) so that informations about the
multi Taub-NUT space can be captured\footnote{Note that the multi Taub-NUT geometry is deformed due to the
backreactions of branes and fluxes in the background.}.
The other $f_i$ would definitely be functions of $\vert u\vert$ but could
have dependences on other coordinates too. The $V_i$'s now specify the harmonic functions for all the wrapped
D5-$\overline{\rm D5}$ pairs on the multi Taub-NUT two-cycles.

In the above metric we can go to either the conformal or the non-conformal limits. The conformal limits will be
given by some special re-arrangements of the 24 seven-branes whose individual contributions appear in \eqref{totla}.
The non-conformal limits are of course any generic distributions of the seven-branes in \eqref{totla}. In addition, this limit can be probed by adding fractional D5-branes as will be explained in the next section. Each of the
two limits would also have their individual UV and IR behaviors. The IR behavior for both the conformal as well as the
non-conformal limits shouldn't be too difficult to determine from the above form of the metric \eqref{totla}.

At IR we expect that all informations about the multi Taub-NUT space will be washed out (because we are probing
the system with wavelengths larger than the resolutions of the Taub-NUT singularities). This means at IR the
system is probed by D3-branes. We also expect
\bg\label{ko}
k ~ \equiv ~ \beta\left[{1\over \vert u\vert^2 + r^2} - {1\over \vert u\vert^2 + (r- \epsilon)^2}\right] ~ = ~ 0
\nd
in \eqref{totla}, as the ${\rm D5}$-$\overline{\rm D5}$ pairs would effectively overlap, and so there would be no $dx_0dx_1$ cross-terms
in the metric. The metric then takes the following form:
\bg\label{dhongsho}
ds^2 = && {1\over \sqrt{V_2}}\Big(-f_1 V_1^{-1} dx_0^2 + f_3 dx_1^2 +
f_2 dx_2^2 + f_4 dx_3^2\Big)
+ \sqrt{V_2}\Bigg[f_5V_1^{-1}\left(dx^6 + \sum_\sigma F^\sigma_i dx^i\right)^2 \nonumber\\
&& ~~~~~ + f_7 dx_7^2 + f_8 (dx_8^2 + dx_9^2)
+ f_6 \tau_2(u) \Bigg\vert \eta^2(\tau(u))\prod_{i=1}^{24} {du\over (u - u_i)^{1/12}}\Bigg\vert^2 \Bigg],
\nd
which is very suggestive of the multi D3-brane metric provided certain conditions are imposed on
($f_1, \cdots,  f_4$) at IR. The condition that we want for our case would be the following obvious one:
\bg\label{obvious}
f_1 V_1^{-1} ~ \approx ~ f_i
\nd
with $i = 2, 3, 4$. This is not too difficult to show. In the far IR, as we discussed above, the system is described
by D3-branes probing the geometry instead of the ${\rm D5}$-$\overline{\rm D5}$ pairs. This means that the
Taub-NUT geometry is essentially decoupled from the D3-brane geometry implying that the metric seen along
the D3-brane directions is given by the first line of \eqref{bakkarch} with $k = 0$ i.e $f_1 = f_i$. Additionally the
other warp factor $V_1$ is defined in terms of the fundamental strings, $\alpha_1$, as shown in \eqref{dista}. We can
always make a Lorentz transformation to go to a frame of reference where only world-volume magnetic fields,
$F_{67}^{(1, 2)}$, are turned on
and the electric field, $F_{0i}$, is zero (see also footnote 19). Thus $\alpha_1 \to 0$ and $V_1 \approx  1$, so that
\eqref{obvious} is satisfied.
Therefore our {\it ansatze} for the IR metric will take the following form:
\bg\label{irgeometry}
ds^2_{\rm IR} = && {\cal F}_1^{-1/2} ds^2_{0123} + {\cal F}_1^{1/2} ds^2_\perp =
{\cal F}_1^{-1/2} (-dx_0^2 + dx_1^2 + dx_2^2 + dx_3^2) + {\cal F}_2 \vert d\vec{w}\vert^2 \nonumber\\
&& + {\cal F}_3 \left(dx^6 + \sum_\sigma F^\sigma_i dx^i\right)^2 + {\cal F}_4
\tau_2(u) \Bigg\vert \eta^2(\tau(u))\prod_{i=1}^{24} {du\over (u - u_i)^{1/12}}\Bigg\vert^2,
\nd
where ${\cal F}_i$ are related to each other by supergravity EOMs. Thus the non-conformal IR limit does not have any
immediate simplification. But if we go to the special arrangements of the seven-branes where we expect
constant coupling scenarios \cite{senF, DM1} then the EOMs connecting ${\cal F}_i$ should simplify to give us
the near-horizon $AdS_5$ geometry.

On the other hand, our background \eqref{totla} could also tell us about the UV geometry. At UV we cannot ignore the
Taub-NUT singularities and therefore the ${\rm D5}$-$\overline{\rm D5}$ pairs would wrap various vanishing cycles of the multi
Taub-NUT geometry. The D5-brane charges cancel, but as we discussed before, the D3-brane charges add up and in fact
the D3-branes are {\it delocalized} along the $x^{6,7}$ directions. From the above discussions, we now expect the UV
metric to be given by:
\bg\label{ltubaz}
ds^2_{\rm UV} = && {1\over \sqrt{V_2}}\Bigg[-f_1 V_1^{-1} dx_0^2 + f_3 dx_1^2 +
f_2 dx_2^2 + f_4 dx_3^2 + f_7 V_2dx_7^2
+ f_5 V_2 V_1^{-1}\left(dx^6 + \sum_\sigma F^\sigma_i dx^i\right)^2\Bigg] \nonumber\\
&& + \sqrt{V_2}\left[f_8 (dx_8^2
+ dx_9^2)
+ f_6 \tau_2(u) \Bigg\vert \eta^2(\tau(u))\prod_{i=1}^{24} {du\over (u - u_i)^{1/12}}\Bigg\vert^2 \right].
\nd
Therefore the UV physics is now captured not by
a four-dimensional spacetime, but by a six-dimensional spacetime! In the constant coupling scenario of \cite{senF, DM1}
the near-horizon geometry should give us an $AdS_7$ spacetime. It would be interesting to compare the UV and IR
limits with \cite{GM} (and the earlier work of \cite{poln3}).

Before moving further, let us make two comments on the IR metric of \eqref{irgeometry}. This will help us to compare
our F-theory constructions with the brane constructions in type IIA \cite{wittenM} and the brane network in type
IIB \cite{bbt}.

\begin{itemize}
\item The above metric \eqref{irgeometry} {\it cannot} come from a type IIA brane configuration with
NS5, D4 and D6-branes. In fact even in the so-called delocalized limit the form \eqref{irgeometry} cannot be
recovered. In particular it is not possible to see how the second term in the second line of \eqref{irgeometry} could
appear from D6-branes of type IIA\footnote{One might observe that a T-duality along the isometry direction of the
Taub-NUT space i.e along the $x^6$ direction, naively leads to a NS5-brane {\it delocalized} along the
$x^6$ direction. In \cite{ghm} this issue has been addressed in great details and the final
answer reveals an additional dependence of the NS5 harmonic function along the angular $x^6$ direction (see also
\cite{tong}). However similar analysis have not been attempted for the seven-branes, and at this stage it is
not {\it a-priori} clear to us how this T-duality should be taken to allow for localized gravitational solutions.}.
\item As we discussed in the previous subsubsection \ref{subsecNetwork}, a T-duality along $x^4$ or $x^5$ to get the brane
network model of \cite{bbt} is {\it not} possible because the metric \eqref{irgeometry} has non-trivial
dependence along the $u$-plane! If we delocalize along these directions then we can recover the brane network of
\cite{bbt} but will lose all non-trivial information on the $u$-plane. Therefore the F-theory picture captures
more information than the brane network of \cite{bbt}.
\end{itemize}

Thus from the above comments we see that the F-theory models are in some sense better equipped to capture non-trivial
informations of the corresponding gauge theories as the probe branes have direct one-to-one connections to the
corresponding gauge theories. The only restriction that we could see in our models has to do with the {upper-bound}
on the {\it number} of seven-branes. F-theory tells us that the number of seven-branes have to be at most 24 otherwise
the singularities on the $u$-plane will be too drastic to have a good global description \cite{vafaF}. This restriction on the
number of seven-branes (or to the global symmetries of the corresponding gauge theories) should not be too much of
an issue because one may resort to only local F-theory description assuming that the global completions may be done
by introducing anti-branes that would preserve ${\cal N} = 2$ supersymmetry up to certain energy scales (see also
\cite{kleban}).
The energy scale
may be chosen in such a way that all the above discussions may succinctly fit in. The global symmetries in these
theories may then be made arbitrarily large so as to encompass most of the Gaiotto's models. It would of course be
an instructive exercise to explicitly demonstrate a concrete example with a large global symmetry that, in the
Seiberg-Witten sense, remains
{\it integrable}. Once there, the far UV picture of this model should be interesting to unravel from our set-up.

One final thing before we end this subsection is to analyze the background fluxes. At the far IR the six-form
charges should cancel completely but at UV they should appear as dipole charges\footnote{For
${\rm D5}$-$\overline{\rm D5}$ pairs,
the tachyonic behavior emerges at distances of
order $\sqrt{\alpha'}$ or less \cite{dhhk}. (See appendix \ref{tachyonappendix}.)
In a tachyon-free system, we expect D5-$\overline{\rm D5}$ to be separated by a distance larger than that, creating a dipole moment in the system.
}.
The four-form charges should be
quantized and should be proportional to the number of ${\rm D5}$-$\overline{\rm D5}$ pairs. In addition to that there would be a
background axio-dilaton $\tau(u)$ that is a function on the $u$-plane, and  NS and RR two-forms field with the
required three-form field strengths. For the conformal cases we expect $\tau(u)$ to take one of the values given in
\cite{senF, DM1}. These fluxes and branes deform both the Taub-NUT and the seven-brane geometries and together
they preserve the required supersymmetry for our case.

\subsubsection{Mapping to the conformal cases \label{ConfGMap}}

The ${\rm D5}$-$\overline{\rm D5}$-brane pairs at the Taub-NUT singularities also tell us what the UV gauge symmetry should be for our case.
Imagine we have a $m$ multi-centered Taub-NUT geometry, then the $N$ ${\rm D5}$-$\overline{\rm D5}$ brane pairs wrapped around the
$m$ vanishing cycles lead to $mN$ fractional D3-branes where each of the $N$ fractional D3-branes carry a total RR
charge of $N/m$ in appropriate units.
Since there are $m$ copies of this, there is a total charge of $N$ D3-branes, leading us to
speculate the UV gauge symmetry to be $m$ copies of $SU(N)$, i.e:
\bg\label{uvgs}
SU(N) \times SU(N) \times SU(N) \times \cdots \times SU(N)  .
\nd
Once the wrapped ${\rm D5}$-$\overline{\rm D5}$ pairs are decomposed in terms of fractional D3-branes\footnote{Recall that there are no
D5-brane charges in the background.},
these fractional branes can now freely move along the F-theory $u$-plane, i.e
the Coulomb branch of the theory\footnote{Recall that if these D3-branes move along the Taub-NUT directions (i.e
the Higgs branch) they become fractional instantons.}.
This is illustrated in {\bf figure \ref{wrappedTN}}. However even the individual
set of $N$ fractional branes may separate by further Higgsing to $U(1)^N$. In that case the individual fractional
D3-brane carry a net RR charge of ${1/m}$ in appropriate units.

It is now interesting to see how supersymmetry
and global symmetries would constrain the underlying picture. Since the
D5-brane charges cancel, the model only has fractional D3-branes and therefore the fractional-D3 and seven-branes
preserve the required supersymmetry as we discussed before. However the global symmetries are crucial. So we
should look for various arrangements of the seven-branes that allow $\tau(u) = $ constant in the $u$-plane. These
arrangements should be related to the models studied by Gaiotto \cite{gaiotto}.
\begin{figure}[htb]
        \begin{center}
\includegraphics[height=3cm]{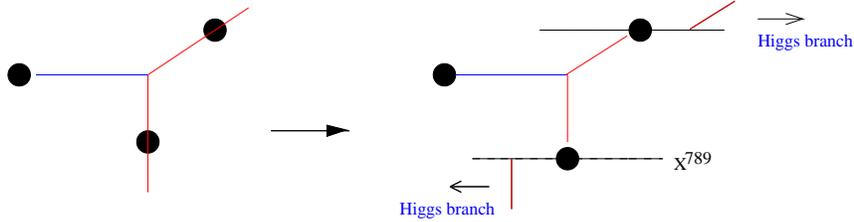}
        \end{center}
        \caption{The brane junction that we get in the left of the figure may further be truncated by
breaking the NS5 and the ($p, q$) five-brane and moving along the Higgs branch. However as we argue below,
this may not be required to see the underlying dualities.} \label{bbtreq}
        \end{figure}
One interesting example is related to the construction that we had in {\bf figure \ref{TN2I}} wherein we showed how
the simplest T-dual brane network may come out from our scenario. The D5-brane ends on the seven-brane, and so would
any horizontal (i.e along $x^6$) D5-branes in this scenario. However the NS5-branes and the ($p, q$) five-branes
have to intersect the seven-branes as shown in the left of {\bf figure \ref{bbtreq}}. From \cite{bbt} we might
expect the figure on the right where parts of the NS5 and the ($p, q$) five-branes have been moved away along the
Higgs branch. This configuration is in principle rather non-trivial to get from the Taub-NUT scenario, but
there is no reason for an exact one-to-one correspondence \cite{bbt} as we argued earlier.
The ${\cal N} = 2$ dualities should in principle be seen
as long as we have the D5-brane configurations right.

\begin{figure}[htb]
        \begin{center}
\includegraphics[height=3cm]{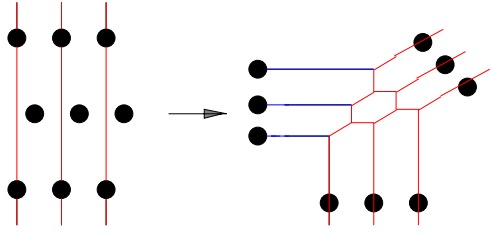}
        \end{center}
        \caption{On the left is a non-susy configuration that appears from naive T-duality of the
Taub-NUT model, as the branch cuts of the seven-brane (shown as solid black circles) would modify the
parallel NS5-branes configuration from the local axio-dilaton charges. This is similar to the deformation of a
string-network from background axion as shown in \cite{DMstring}. The arrangement of the seven-branes along
$x^6$ direction come from the original framework of the wrapped seven-branes with $A_6$ switched on. T-duality convert
$A_6$ to $x^6$ shown on the left.
On the right is the susy configuration by moving the seven-branes across the
``bent'' NS5-branes.} \label{asmodel}
        \end{figure}

To see further how this is implemented let us consider
our UV configuration of a three
set of three seven-branes wrapping three-centered Taub-NUT manifold with no fractional D3-branes. One immediate
advantage of this is that, since there are no fractional D3-branes to start with, a T-dual map to \cite{bbt} will be
easier.
The Weierstrass
equation governing the background at a given point is given by:
\bg\label{weq1}
y^2 ~ = ~x^3 + x (c_0 + c_1 z) + b_0 + b_1z + b_2 z^2
\nd
where we are choosing the {\it split} case of the Tate algorithm \cite{tate} where the choices of ($c_i, b_i$)
can be read off from eq. (4.7) of \cite{bikmsv}. This means that the discriminant locally is of the form
\bg\label{disloc}
\Delta ~ \sim ~ z^3
\nd
and so if we have three copies of this on the $u$-plane, we are
guaranteed that we will have {\it no} gauge symmetry but only a global symmetry of
\bg\label{glon}
 SU(3) \times SU(3) \times SU(3).
\nd
In one set of three seven-branes we can first switch on constant $A_6$ fields so that a T-duality along $x^6$
may lead to an arrangement of the seven-branes shown in the LHS of {\bf figure \ref{asmodel}}. Note however that
due to the background axio-dilaton, the T-dual NS5-branes will {\it not} remain straight. The background axio-dilaton
will affect the NS5-branes and they will in turn get bent. This phenomena is exactly what we see for string
networks. In \cite{DMstring} it was shown how a network of ($p, q$) strings get bent in the presence of axio-dilaton.

Once this is taken care of, we can
switch on a time-varying gauge field on the same set of seven-branes in exactly the similar way we discussed
earlier. This would create fractional D3-branes to cancel the gauge anomalies which, in the T-dual framework, is
given by the RHS of {\bf figure \ref{asmodel}}. The other two sets of three seven-branes\footnote{Clearly not all the
seven-branes are D7-branes, as we would need [$p, q$] seven-branes for consistency with Gauss' law.}can be arranged to
intersect the NS5 and the ($p, q$) five-branes. This is also exactly the configuration studied in \cite{bbt} (with
mild differences).

Note that the above configuration is in principle {\it different} from the configuration of
three fractional D3-branes probing seven-branes background where the seven-branes wrap
multi-centered Taub-NUT geometry. The T-dual of the $m$-center Taub-NUT space would be $m$ parallel NS5-branes as above.
The UV gauge group will be determined as \eqref{uvgs} but we may only consider
the low energy limit where the Taub-NUT singularities are not prominent. This however doesn't mean that we have
recovered the above model because there would still be a remnant gauge symmetry in the model even at far IR. We may
play the same game of removing fractional D3-branes by
switching on time-varying gauge field on each of the Taub-NUT cycles but the model will not be similar to our earlier
case and the gauge theory dynamics will be different.

Coming back to our model, we can now rearrange the seven-branes using F-theory Weierstarss equation to go to another
limit with a different global symmetry. This time the Weierstrass equation can be changed from \eqref{weq1}
to the following local form:
\bg\label{weq2}
y^2 ~ = ~ x^3 + z^4,
\nd
implying a global $E_6$ symmetry on the gauge theory side. From F-theory side the discriminant locus and the
underlying four-fold ${\cal M}_8$ will become respectively:
\bg\label{disffold}
\Delta ~ \sim ~ z^8, \qquad {\cal M}_8 ~ = ~ {\mathbb R}^4/{\bf Z}_3 ~ \times ~ {\bf TN}_3.
\nd
Observe that globally the K3 manifold has degenerated to its ${\bf Z}_3$ orbifold limit with a full global symmetry
of $E_6^3$ \cite{DM1}. For this global symmetry we are indeed at the constant coupling
point \cite{DM1} (see also \cite{minahan1, minahan2}).

In our Taub-NUT picture we have now redistributed the seven-branes now as three sets with {\it eight} seven-branes
in each set. We can move the two set of sixteen seven-branes in the $u$-plane so that we only allow a global symmetry
of $E_6$. Our picture can also be supported by the T-dual brane network of \cite{bbt}. This then would realize the
Argyres-Seiberg duality \cite{sa}.

Yet another example to consider would be to view the $SU(3)$ global symmetry to come from an $SU(4)$ symmetry by
Higgsing the ${\bf 4}$. This means we are bringing in another set of seven-branes so that the overall configurations wrap
a Taub-NUT with four singularities. The local Weierstrass equation now will be:
\bg\label{weq3}
y^2 ~ = ~ x^3 + x(c_0 + c_1 z) + b_0 + b_1z + b_2z^2 + b_3z^3
\nd
with special relations between ($c_i, b_i$) such that we are at the {\it split} ${\bf A}_3$ case \cite{tate}.
These relations are
worked out in \cite{bikmsv} which the readers may look up for more details. The discriminant locus is as expected:
\bg\label{dlo}
\Delta ~ \sim ~ z^4,
\nd
so that we have a global $SU(4)$ symmetry. As before if we make three copies of this we will have the required
global symmetry of $SU(4)^3$. This configuration maps directly to the brane network studied in \cite{bbt} so we
don't have to go through the details. It suffices to point out that the rearranged seven-branes may now give a
global symmetry of (see also \cite{DM1})
\bg\label{rags}
E_7 ~\times ~E_7 ~\times ~ SO(8),
\nd
so that the axio-dilaton remains constant throughout the $u$-plane. Locally near one of the $E_7$ singularity the
F-theory manifold is typically an orbifold of the form:
\bg\label{orbF}
{\mathbb R}^4/{\mathbb Z}_4 ~ \times ~ {\bf TN}_4,
\nd
 which means that our K3 has become a ${\mathbb Z}_4$ orbifold of the four-torus.

The above decomposition of the underlying K3 manifold into its various orbifold limits give us a hint what the next
configuration would be. This would be the ${\mathbb Z}_6$ orbifold of the four-torus so that the conformal global symmetry
should be \cite{DM1}
\bg\label{cgssb}
E_8 ~\times ~ E_6 ~\times ~ SO(8).
\nd
Now since the ${\mathbb Z}_6$ orbifold creates a deficit angle\footnote{Recall that for a singularity to be of an
orbifold type the deficit angle has to be $2\pi \left(1 - {1\over n}\right)$ for a fixed point of order $n$.}
of at most ${5\pi \over 3}$ we know that this is an orbifold with a fixed point
of order 6. Therefore our starting point would be to put three copies of six seven-branes wrapping a Taub-NUT
with six-singularities leading to an $SU(6)^3$ global symmetry. This then clearly enhances to \eqref{cgssb} with the
local F-theory four-fold given by:
\bg\label{orbF}
{\mathbb R}^4/{\mathbb Z}_6 ~ \times ~ {\bf TN}_6.
\nd
The above set of configurations were studied without incorporating any ${\rm D5}$-$\overline{\rm D5}$-branes in the background. Once we
introduce the probes we will not only have global symmetry, but also gauge symmetry. A special rearrangement of the
seven-branes may help us to study the conformal theories leading to other Gaiotto dualities. We will discuss a
more detailed mappings to these cases in the sequel.

\subsubsection{Beyond the conformal cases \label{subsecNonConf}}

\begin{figure}[htb]
        \begin{center}
\includegraphics[height=6cm]{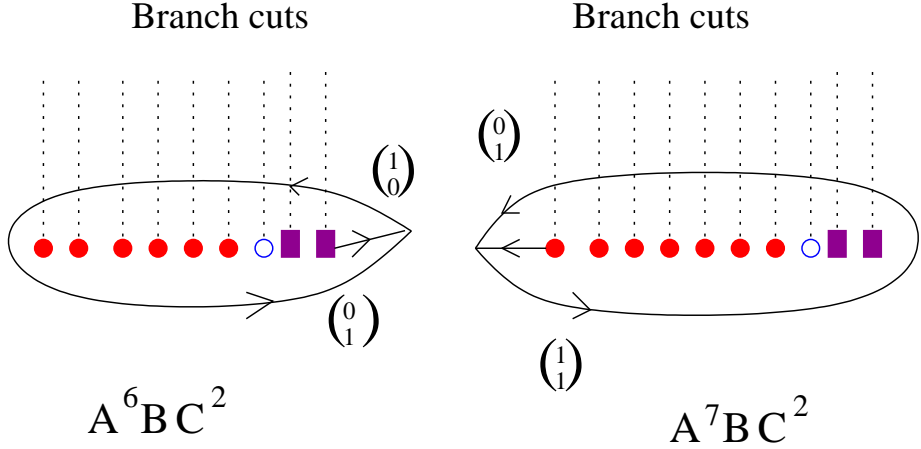}
        \end{center}
        \caption{A simple non-conformal deformation of the theory where the
multiple ${\rm D5}$-$\overline{\rm D5}$ branes are probing a $E_7 \times E_8$ singularity. The
global symmetry will have additional $U(1)$ that are not shown in the figure.
In the language of \cite{gaberdiel} the filled circles are $A$-branes, the
empty circles are $B$-branes and the filled squares are $C$-branes.} \label{nonconformal}
        \end{figure}

Since our model is a direct construction in F-theory, all informations of the type IIB background
under non-perturbative corrections are transferred directly to the D3-brane probes. This in particular means that
arrangements of the seven-branes that lead to non-trivial axio-dilaton backgrounds would also be transferred to the
D3-brane probes, except now they would appear as non-conformal theories on the D3-branes. A simple non-conformal
deformation, with our set-up discussed in the previous subsection, is given in {\bf figure \ref{nonconformal}}. This
could be generated from $SO(8)$ in \eqref{rags}
breaking completely to $SU(2)$ by first going to $SO(7)$ and then $SO(7)$ breaking to
$SU(2)\times SU(2)\times SU(2)$. Recall from \cite{gaberdiel} that\footnote{In the following ${\bf A}, {\bf B}$ and
${\bf C}$ are the three monodromy matrices given as:
$${\bf A} ~ \equiv ~ \begin{pmatrix} 1&{1}\\ {0} & 1 \end{pmatrix}, ~~~~
{\bf B} ~ \equiv ~ \begin{pmatrix} ~4&{~9}\\ {-1} & ~2 \end{pmatrix}, ~~~~
{\bf C} ~ \equiv ~ \begin{pmatrix} ~2&{~1}\\ {-1} & ~0 \end{pmatrix}$$
These monodromy matrices are derived from the monodromies around D7 and the two ($p, q$) seven-branes in {\bf figure 8}
respectively. For more details the authors may refer to \cite{gaberdiel}.}
\bg\label{so8}
SO(8) ~ \equiv ~ {\bf A}^4 {\bf B} {\bf C},
\nd
so that the perturbative pieces generate the subgroup $SU(2) \times SU(2)$. Separating the [$0, 1$] and the [$1, -1$]
seven-branes from the bunch of the six seven-branes allow us to achieve this. Once we further break the other $SU(2)$,
we can easily generate the ${\bf 248}$ of $E_8$ from \eqref{rags} via:
\bg\label{gene8}
{\bf 248} ~ = ~ ({\bf 3}, {\bf 1}) ~ \oplus~ ({\bf 1}, {\bf 133}) ~ \oplus~ ({\bf 2}, {\bf 56}) .
\nd
\noindent This is clearly a non-conformal deformation in the Taub-NUT background as the axio-dilaton
values are no longer
constant in the $u$-plane. It is interesting to note that if we take other model \eqref{cgssb} then there exist a
limit where the non-conformal deformation in this model is precisely the non-conformal deformation of the earlier case.
This is when $SO(8)$ in \eqref{cgssb} is completely broken to $U(1)$ by moving all the ${\bf A}, {\bf B}$ and
${\bf C}$ branes except one ${\bf A}$ brane. Under this circumstances the ${\bf 56}$ of $E_7$ is easily generated from
\eqref{cgssb} for the D3-brane probes to see identical physics as the earlier case:
\bg\label{56e7}
{\bf 56}~ = ~ {\bf 1} ~ \oplus ~{\bf 1} ~ \oplus ~{\bf 27} ~ \oplus ~ {\overline{\bf 27}}.
\nd
Finally, to see similar non-conformal deformation from the first $E_6^3$ model that we studied above, we can go back to
the {\it unenhanced} case for one of the $E_6$ group, namely the $SU(3)^3$ global symmetry with the particular
arrangements of the seven-branes as in {\bf figure \ref{asmodel}}. For this case the ${\bf 56}$ of $E_7$ is
generated as \eqref{56e7}, but the ${\bf 248}$ of $E_8$ is now generated via:
\bg\label{e8e7ks}
{\bf 248} ~ = ~ ({\bf 8}, {\bf 1})~ \oplus ~({\bf 1}, {\bf 78})~ \oplus ~({\bf 3}, {\bf 27})~ \oplus
~({\bar{\bf 3}}, {\overline{\bf 27}}),
\nd
provided of course that the remnant subgroup of $SU(3) \times SU(2) \times U(1)$ is completely broken. Only under this
case the physics seen by the D3-brane probes will be identical.

The above non-conformal deformations were extensions of the conformal theories with exceptional global symmetries. They
aren't the simplest non-conformal models that we could study here. There exist simpler models if we introduce, in
addition to the ${\rm D5}$-$\overline{\rm D5}$-brane probes, some {\it additional} D5-branes wrapping vanishing 2-cycles of the Taub-NUT
space.

Let us take a concrete example where we have $k$ ${\rm D5}$-$\overline{\rm D5}$-brane pairs at a point in the Taub-NUT space with $m$
singularities. In addition to these probes, let us also introduce $M_i$ (with $i = 1, \cdots, m$)
D5-branes wrapping the $m$ vanishing 2-cycles
of the Taub-NUT space. It is immediately clear that the gauge symmetry now will change from \eqref{uvgs} to the
following:
\bg\label{gsnwb}
\prod_{i = 1}^m SU(k + M_i) ~ \equiv ~ SU(k + M_1) ~\times ~SU(k + M_2) ~\times ~\cdots~ \times ~SU(k + M_m).
\nd
The above theory is obviously non-conformal as the additional wrapped D5-branes break the conformal invariance
already in the absence of any flavor symmetry. If we take the Taub-NUT and wrap $M$ D5-branes
on the vanishing 2-cycle,
then the gauge group
will be special case of \eqref{gsnwb}, namely
\bg\label{kebot}
SU(k + M) ~ \times ~ SU(k).
\nd
This brings us exactly to the cascading models of \cite{poln2, poln3, aharonyn2, benini2, benini3} where the authors
have argued cascading behavior in this model (see \cite{benini2, benini3} for a more recent study)! It is then
clear that our model can have an even more interesting cascading dynamics because there is an option of having a
 much bigger gauge group as can be seen from \eqref{gsnwb}.
Furthermore due to the presence of seven-branes, the cascading model is of the
Ouyang type \cite{ouyang1, ouyang2}. This means
there is a chance that cascades would be {\it slowed} down by the presence of
fundamental flavors, much like the one
studied in \cite{ouyang1, ouyang2}.

The connection to ${\cal N} = 1$ cascade \cite{KSks} is now clear: we can break the ${\cal N} =2$ supersymmetry
by non-trivially fibering the Taub-NUT over the compactified $u$-plane. Writing $u \equiv z_4 = x^4 + ix^5$, the
fibration is explicitly:
\bg\label{fibex}
z_1^2 ~+~ z_2^2 ~+~ z_3^2 ~= ~-z_4^2 ~ \equiv ~ -u^2,
\nd
which is an ALE space with coordinates ($z_1, z_2, z_3$)
fibered over the $u$-plane. Near the node $x^4 = x^5 = 0$ the geometry is
our familiar Taub-NUT space, and the equation \eqref{fibex} is a conifold geometry.
This is dicussed in the next section in
more details after \eqref{dhop}. Once this is achieved the ${\cal N} =1$ cascading behavior can take over with the
termination point governed by {\it either} a confining theory, or a conformal theory depending on the choice of
the flavor symmetry. We will provide a more detailed exposition of these ideas pertaining to the cascading model
in the sequel.

\begin{figure}[htb]
        \begin{center}
\includegraphics[height=6cm]{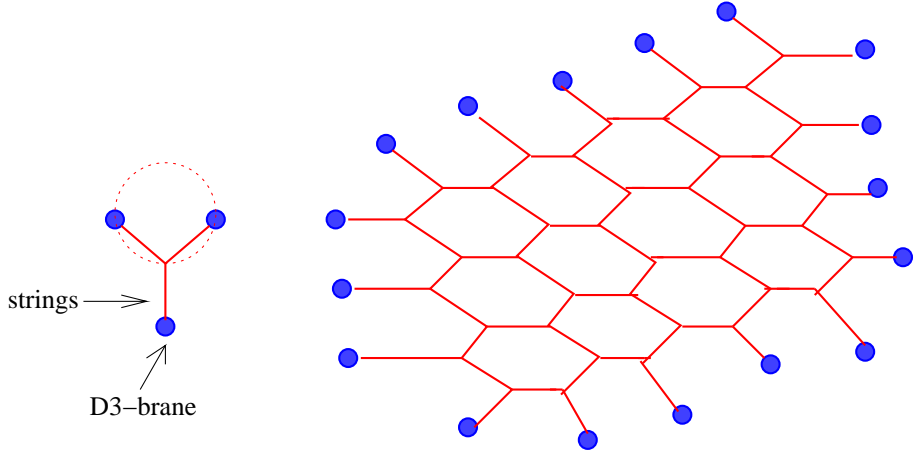}
        \end{center}
        \caption{New D3-brane states in the theory. The D3-branes are connected by either a 3-string junction or a string-network. These are stable states at IR and exist outside the corresponding curves of marginal stability.} \label{D3network}
                \end{figure}

There is yet another avenue that may open up if we go beyond the simple probe analysis.
One such scenario is depicted in {\bf figure \ref{D3network}}. In the F-theory framework there could be new
states if we can connect multiple D3-branes via a string network (much like the ones discussed in \cite{neksav}).
When such a network of D3-branes are made
to probe our Taub-NUT background, the IR physics would still remain ${\cal N} = 2$ supersymmetric.
Once we are away from the curves
of marginal stability (shown on the left of {\bf figure \ref{D3network}}) the dynamics of the D3-branes will be
constrained by the network and so it would be interesting to unravel the world-volume dynamics.

Before we end this section, let us take a
slight detour and lift the simplest
IR configuration to M-theory (that involves lifting the type IIB D3-branes with no network between them and no wrapped
D5-branes to break the conformal invariances, to M-theory).
This will in fact help us to go smoothly to the
next model. In the following we will give a brief discussion and more details will be relegated to the next section.

When we lift our configuration to M-theory we have a configuration of $k$ M2 branes at a point on the
{\it non-compact} four-fold ${\rm K3} \times {\bf TN}_n$, where $k$ is any integer.
Since the four-fold is non-compact there is no anomaly or charge cancellation condition, and therefore any
numbers of M2-branes can be added to the system. The various arrangements of the seven-branes go to the
singularities of the torus inside the K3 manifold. The orientation of this torus gives us the
quantum corrected type IIB axio-dilaton. Interestingly
in M-theory, as we will see later,
one may {\it trade off} some
of the M2-branes with background $G$-fluxes (see also \cite{becker4}, \cite{DRS}). This will be related to an
interesting interplay between {\it abelian} instantons and branes.
The $G$-fluxes appear as NS and RR fluxes in
the type IIB side. The NS fluxes could be gauge transformed to F-fluxes on the seven-branes\footnote{Note that
as we are {\it not} at the orientifold point the $B_{\rm NS}$ fluxes could be oriented parallel to the seven-branes.}.
Of course the physics of these fluxes is different from the physics of the self-dual fluxes \eqref{selfdual} that we
used to understand Gaiotto-type dualities.

This basically concludes our discussion about the connection between a class of Gaiotto models and F-theory with a multi Taub-NUT 
space. In the following section we will further modify the Taub-NUT space and study the subsequent implications. 

\newpage

\section[Model 3: Multiple D3's probing seven-branes on a K3 background]{Model 3: Multiple D3-branes probing seven-branes on a K3 background \label{model3section}}

All the above consequences with Taub-NUT spaces have rather straightforward explanations. An interesting question to
ask now is what happens if we compactify the Taub-NUT space? Any generic compactification will break
supersymmetry, but if we replace the Taub-NUT space by a ${\mathbb Z}_2$ orbifold of ${\bf T}^4$ or by a K3 manifold
then supersymmetry will be restored. However now due to the compactness of the internal space, new complications
will arise. This can be easily seen by lifting the configuration to M-theory as before. We now have M-theory on
either ${\rm K3} \times {\bf T}^4/{\mathbb Z}_2$ or ${\rm K3} \times {\rm K3}$
manifold\footnote{A more non-trivial manifold to consider will presumably be
${({\bf T}^4 \times {\bf T}^4)/\left({\mathbb Z}_2 \times {\mathbb Z}_2\right)}$ with the two ${\mathbb Z}_2$
actions acting in various
different ways on the two ${\bf T}^4$.
In this paper we will only take the simplest ${\mathbb Z}_2$ actions whose
blow-ups are given by ${\rm K3} \times {\rm K3}$. More non-trivial actions of the ${\mathbb Z}_2$ groups will be
investigated in
the sequel.}.
In the absence of fluxes this system can only be
probed by 24 M2-branes. One may be able to reduce the number of M2-branes by switching on
$G$-fluxes \cite{SVW}, \cite{becker4}, \cite{DRS},
so in type IIB we can only have 24 D3-branes probing the background in the absence
of three-form fluxes, or lower number of D3-branes if we switch on the three-form fluxes. It is also important to
note that the D3-branes, in the presence of K3 manifold, cannot dissociate into ${\rm D5}$-$\overline{\rm D5}$ pairs
because such configurations will break supersymmetry. This also means that F-theory on ${\rm K3} \times {\rm K3}$
do have an
orientifold limit probed by the D3-branes, a fact well known from earlier works \cite{DRS}.

Our next question would be to see if we can argue for brane creation here also. With a little effort one may convince
oneself that there would be no $1+1$ dimensional chiral anomaly here, and therefore no reason for creating any
D3-brane sources. However this model has {\it two} different dualities.
The first one stems from the fact that the system is defined at an
orientifold point. This duality will take us to the heterotic theory where the 24 D3-branes would become
24 small instantons. The second one stems from the situation when the system is away from the orientifold point.
Now the background can be probed by multiple ${\rm D3}$-$\overline{\rm D3}$ pairs, but this time the duality will lead to
strongly coupled type IIB
on another non-K\"ahler manifold or even to an interesting M(atrix) theory!
We will start by elaborating the first case.

\subsection{F-theory/heterotic dualities and a supergravity picture \label{FhetSugra}}

There are two distinct cases that we need to understand from the heterotic dual. The first case is when the type
IIB background is probed by 24 D3-branes, and the second case is when the type IIB background is probed by $r$
D3-branes where $r < 24$. These two cases could also be analyzed from M-theory, as M2-branes probing a four-fold
geometry. In past such scenarios in M-theory were addressed with fluxes on a four-fold \cite{becker}. However the
scenario of \cite{becker} with fluxes may be replaced by branes using a simple trick.
To illustrate this, consider M-theory
compactified on a four-fold ${\cal M}_8$. This background becomes
unstable, unless fluxes satisfying
$$
\int_{{\cal M}_8}G \wedge G = {\chi \over 24}
$$
are included. Here $\chi$ is the Euler characteristic of ${\cal
M}_8$. The background is described with a warp factor $\Delta$ in the following way \cite{becker}:
\bg\label{beckback}
&&ds^2 = \Delta^{-1} ds^2_{012} +
{\Delta}^{1/2} ds^2_{{\cal M}_8} \nonumber\\
&& G_{012m} = \partial_m
\Delta^{-3/2}, \qquad  \square \Delta^{3/2} = {\rm
sources}.
\nd
The above background was derived in \cite{becker} {\it without} resorting to any branes. Now
imagine that we replace the fluxes by $n=\chi/24$ M2 branes. The
background then is the usual background with M2-branes which, in turn, has the following standard form:
\bg\label{metformtwo}
&& ds^2 = H^{-2/3}ds^2_{012} + H^{1/3}
ds^2_{{\cal M}_8} \nonumber\\
&& G_{012m} = {1\over 2} {\partial_m H \over H^2},
\qquad \square H = {\rm sources},
\nd
where $G$ is the source. The
two backgrounds above, i.e. equations \eqref{beckback} and \eqref{metformtwo},
become equivalent if we identify $H = \Delta^{3/ 2}$.

The above identification is not a big surprise as one could have rederived the result of \cite{becker} using
\eqref{metformtwo}. However a more general statement could be made at this stage which is not that obvious:
if we can replace branes by fluxes (or vice-versa if the system is
compact), the backgrounds for the two systems become identical.

Motivated by the M-theory argument, we can ask if we can play the same game in the
heterotic side. Let us first consider the case when the type IIB background is probed by 24 D3-branes. We will
also take the M-theory four-fold to be ${\rm K3} \times {\rm K3}$.
In the heterotic side this would be related to the
conformal K3 flux background, i.e in
the heterotic side we have ${\cal H}$-fluxes (RR three-form
fluxes for type I). We would like to interpret these fluxes as
heterotic NS5-branes (D5-branes for type I) wrapped on a torus,
exactly as we did for the M-theory case above. The metric for a
NS5-brane at a point of a non-compact K3 and a compact $T^2$ along $x^{4,5}$ is
\bg\label{nsfivemet}
&&ds^2 = ds^2_{012345} + H_5~ds^2_{K3}, \nonumber\\
&& {\cal H}_{\mu\nu\rho} = \sqrt{g}~
\epsilon^\sigma_{~~\mu\nu\rho} ~\partial_\sigma \phi, \qquad e^\phi =
\sqrt{H_5}.
\nd
Under an S-duality \eqref{nsfivemet} is mapped to the D5-brane metric
of type I. By redefining the harmonic function of the NS5-brane
as $H_5 = \Delta^2$ we will reproduce the standard conformal K3 result of \cite{strominger} with dilaton $\phi$ as:
\bg\label{strumi}
ds^2 = ds^2_{012345} + e^{m\phi} ds^2_{\rm K3}
\nd
 with $m = 2$. Thus, in
order to understand how the sources are mapped we express
\eqref{nsfivemet} in complex coordinates
\bg\label{derivofh}
{\cal H}_{ a \bar a b} = \left(g_{a \bar a} g_{b
\bar b} - g_{a \bar b} g_{\bar b a}\right)~g^{b \bar b}
~\partial_b\phi  = \partial_b e^{2 \phi} = \partial_b \Delta^2,
\nd
which agrees precisely with the torsional equation of \cite{stromtor}!

Since the internal manifold $K3 \times T^2$ is assumed to be large in appropriate stringy scales, the $3+1$ dimensional
theory will be weakly coupled\footnote{The four-dimensional coupling is inversely proportional to
the volume of the internal space. Although for us only a large $K3$ may suffice; for other reasons elaborated
in \cite{carluest} one would prefer large sized $K3\times T^2$.}.
The absence of cross terms in the
heterotic metric \eqref{strumi} implies that there is no $H_{NS}$ flux.
The above arguments also give us a simple way to verify that
supersymmetry is preserved on the type IIB side.

Let us now turn to the second case where we probe the type IIB background with
$r < 24$ D3-branes. A lift to M-theory will tell us that to cancel global charges and anomalies
we have to switch on fluxes. In the type IIB side this means that we should have three-form fluxes as
well as D3-branes.

This situation is more non-trivial and has not been addressed in the literature. If $r = 0$ then the heterotic dual is
given by the well known non-K\"ahler manifold discussed in \cite{DRS, becker1, becker2, becker3, becker4}. However
if we have {\it both} branes and fluxes then the non-K\"ahlerity will be further effected. In the following therefore
we will assume that, motivated by the equivalence between \eqref{beckback} and \eqref{metformtwo}, the warp factor
in the M-theory lift of this case will capture {\it both} the information of the fluxes as well as the M2-branes in the
limit where the M2-branes are probes. This will immediately imply that the type IIB metric is given by:
\bg\label{becmet}
ds^2 = \Delta^{-1} ds^2_{0123} + \Delta~ds^2_{K3} +
\Delta~ds^2_{P^1},
\nd
with all the internal directions proportional to the
same power of the warp factor $\Delta$. This warp factor is {\it not}
just the harmonic function of a D3-brane, i.e $\Delta \ne \sqrt{H_3}$.
On
the dual heterotic side we will have NS5-branes wrapping the
fiber of the non-K\"ahler manifold. This is different
from the discussion for the earlier case where we had heterotic NS5-branes wrapping a
torus. It would then seem that the crucial difference between the two cases come from the wrapping modes of the
heterotic five-branes. However the situation at hand is little more subtle than that.

First we should ask what it means to introduce ${\chi\over 24}$ number of M2-branes at a point on the
four-fold geometry. One simple way to interpret this would be to go back to the Chern-Simons coupling
$\int C \wedge G \wedge G$ and demand that:
\bg\label{gfla}
G \wedge G ~ = ~ {1\over 24}~ \chi ~ \delta^8(y - y_0)
\nd
i.e switching on localized $G$-fluxes in M-theory. This immediately implies that in the type IIB side we have
24 D3-branes as small instantons on the seven-branes, as the localized $G$-fluxes in M-theory map to the
seven-brane gauge fields \cite{DRS, becker2}. They in turn are obviously the heterotic small instantons!

On the other hand one may not demand \eqref{gfla} but still switch on a source of $\int C$ via switching on
non-zero $\int G \wedge G$ over the four-fold. These {\it cannot} be the heterotic small instantons as there
are no small instantons in the type IIB side. All informations about the small instantons now go as
{\it torsion} into the
heterotic side to deform the three-fold into a non-K\"ahler manifold. Therefore replacing branes by fluxes
in M-theory would amount to replacing abelian $C$-field small instantons by deformed instantons. These deformed
instantons can be viewed as the small instantons being fattened to a finite size. Or more appropriately,
in type IIB theory
this would mean that the D3-branes are dissolved in the seven-branes, and then the instanton sizes are increased
such that:
\bg\label{iibhap}
C_4 \wedge F \wedge F ~ \longrightarrow ~ C_4 \wedge H_{NS} \wedge H_{RR}.
\nd
The equivalent picture
on the heterotic side will be that of a
conformally K\"ahler manifold with small instantons going to a non-K\"ahler manifold with
torsion replacing the instantons. A somewhat similar picture was recently proposed in a slightly different scenario
involving ${\cal N} = 1$ dualities in the heterotic side in \cite{chen1, chen2}.

The above considerations related to the three-form background in the
heterotic/type I case may also be verified using the analysis done
in \cite{gauntlett}. Here it was shown that the heterotic three-form is
determined in terms of generalized calibrations of \cite{papado},
related to the G-structures of \cite{gauntlett}. According to these
references, there exist a generalized calibration-form $\Sigma$,
which determines the possible three-form background ${\cal H}$ in
the heterotic theory via the relation \cite{gauntlett}:
\bg\label{relofgaun}
\ast {\cal H} = d \Sigma +  \Sigma \wedge d\phi
\nd
where $\phi$ is the dilaton. For the six dimensional non-K\"ahler
manifold studied in \cite{sav, becker1, becker2} this would reproduce the
relation that we derived in \cite{becker3} via the superpotential. For the
present case of conformal K3, the first term of \eqref{relofgaun}
would vanish.
Further confirmation of the choice of the metric that is taken
here (i.e the conformal K3) comes from analyzing the torsion
classes \cite{carluest,louisL,gauntlett}. For the conformal K3 case,
we get \bg\label{otrclass} {\cal W}_1 = {\cal W}_2 =  {\cal W}_3  = 0,
\qquad {\cal W}_4 =  {\cal W}_5  =  2~d \phi,
\nd
where ${\cal W}_i, ~ i
= 1,\cdots, 5$ are the five torsion classes. Since all these
details have already appeared in \cite{carluest,louisL,gauntlett} we
will not repeat them here. It is now well known that taking a five brane $-$ wrapped on a
calibrated cycle of a given manifold $-$ we can reproduce the
torsional constraints in any dimensions. We have simply shown that the
${\cal N} = 2$ examples that we study in this paper also fall in this general category.
It is a happy coincidence that all these way of deriving the
results are mutually consistent.

\subsection{D3 brane-antibrane probes and a class of type IIB duals \label{subsecModel3D3IIB}}

The above connection between abelian $C$-field instantons and non-localized $G$-fluxes lead to another interesting
extension of the present model. In this model we can allow $(24 + r)$ D3-branes and $r$ ${\overline {\rm D3}}$-branes to probe the
F-theory geometry, where $r$ could be any large or small integer. It should be clear from our discussions for Model 2,
the tachyons between D3 and $\overline{\rm D3}$ can be made massless by switching on appropriate fluxes on the
world-volume of the three-branes. The system could preserve some supersymmetry, and for special arrangement of fluxes
even ${\cal N} = 2$ supersymmetry can be preserved. However due to the world-volume fluxes, the system has to
be away from the orientifold point, as orientifolding will project out all fluxes that have  even number of components along the orbifolding directions.
This in particular means converting $m^2 = 0$ tachyon to
$m^2 < 0$, resulting in the annihilation of all ${\rm D3}$-$\overline{\rm D3}$ pairs. The final heterotic dual would be what
we studied in the above subsection \ref{FhetSugra}.

The situation can be improved if we are away from the orientifold point. Now all world-volume fluxes are allowed, and
we can ask what sort of dualities can be constructed from this scenario. The type IIB metric for $r$ multiple
D3 $\overline{\rm D3}$-brane probing a ${\rm K3} \times {{\mathbb R}^2}/{{\mathbb Z}_2}$ is given by the following metric
written in a slightly suggestive way (and using the notations of \cite{townsend1, townsend2, bk}):
\bg\label{tf+r}
ds^2 &= & {1\over \sqrt{V}} \left(-{dt^2\over U} + dx_3^2\right) + \sqrt{V} \left({dx_1^2\over U} + dx_2^2 +
ds^2_{{\rm K3} \times {{\mathbb R}^2}/{{\mathbb Z}_2}}\right)\nonumber\\
 & = &{1\over \sqrt{V}}\left(-{dt^2\over U} + {Vdx_1^2\over U} + V dx_2^2 + dx_3^2\right)
+ \sqrt{V} ds^2_{{\rm K3} \times {{\mathbb R}^2}/{{\mathbb Z}_2}}.
\nd
The first line of \eqref{tf+r} should be thought of as a D-string metric in the far IR. As we go to higher energies
the delocalization becomes more apparent and we start seeing a four-dimensional metric as shown in the second line of
\eqref{tf+r}. Note that this situation is different from the ${\rm D5}$-$\overline{\rm D5}$ pairs as the ${\rm D3}$-$\overline{\rm D3}$ pairs
are left unwrapped. Therefore the four-dimensional physics emerges even at low energies.
Furthermore,
this picture is somewhat similar to \cite{vafaBA}
and therefore we should expect a gauge group of the form
$SU(r\vert r)$ (see also \cite{ABSV08}).
Our system is stable and all bound branes are delocalized along the
world-volume of the three-branes. Therefore, as mentioned
above, the physics is captured by four-dimensional gauge theories.

Another interesting point is when we partially break the gauge group. We can break the gauge group by pulling some
numbers of branes or anti-branes to infinity.
If we define a quantity called $k$ as in \eqref{dista} then breaking the gauge group
would amount to adding the following metric to \eqref{tf+r}:
\bg\label{addm}
ds^2_{\rm add} = {1\over U\sqrt{V}} \left(-k^2 dx_1^2 + 2k dt dx_1\right),
\nd
somewhat similar to what we saw in \eqref{totla}. In the usual case, as we saw before, typically $k$ vanishes when the
full supergroup is realized.

To lift this configuration to M-theory, we will
continue keeping ${{\mathbb R}^2}/{{\mathbb Z}_2}$ as the non-compact orbifold limit of
${\bf P^1}$. The M-theory lift is then straightforward:
\bg\label{mlift}
ds^2 = U^{1/3}\left(-{dt^2\over UV} + {dx_1^2\over U} + dx_2^2\right) + U^{1/3}\left[ds^2_{\bf \Sigma} +
d\widetilde{x}_3^2 + {V\over U}\left(dx_{11} + {V-1\over V} dt\right)^2\right],
\nd
where ${\bf \Sigma} = {\rm K3} \times {{\mathbb R}^2}/{{\mathbb Z}_2}$ is the six-dimensional internal space and $\widetilde{x}_3$ is the
dual compact cycle. There is also a three-form field coming from the electric field switched on the ${\rm D3}$-$\overline{\rm D3}$
pairs\footnote{There are expectedly
no M2-brane sources in the background \eqref{mlift}. However in the symmetry breaking
scenario of \eqref{addm} there would exist sources of the form $$C ~ = ~ -{k\over U} ~ dt \wedge dx_1 \wedge dx_2,$$
where $k$ is defined in \eqref{dista}. Clearly these sources vanish for $k \to 0$.}:
\bg\label{dakba}
C ~ = ~ (1+U^{-1}) dt \wedge dx_1 \wedge dx_{11}.
\nd
Let us now consider the case where we compactify the orbifold ${{\mathbb R}^2}/{{\mathbb Z}_2}$ to ${\bf P^1}$. Clearly anomaly
cancellation condition in M-theory will amount to adding 24 M2-branes to the system \cite{SVW}. To add 24 M2-branes
to the existing set of $r$ ${\rm M2}$-$\overline{\rm M2}$-brane pairs, we will make the following ansatz for the metric:
\bg\label{add24}
ds^2 =  W^{-2/3}\left(-{dt^2\over \widetilde{U}^{2\over 3}\widetilde{V}}
+ {dx_1^2\over \widetilde{U}^{2\over 3}} + \widetilde{U}^{1\over 3}dx_2^2\right)
+ W^{1/3}\left[\widetilde{U}^{1\over 3}\Big(ds^2_{\bf \Sigma} +
d\widetilde{x}_3^2\Big) + {\widetilde{V}\over \widetilde{U}^{2\over 3}}\left(dx_{11}
+ {\widetilde{V}-1\over \widetilde{V}} dt\right)^2\right]\nonumber\\
\nd
in addition to allowing for a three-form flux $C$ originating from the sources of the M2-branes. In the language of the
previous sub-section the M2-branes can be regarded as localized sources of $G$-fluxes in the internal space, much like
\eqref{gfla}.

Note the key differences between \eqref{add24} and \eqref{mlift}. A new warp factor $W$ is introduced that appears in
the metric \eqref{add24} exactly as in a M2-brane metric. The other warp factors ($U, V$) have been replaced by their
new values ($\widetilde{U}, \widetilde{V}$). They are determined from the backreactions of the extra M2-branes and the
world-volume fluxes
added to the system (much like the analysis discussed around \eqref{bakkarch} earlier), and can be written as:
\bg\label{uvcorr}
\widetilde{U} ~=~ U + {\cal O}(W), ~~~~~~~ \widetilde{V} ~=~ V + {\cal O}(W).
\nd
Let us now discuss the supersymmetry of the system. From type IIA perspective we have D2-branes with
electric and magnetic fields $F_{10}$ and $F_{21}$ respectively as well as free D2-branes. For the brane anti-brane
system, the supersymmetry condition is the following (see also \cite{superd}):
\bg\label{susyreqop}
{\sqrt{{\rm det}~g}\over \sqrt{{\rm det}(g + 2\pi\alpha' F)}} \left(\gamma_{012} + F_{10} \gamma_1\gamma_{11} +
F_{21}\gamma_0\gamma_{11}\right)\epsilon ~ = ~ \pm \epsilon,
\nd
where we will take for simplicity $g_{\mu\nu} = \eta_{\mu\nu}$ and $\pm$ in \eqref{mlift} refer to the brane
and the anti-brane respectively. On the other hand a D2-brane without any electric or magnetic fields on its
world-volume will satisfy:
\bg\label{d2b}
\gamma_{012} \epsilon ~ = ~ \epsilon .
\nd
In the limit when we take a critical electric field of the form $\vert F_{01}\vert^2 = 1$ on the D2 anti D2-branes, the
supersymmetry equations \eqref{susyreqop} and \eqref{d2b} simplify to the following constraint equations:
\bg\label{cone}
\gamma_{012} \epsilon ~ = ~ \epsilon, ~~~~~~~ \gamma_{01} \epsilon ~ = ~ - \gamma_{11} \epsilon, ~~~~~~~
\gamma_0\gamma_{11} \epsilon ~ = ~ \epsilon ,
\nd
which are the supersymmetry constraints for fundamental strings along $x^1$, D2-branes along $x^{0,1,2}$ and
D0-branes. This system is in general non-supersymmetric {\it unless} the F1 and the D0-branes are dissolved inside
the D2-branes by switching on appropriate electric and magnetic fields. Of course this boils down to the statement
that the 24 D2-branes (or M2-branes from M-theory perspective) should also come with appropriate electric and
magnetic fields switched on\footnote{Note that in this case there would be net M2-brane charges, and there would
be no anti-branes to cancel the 24 extra branes. This in turn would explain the choices of the warp factors in
\eqref{add24}. The warp factor $W$ justifies the existence of M2-brane charges, and the deformed warp factors
($\widetilde{U}, \widetilde{V}$) justifies the backreactions from the world-volume fluxes.}.

\begin{figure}[htb]
        \begin{center}
\includegraphics[height=4cm]{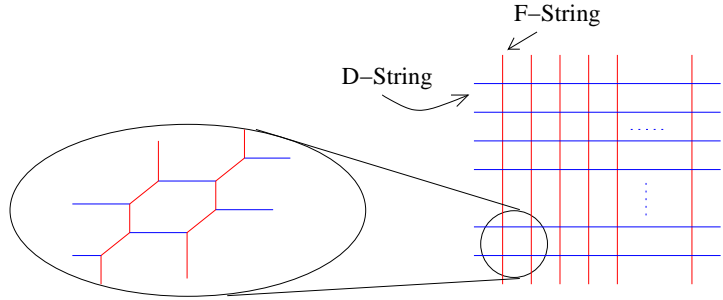}   \end{center}
        \caption{The string network that lies in the world-volume of the D3-branes parallel to the seven-branes. This
picture is only at a special point in the F-theory $u$-plane. At a generic point in the
Coulomb-branch the network may break into
smaller subspace of networks that are spread on the $u$-plane.}
        \label{matrix1}
        \end{figure}

{}From type IIB point of view, the ${\rm D3}$-$\overline{\rm D3}$-brane pairs come equipped with appropriate electric and magnetic
fields. This in turn means that the supersymmetry condition is the {\it same} as the supersymmetry
condition for a D-string and a fundamental string intersecting at the point, i.e as a string network of the kind
studied in \cite{DMstring, senstring}.
Therefore adding extra D3-branes means that these branes come with
string network dissolved in their world-volume. This is illustrated in {\bf figure \ref{matrix1}}.
Supersymmetry is then naturally preserved.

\subsection{A possible duality to M(atrix) theory on a four-fold \label{MatrixTheory}}

There is yet another perspective for this scenario that may appear from the discussions about abelian instantons and
fluxes that we had before. This has to do with the anomaly (or charge) cancellation condition for compact manifolds in
either type IIB or M-theory. Instead of cancelling the three-form charges using 24 D3 or M2-branes in type IIB or
M-theory respectively, we may want to cancel the charges by switching on appropriate fluxes.

The 24 M2-branes that we switched on are sources of localized $G$-fluxes in M-theory. Similarly, the world-volume
magnetic field becomes another source of $G$-flux just like \eqref{dakba}. Finally, both the world-volume electric
field and the spacetime dilaton get absorbed in the geometry. Therefore it makes sense to convert the localized
$C$-field instantons to fluxes. Thus there are two possibilities:
\begin{itemize}
\item The localized $C$-field instantons are {\it converted}
to fluxes by adding non-perturbative terms to the
action.
\item The localized $C$-field instantons are {\it replaced} by fluxes, i.e as though we are going from
\eqref{gfla} to \eqref{iibhap}.
\end{itemize}

Note that the first case is highly non-trivial, and at this stage we don't know how to argue for this when the
number of branes are small. For large number of branes, and for lower supersymmetry, there are some examples where
this could happen. An example is \cite{chen2}. We will discuss this briefly later.

In the following therefore we will concentrate mostly on the second case where we don't have to justify the
transitions between instantons and fluxes. The second case therefore deals with two {\it seperate} scenarios: one
with instantons and the other with fluxes, with no apparent connection between them. Whatever connection there is
or could be, will be speculated later.

Once we go for the second case, we have essentially the familiar situation of M-theory on a four-fold with
fluxes \cite{becker, DMM, GVW, DRS}, albeit the four-fold description is a little more non-trivial than ${\rm K3} \times {\rm K3}$
because of the underlying type IIA D0-branes. In the limit when the number of type IIA D0-branes is very large we
will be in yet another familiar territory: M-theory on ${\rm K3} \times {\rm K3}$ in the infinite momentum frame and in the
presence of fluxes. This is of course the M(atrix) description of M-theory \cite{BFSS}.
However, the M(atrix) description is bit non-trivial because of the background $G$-fluxes. Once we
replace one of the K3 by non-compact ${\mathbb R}^4$ the background $G$-fluxes vanish automatically, leaving us
with M(atrix) theory on K3. This has been studied in \cite{quivers, DOS, raja}.

M(atrix) theory on K3 (or ALE space) is described most efficiently in terms of M(atrix) string theory \cite{DVV}.
The theory
has both a Higgs and a Coulomb branch. The Higgs branch is the usual motion of the matrix string
on the ALE space oriented along
$x^{6,7,8,9}$ directions. The Coulomb branch is when the string is stuck at the orbifold singularity in the internal
space, and moves along the spacetime directions \cite{quivers}.

Once we blow-up the ALE (orbifold) singularities by adding FI D-terms, the metric along the Higgs branch change
in the correct way to account for the underlying K3 space \cite{quivers}.
The difficulty in analyzing the corresponding string theory lies
precisely in the difficulty in constructing a compact metric on the K3 surface. Other than that, the physics remains
identical to the ALE case.

To see how the M(atrix) description works for ${\rm D2}$-$\overline{\rm D2}$-branes on a K3 surface, we can use the
analysis of
\cite{JMyers, quivers} where M(atrix) string theory for type IIB on ALE and K3 spaces have been studied and combine
this with the study by \cite{bk} for ${\rm D2}$-$\overline{\rm D2}$-branes on flat spacetime. In the following we will sketch the idea, and
more details will be presented in the sequel.

M(atrix) theory on K3 manifold is expressed in terms Higgs and Coulomb branches. In the presence of ${\rm D2}$-$\overline{\rm D2}$-branes
with electric and magnetic fluxes the Coulomb branch physics can either be expressed in terms of D-particles or in terms
of T-dual D-strings. In the infinite momentum frame this is either M(atrix) theory or M(atrix) string theory
respectively. It turns out however that for the ${\rm D2}$-$\overline{\rm D2}$-branes M(atrix) theory might be a better way to
capture the physics here. Therefore in the language of type IIB probes, this would be the T-dual description.

The Hamiltonian in the Coulomb branch is succinctly expressed in terms of three coordinates $X_{1, 2,3}$, such that
the Hamiltonian takes the simple BPS form \cite{bk}:
\bg\label{chokra}
{\cal H} = && {1\over 2} {\rm tr}\Big\{(D_0X_1 + i[X_3, X_1])^2 + (D_0X_2 + i[X_3, X_2])^2 + (D_0X_3)^2 +
[X_1, X_2]^2 \nonumber\\
&& ~~~~~~~~~~ + 2i[X_1, X_3(D_0X_1)] + 2i[X_2, X_3(D_0X_2)]\Big\}
\nd
in the absence of any world-volume fluxes. It is clear that the Hamiltonian then is bigger than or at least equal to the
trace of the second line in \eqref{chokra}.

In the presence of world volume fluxes there would be no three-form sources, but there would be fundamental
string sources, D-particle sources and a dipole moment term given as (see for example \cite{myersd}):
\bg\label{dmomemt}
-{1\over 3}\cdot {\epsilon\over (2\pi)^2 g_s l^3_s}\int {d^3\sigma \over UV},
\nd
where $\epsilon$ is the dipole length (i.e the separation between D2 and $\overline{\rm D2}$-branes) and the integral is over the
world-volume of the two-branes. Note that in the Coulomb branch the D-particles can come together to enhance the
gauge symmetry to $SU(N)$ where $N$ is number of ${\rm D2}$-$\overline{\rm D2}$ pairs.

The Higgs branch, on the other hand, can be best understood using M(atrix) string theory \cite{DVV},
i.e from type IIB perspective
where we have ${\rm D3}$-$\overline{\rm D3}$ pairs. Let $X^i$ denote all the coordinates that specify the spacetime directions
$x^{0, 1, 2, 3}$
as well as the F-theory $u$-plane spanning, from the IIA D-particle point of view, the Coulomb branch that we discussed
above. The Higgs branch is parametrized by $Y^a$ with $a = 6, 7, 8, 9$ i.e the directions along the ALE space. The
local ${\mathbb Z}_2$ projection is represented by a unitary operator ${\bf U}$ as:
\bg\label{z2}
 {\bf U} Y^a {\bf U}^{-1} ~ = ~ - Y^a.
\nd
The solution for the above equation is represented in terms of $2N \times 2N$ matrices with two arbitrary set of
$N \times N$ matrices such that $Y^a$ and
\bg\label{za}
Z^a ~ \equiv ~ Y^a \begin{pmatrix} 0&{\bf 1}_{N \times N}\\ {\bf 1}_{N \times N} & 0 \end{pmatrix}
\nd
are traceless. Under these conditions
$Y^a$ then enters the following world-sheet lagrangian:
\bg\label{wslag}
{\cal L} ~ = ~ \eta_{ij} \partial X^i \partial X^j ~ + ~ g_{ab} \partial Y^a \partial Y^b + \cdots,
\nd
where $g_{ab}$ is the metric of the ALE space and the dotted terms are the fermionic completions. The fact that the full
A-D-E kind of singularities may appear in \eqref{wslag} has been discussed in some details in the literature. (See for
example \cite{JMyers}.)

This story gets a little more involved when we add FI parameters to blow up the A-D-E singularities. Once compactified
this leads to K3 manifold and therefore we will now have M(atrix) description on K3 $\times$ K3 manifold. Due to
Gauss' law, K3 $\times$ K3 compactification is inconsistent unless fluxes are added to compensate for the free three-form
charges. This means we have a more non-trivial M(atrix) description for the system. Our conjecture then is that the
theory on the ${\rm D3}$-$\overline{\rm D3}$ probes is captured, in the limit where the number of probe pairs go to infinite, by
M(atrix) theory on K3 $\times$ K3 (with additional fluxes).

\subsection{${\cal N} = 1, 2$ examples and geometric dualities \label{subsecModel3N12geomDual}}

In the previous subsection we briefly touched upon the issue of branes versus fluxes and how to go from one picture to
another. Herein lies the heart of gauge/gravity duality: in one side of the duality we have D-branes and on the other
side the branes have vanished and are replaced by fluxes. In the decoupling limit, the gauge theory on the branes is
captured by geometry, i.e a gravitational background with fluxes.
\begin{figure}[htb]
        \begin{center}
\includegraphics[height=3cm]{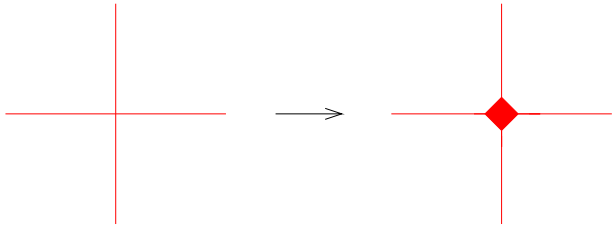}   \end{center}
        \caption{Under non-perturbative corrections the intersection region of two M5-branes may blow to form a
diamond type configuration. In type IIB theory this means that the branes (LHS of the figure where they have
shrunk to zero at the intersection) may be replaced by geometry with fluxes (RHS where the diamond configuration
exists).}
        \label{blowup}
        \end{figure}
An example where branes are replaced by fluxes has appeared in \cite{DOT} and is shown in {\bf figure \ref{blowup}}. A
configuration of large $N$ wrapped
D5-branes probing the resolved conifold geometry can be T-dualized to an intersecting
NS5-brane configuration where the NS5-branes are orthogonal to each other but separated along a different
direction compared to an equivalent configuration for the T-dual of a conifold geometry. The $N$ wrapped D5-branes
T-dualize to D4-branes stretched between the two separated NS5-branes such that they are suspended between them.
This configuration is the far IR of the geometric transition set-up and is discussed in
details in \cite{DOT, DOT2, DOT3}. In the limit where the cycle size shrinks to zero, the M-theory lift of
this is when the suspended M5-branes between the two orthogonal M5-branes shrink to zero size. But now the
non-perturbative corrections\footnote{For example switching on an
Affleck-Dine-Seiberg \cite{ADS, ADS2} type superpotential to the system.} can change the picture by blowing up the
intersecton region into a diamond
as shown in \cite{DOT} and as depicted here in {\bf figure \ref{blowup}}. In type IIB side, this blowup
of the intersection region amounts to dissolving the wrapped D5-branes to geometry and fluxes! This is the essence of
gauge/gravity duality here.

A somewhat similar story also unfolds in the heterotic theory \cite{chen2}. The NS5-branes in the heterotic theory
are the small instantons and therefore in the limit where the number of small instantons become large
the system should be
describable in terms of pure geometry and fluxes, exactly in the way we explained above. The ADHM sigma model \cite{ADHM}
of the theory should allow for a superpotential term of the Affleck-Dine-Seiberg \cite{ADS, ADS2} form that would
allow for instanton transitions by blowing them up from zero sizes to finite sizes. In this limit these instantons are no
longer
described by a world-volume theory of heterotic five-branes but by a background gravitational solution with
torsion \cite{chen2}. This torsional background turned out to be a specific deviation of the Maldacena-Nunez \cite{MNun}
geometry that allows for the three-form field strength to be non-closed (see footnote below).
\cite{chen2}.
The rationale for this is that under small instanton transitions the initial gauge group of
\bg\label{ingauge}
Sp(2N) ~\times~ SO(32)
\nd
breaks completely\footnote{Even before transition the $SO(32)$ gauge group may be broken to smaller subgroups by Wilson
lines. However once the gauge group \eqref{ingauge} breaks completely the standard embedding is no longer possible,
and so the three-form has to change from the Maldacena-Nunez \cite{MNun} solution, which is a {\it closed} three-form,
to a new one that is {\it non-closed} and cancels the
heterotic anomalies. This non-closure of the three-form is therefore essential to resolve all the issues encountered
in constructing a gauge/gravity duality in the heterotic theory.
These details are explained in \cite{chen2}.}
so that the physics is captured by a non-K\"ahler background with torsion \cite{chen2}. Again, as we discussed above,
this seems to be the generic essence of gauge/gravity duality here too.

The above two examples were both for the minimally supersymmetric ${\cal N} =1$ theories, so
the question now is whether such transitions may be observed with abelian instantons for our ${\cal N} = 2$ theories.
The localized $G$-fluxes in
M-theory are the seven-brane gauge fields \cite{DRS} and in the limit the instantons on the seven-branes are
zero size, they become the type IIB D3-branes. We can move the D3-branes along the Coulomb branch so that the
matter multiplets on them become heavy.
Therefore, below a certain scale, these states could then be integrated out.

For a compact internal four-fold and with small Euler number, this is hard to achieve. So the M(atrix) description may be
the simplest and possibly the best dual description that we could provide under the given circumstances. For the
non-compact case there is a well-known example of M-theory on a four-fold of the form:
\bg\label{dhop}
{\bf T}^2 ~ \ltimes ~ \left({\bf ALE} ~\times ~ {\mathbb R}^2\right),
\nd
which can be described by a dual gravitational description when probed by a large number of M2-branes. On the other hand,
if we blow-up the ALE space to
\bg\label{bup}
z_1^2 ~ + ~ z_2^2 ~ + ~ z_3^2 ~ = ~ \mu^2,
\nd
where $z_i$ are complex coordinates and $\mu$ is a complex deformation, then we know of one case where there exists
another gravitational dual for the probes. This is
again the minimally supersymmetric case where the coordinate $z_4$ parametrises
${\mathbb R}^2$ in \eqref{dhop}. A curve of the form
\bg\label{z4}
z_4 ~ -~ i\mu ~ = ~ 0
\nd
intersects the curve
\eqref{bup}\footnote{This intersection not only reduces the
supersymmetry but also removes the Coulomb branch of the theory.}.
The four-fold resulting from this intersection,
which is a torus fibration over a singular conifold, leads to the Klebanov-Witten model \cite{KeWi}.
The conformal field theory on the probes is captured by an AdS${}_5$ geometry.

The Klebanov-Witten model however doesn't explain {\it how} this transition happens. The abelian instantons
in M-theory have been replaced by delocalized $G$-fluxes. Once the base of the four-fold is further
deformed as
\bg\label{furbup}
z_1^2 ~ + ~ z_2^2 ~ + ~ z_3^2 ~ + ~ z_4^2 ~ = ~ t z_0^2
\nd
for any $t$, the delocalized $G$-fluxes lead to the NS and RR three-forms being
switched on the type IIB side \cite{anke3} instead of just the five-form as in the previous
case. These lead to
another set of dual descriptions that are suited for {\it non-conformal} theories!

Clearly the picture is not yet complete as the main reason for transition has not yet been clarified.
For higher supersymmetry like ${\cal N} = 2$ the cascading dynamics that we studied in the previous section using
ALE space may be the closest in realizing phenomena similar to ${\cal N} = 1$, although the IR dynamics may not
necessarily lead to a confining theory. For typical ${\cal N} = 1$ theories,
it could be that
similar arguments like switching on an Affleck-Dine-Seiberg type superpotential term is responsible in converting
the abelian instantons to fluxes. A somewhat similar idea is hinted in \cite{KSks} where an
Affleck-Dine-Seiberg superpotential is responsible for blowing up the conifold singularity to a deformed conifold. 

The ${\cal N} = 1$  story could be extended more, but we will not do it so here. Instead we will go to the next section where we will present our 
final set of configurations that realise a different class of ${\cal N} = 1$ theories.

\newpage

\section[Model 4: Multiple D3's probing intersecting seven-brane backgrounds]{Model 4: Multiple D3-branes probing intersecting seven-brane backgrounds \label{model4section}}

In the above subsection \ref{subsecModel3N12geomDual} for Model 3 we briefly discussed the ${\cal N} = 1$ case which comes from changing the
underlying four-fold geometry from ${\rm K3} \times {\rm K3}$ to another four-fold whose construction can be alternatively motivated using the following simple algebraic
geometric construction \cite{anke3}.

Consider F-theory on an
elliptic fibered Calabi-Yau four-fold ${\bf X} \hookrightarrow {\bf B}$ over a three-fold base ${\bf B}$. Suppose that ${\bf B}$ contains a smooth curve
${\bf E} \simeq {\bf P}^1$ with normal bundle ${\cal O}(-1) \oplus
{\cal O}(-1)$. Then there is a simple way to construct our required four-fold that we briefly alluded to in the previous sections: make a conifold transition
from ${\bf B} \rightarrow {\bf B}'$ obtained by contracting the ${\bf P}^1$ to a
point and then smoothing. This gives another elliptically fibered
Calabi-Yau four-fold:
\bg\label{4fol}
{\bf X}'~ \hookrightarrow ~ {\bf B}',
\nd
which is the required manifold that, in the presence of both RR and NS fluxes
in type IIB would deform in the right way to give us a conformally
K\"ahler manifold which is topologically
the same as ${\bf B}'$ \cite{anke3, chen1}. This
construction yields the same result as the one discussed above in
\eqref{furbup}\footnote{Although in more general cases the gravity duals tend to be
non-geometric. This aspect has been elaborated in great details in \cite{chen1, chen2}.}.

The above construction is interesting in the sense that it gives us a class of ${\cal N} = 1$ dual descriptions that
come from deforming certain ${\cal N} = 1$ conformal theories. These ${\cal N} =1$ CFTs are defined in terms of RG
flows that terminate on $1_{\mathbb C}$-dimensional fixed surfaces. The exactly marginal coupling associated with the quartic
superpotential $W$ helps us to navigate the fixed point surface (for continuous choice of the coefficient in $W$).

The model that we are interested in also has a similar conformal surface: it is a $1_{\mathbb C}$-dimensional curve in the
space of three couplings. This model will be realized by multiple D3-branes probing intersecting seven-brane
configurations and will fall in the class of models discussed in \cite{LS, tacwecht1, tacwecht2}.
We will see how dualities in these models could be understood in terms of non-K\"ahler manifolds.

\subsection{The heterotic dual with fluxes and five-branes \label{subsecModel4HetFlux5}}

The model that we 
investigate here is generated by multiple D3-branes probing  intersecting seven-branes
or intersecting orientifold-planes backgrounds. The gauge theory on $k$  D3-branes is $Sp(2k) \times Sp(2k)$
in the presence of
one set of intersecting orientifold-plane background. If we allow for local charge
cancellation, then there is an additional global symmetry given by $U(4) \times U(4)$ where an intersecting set
of four D7-branes are placed perpendicular to each other. The various multiplets are given by 3-3 as well as
3-7 strings as ($A, B$) and ($Q, q, {\bar Q}, {\bar q}$) respectively. The representations of various
fields in terms of both local and global groups are given in {\bf table \ref{representations}}. With this it is easy to construct the
superpotential, which is given by:
\bg\label{spoco}
W ~ = ~ \lambda(Q A {\bar Q} ~ + ~ q B {\bar q}),
\nd
so that the anomalous dimensions $\gamma_{g_i}$ for the two
gauge groups $g_i$ and the anomalous dimension
$\gamma_\lambda$ for the superpotential coupling $\lambda$ lead to
a two-dimensional surface
of fixed points where all RG flows terminate\footnote{This line passes through $g_i \to 0$ point. Note also that
naively using the arguments of \cite{LS} would lead us to a $1_{\mathbb R}$-dimensional conformal surface. This would
be a contradiction because the underlying supersymmetry requires an even dimensional conformal surface.
The point then is to include the theta angles for the gauge groups, and the phase
parts of the marginal superpotential couplings. This will essentially double-up the number of constraints (the
equations are already doubled-up), so that the conformal surface becomes even dimensional. This has been explained
recently in \cite{tacwecht1, tacwecht2}. We thank Yuji Tachikawa for clarifying the picture for us.}.
As before, continuous set of marginal couplings allow us to
navigate on this surface.

\begin{table}[h!]
 \begin{center}
\begin{tabular}{|c|c|}\hline Fields & $Sp(2k) \times Sp(2k) \times U(4) \times U(4)$ \\ \hline
$Q$ & ($2k, 1, 4, 1$) \\  \hline
${\bar Q}$ & ($1, 2k, {\bar 4}, 1$) \\  \hline
$q$ & ($2k, 1, 1, 4$) \\  \hline
${\bar q}$ & ($1, 2k, 1, {\bar 4}$) \\  \hline
$A, B$ & ($2k, 2k, 1, 1$) \\  \hline
  \end{tabular}
\end{center}
  \caption{The representations of various fields in our theory for both local $Sp(2k) \times Sp(2k)$ and
global $U(4) \times U(4)$ symmetries.}
  \label{representations}
\end{table}

The $U(4) \times U(4)$ is relevant for the constant coupling scenario and is equivalent to the $D_4$ singularity
that we discussed for Model 1. We will discuss soon how the $U(4)$ singularity may come about here.
Here there are more choices once we go away from the orientifold limit. The Weierstrass
equation governing the F-theory axio-dilaton is given by:
\bg\label{wseop}
y^2 ~ = ~ x^3 + f(u, v)x + g(u, v),
\nd
where the coordinate
$u$ is the usual $u$-plane discussed earlier, $v \equiv x^8 + i x^9$ is another two-dimensional complex plane,
$f(u, v)$ is a polynomial of degree ($8, 8$) and $g(u, v)$ is a polynomial of degree ($12, 12$). To determine
$f$ and $g$ let us define:
\bg\label{deffg}
f_k(u) & = & \prod_{i= 1}^8 A_{ik} (u - a_{ik}) ~ + ~ \prod_{i= 1}^4 B_{ik} (u - b_{ik})^2
~ + ~ \prod_{i= 1}^2 C_{ik} (u - c_{ik})^4 ~ + ~
D_k (u - d_k)^8, \nonumber\\
g_k(u) & = & \prod_{i= 1}^{12} M_{ik} (u - m_{ik}) ~ + ~ \prod_{i= 1}^6 N_{ik} (u - n_{ik})^2
~ + ~ \prod_{i= 1}^4 S_{ik} (u - s_{ik})^3
~ + ~
\prod_{i= 1}^3 P_{ik} (u - p_{ik})^4 \nonumber\\
&& ~~~~~~~~ + ~ \prod_{i= 1}^2 Q_{ik} (u - q_{ik})^6 ~ + ~ R_k (u - r_k)^{12},
\nd
where the coefficients are allowed to take values determined by the underlying dynamics of F-theory. The polynomials
$f_k(u)$ and $g_k(u)$ are now used to determine $f(u, v)$ and $g(u, v)$ as:
\bg\label{fvgv}
f(u, v) ~ \equiv ~ f_1(u) f_2(v), \qquad g(u, v) ~ \equiv ~ g_1(u) g_2(v).
\nd
These polynomials give us the physics not only at the orientifold point, but also away from it. In fact at the
orientifold point the description can be made a little simpler by choosing following set of polynomials in $u$ and
$v$:
\bg\label{gdass}
&&Z_i ~ \equiv ~ (u - \hat{u}_i)(v - \hat{v}_i),\nonumber\\
&&W_i ~ \equiv ~ (u - {u}_i)(v - {v}_i),\nonumber\\
&&U_i ~ \equiv ~ (u - \widetilde{u}_i)(v - \widetilde{v}_i).
\nd
Using these variables we can define $f(u, v)$ and $g(u, v)$ now in the following way that captures the physics at the
orientifold point:
\bg\label{fgndef}
f(u, v)  =  \prod_{i=1}^8 {\cal A}_1^i W_i + \prod_{i=1}^4 {\cal A}_2^i Z_i^2, \quad
g(u, v) =  \prod_{i=1}^{8}  \prod_{j=1}^{4} {\cal A}_3^{ij} W_i Z_j + \prod_{i=1}^4 {\cal A}_4^i Z_i^3 +
\prod_{i=1}^{12} {\cal A}_5^i U_i,
\nd
with ${\cal A}^i_k$ being the coefficients that could be related to the coefficients in \eqref{deffg}. The above choice
of polynomials \eqref{fgndef} is somewhat equivalent to the discussion in \cite{becker4}, to which one may look for
more details\footnote{We correct a typo in \cite{becker4}.}. The discriminant locus then spans the following curve:
\bg\label{dloka}
\sum_{n, p} {\cal C}_{np} \prod_{i, j, k}  Z_i^{6-2n-3p} W_j^n U_k^p ~ = ~ 0,
\nd
 {}from which one can factorize a simpler hypersurface governed by the curve \cite{becker4}
\bg\label{simcur}
{\cal C}_{20} {\cal F} ~ + ~ {\cal C}_{01} {\cal G} ~ = ~ 0.
\nd
The polynomials ${\cal F}, {\cal G}$ are of order 16 in ($u, v$). Exactly under this condition the type IIB theory
has a heterotic dual given by heterotic compactification on a K3 manifold. On the other hand the
curve\footnote{Note that ${\cal C}_{mn} \ne {\cal C}_{nm}$. This can be justified easily from our polynomial
expansion \eqref{dloka}.}:
\bg\label{lkeyes}
{\cal C}_{10} {\cal F}_1 ~ + ~ {\cal C}_{11} {\cal F}_2 ~ + ~ {\cal C}_{02} {\cal F}_3 ~ + ~ {\cal C}_{00} {\cal F}_4
~ + ~ {\cal C}_{30} {\cal F}_5 ~ = ~ 0
\nd
with ${\cal F}_i$ being the required polynomial factorized from \eqref{dloka}, should be interpreted as
[$p, q$] seven-branes that form a non-dynamical orientifold plane. Since our concern is \eqref{simcur}, we will ignore
the physics behind the curve \eqref{lkeyes} for the time being.

What happens when we allow $k$ D3-branes to probe the orientifold point? The $k$ D3-branes see an orthogonal
geometry of intersecting seven-branes and planes wrapping orthogonal 2-cycles of a K3 orientifold. The
orientifold action is similar to the one discussed for Model 1, but now the space involutions are done via
Nikulin involutions \cite{nikulin}. This scenario is different from Model 3 because in Model 3 the D3-brane probes
were at a point in ${\rm K3} \times {\bf P}^1$ manifold and so Gauss' law was a constraint there. But now there
are no such constraints. 

The D3-branes at the intersecting orientifold point now dualize to small instantons wrapping 2-cycles of the K3
manifold in the heterotic side. Our first
conjecture 
is that this brings us precisely to the intrinsic torsion and $G$-structures of
\cite{gauntlett}. Recall that for the manifold with an $SU(2)$ structure we have the two usual conditions:
\bg\label{su2str2u}
0 ~ = ~ d(e^{-2\phi}\Omega)~ = ~ d(e^{-2\phi}\ast J),
\nd
whose solutions tell us how the torsion classes ${\cal W}_i$
are represented for this case with $J$ being the fundamental form,
$\Omega$ is the ($3, 0$) holomorphic form and $\phi$ is the dilaton.
Explicitly the solutions are as given earlier in \eqref{otrclass}, with the torsion ${\cal T}$ lying in:
\bg\label{tarkadal}
{\cal T} ~ \in ~ {\cal W}_4 ~\oplus ~ {\cal W}_5 .
\nd
The situation at hand however is more involved now. In the {\it absence} of the probe D3-branes, the heterotic
dual is indeed given by a conformal K3 manifold as was discussed in \cite{becker4}. However in the presence of
the D3-probes, the situation is different. In the following we will argue that the heterotic dual is given by
a non-K\"ahler K3 manifold that is {\it not} a conformally Calabi-Yau manifold. However even before we go
about elucidating this, we should ask how to see the $Sp(2k) \times Sp(2k)$ gauge symmetry directly from the
heterotic side.

To answer both the questions we will have to study the orientations of various branes in the type IIB side. The two
${\mathbb Z}_2$ transformations that we are interested in are generated by \cite{sengimon, sengimon2}:
\bg\label{gimgen}
g ~ = ~ (-1)^{F_L} \cdot \Omega \cdot {\cal I}_{45}, \qquad h ~ = ~ (-1)^{F_L} \cdot \Omega \cdot {\cal I}_{89},
\nd
where, as before, ${\cal I}_{ab}$ denote orbifold action along ($x^a, x^b$). In a more general setting with a
slightly different ${\mathbb Z}_2$ actions \cite{gopamu, dabholkarP, blumZ},
the type IIB manifold is an orientifold of
a K3 manifold (see above). The various orientations are given in {\bf table \ref{gpbranes}}.

\begin{table}[h!]
 \begin{center}
\begin{tabular}{|c||c|c|c|c|c|c|c|c|c|c|}\hline Directions & 0 & 1 & 2
& 3 & 4 & 5 & 6 & 7 & 8 & 9 \\ \hline
D7 & --  & --   & --   & --  & $\ast$  & $\ast$ & -- & -- & --
& -- \\  \hline
${\rm D7}^\prime$  & --  & --   & --   & --  & --  & --  & --  & -- &
$\ast$  & $\ast$\\  \hline
D3 & --  & --   & --   & --  & $\ast$  & $\ast$ & $\ast$ & $\ast$ & $\ast$  & $\ast$
\\  \hline
K3 & $\ast$  & $\ast$  & $\ast$  & $\ast$ & --  &
--   & $\ast$  & $\ast$ & --  & -- \\  \hline
  \end{tabular}
\end{center}
  \caption{The orientations of various branes in the intersecting seven-branes set-up. The dashes are the
directions along which we have either the branes or the manifold.}
  \label{gpbranes}
\end{table}

%

To go to heterotic theory, one way would be to T-dualize twice along the $x^{4, 5}$ directions and then S-dualize.
The T-duality transformations will convert \eqref{gimgen} into the following generators:
\bg\label{gimgenow}
(-1)^{F_L} \cdot \Omega \cdot {\cal I}_{45} ~\to ~ \Omega, \qquad
(-1)^{F_L} \cdot \Omega \cdot {\cal I}_{89} ~ \to ~ \Omega \cdot {\cal I}_{4589},
\nd
which means the orthogonal set of seven-branes and orientifold seven-planes will be converted into a system of
five-branes and O5-planes parallel to nine-branes and O9-planes. This
is the well-known Gimon-Polchinski system \cite{gimpol}. The probe D3-branes will then become D5-branes.

\begin{figure}[htb]
        \begin{center}
\includegraphics[height=5cm]{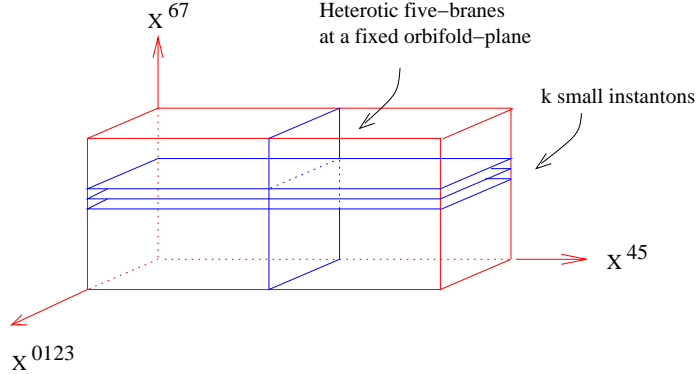}   \end{center}
        \caption{A brane construction to see how in the heterotic theory we recover the $Sp(2k) \times Sp(2k)$ gauge
symmetry. The $k$ D3-branes become heterotic five-branes which we refer them as $k$ small instantons here. The other
type IIB D7-brane
becomes another set of five-branes possibly at a fixed orbifold plane
generating the
two $Sp(2k)$ gauge symmetries via bisection.}
        \label{sp2kgp}
        \end{figure}

The above type IIB system can now be S-dualized so as to be in the same moduli space as an equivalent
heterotic theory\footnote{The following analysis is a rather toy example of a more elaborate duality
that comes from studying the Gimon-Polchinski model \cite{gimpol}. For example it is known that the
GP model is dual to F-theory on a ${\rm CY}_3$ with Hodge numbers (3, 243) that admits an elliptic fibration over
${\bf P}^1 \times {\bf P}^1$ \cite{MV1, MV2}. This in turn is further
dual to $E_8 \times E_8$ heterotic theory on a K3 manifold with
the total number of 24 instantons divided equally into the two $E_8$'s. Since our analysis will only touch the
supergravity aspect of this duality i.e the {\it closed string} sector of the theory,
we will not analyse the gauge symmetry (which is of course the global symmetry
for the world-volume theory
on the probes) and the twisted sector states
carefully in this paper. In particular we will not study the connections
between the $SO(32)$ and the $E_8 \times E_8$ heterotic theories, in the presence of probes,
as well as the intriguing connections to the ${\rm CY}_3$ with Hodge numbers (51, 3) and (3, 51) here.
For more details on this in the absence of
any probe branes, one may refer to \cite{MV1, MV2, ashopin, gopamu, dabholkarP, blumZ}.}.
An S-duality
transformation
will convert the D9/O9 system into equivalent vector bundles, and the other
pair into five-branes stuck along some orbifold plane.
An equivalent type I configuration with the branes arranged as above
would then S-dualize to heterotic theory\footnote{We believe the
following procedure might
explain the duality: the
type I O5/D5 system S-dualizes into an
orbifold five-plane in the heterotic theory \cite{sennbps}. Then the remaining five-branes become heterotic
five-branes. So finally one may get heterotic five-branes stuck to the orbifold plane.}.
On the other hand the $k$ probe D3-branes, which became $k$ D5-branes in type I theory,
will also become heterotic five-branes whom we will refer to as
$k$ small-instantons. These small instantons attain a gauge symmetry of $Sp(2k) \times Sp(2k)$ because of the
O5-plane action in the S-dual type I theory, or because of the five-brane bisections in the heterotic
theory. As the small instantons cannot be broken and moved along the $x^{6, 7}$ directions, the theory will not have
any Coulomb branch as one might have expected from the underlying ${\cal N} = 1$
supersymmetry\footnote{The ${\cal N} = 1$ theory on the probes in this case is rather non-trivial. At a generic point in the moduli space the gauge group may be broken into decoupled set of $k$ $U(1)$'s. For example if we take a single probe D3-brane in this background, the generic gauge group is $U(1)$ and {\it not}
$U(1) \times U(1)$. At the orientifold intersection point the gauge symmetry
enhances to $Sp(2) \times Sp(2)$ that share the same $U(1)$. In the presence of quantum corrections, as one might have expected, the $Sp(2) \times Sp(2)$ gauge symmetry is no longer restored near the region of intersection of the two orientifold planes. The full theory is then given by an ${\cal N} = 2$ SCFT with
$U(1)$ vector multiplet and a massless charged hypermultiplet \cite{sengimon, sengimon2}. Moving a D3-brane along $u$, the
massless
charged hypermultiplet can be interpreted as the monopole/dyon point for one of the $Sp(2)$ group. Similarly moving the
D3-brane now along $v$, the same massless charged hypermultiplet may now be interpreted as the monopole/dyon point of
other $Sp(2)$ gauge group. Thus the non-perturbative effects in this model convert the monopole/dyon point of one
$Sp(2)$ gauge group to the monopole/dyon point of the other $Sp(2)$ gauge group. For more details on this
phenomena the readers may refer to \cite{gimpol,sengimon,sengimon2}.}.
This is illustrated in {\bf figure \ref{sp2kgp}}.

To analyze further we will then assume that we have $k$ small instantons (or $k$ heterotic five-branes) oriented along
$x^{0, 1, 2, 3, 4, 5}$ directions and another (small) set of five-branes oriented along $x^{0, 1, 2, 3, 6, 7}$
directions. The $k$ small instantons wrap 2-cycles of a K3 manifold which, in turn, is oriented along
$x^{4, 5, 8, 9}$. It is easy to see that in the {\it absence} of the $k$ small instantons, the other set of
heterotic five-branes are all located at a point on the K3 manifold. This is the scenario studied in \cite{becker4},
and the metric ansatz for this case is the conformal K3 ansatz, namely:
\bg\label{nikara}
ds^2 ~ = ~ ds^2_{012367} ~ + ~ \Delta^m ds^2_{K3}, \qquad e^\phi ~ = ~ \Delta
\nd
where $\Delta$ is the warp factor and $m$ is an integer. For our case $m = 2$ which appears from the consistency
conditions given in \cite{gauntlett}.

In the presence of $k$ small instantons the situation at hand is no longer that straightforward. The wrapping of the
small instantons on 2-cycles of K3 changes much of the result. However the result can be argued using the
metric for intersecting branes. To use this, let us assume that the two set of five-branes come with warp
factors $h_i = h_i(x^8, x^9)$ with $i = 1, 2$ such that $h_1$ denotes the warp factor for $k$ small instantons
and $h_2$ denotes the warp factor for the other set of five-branes.
As usual we consider the delocalized limit (as this is the only limit
under which we can study the backreactions in a controlled setting).
The above considerations then immediately reproduces the
background for us:
\bg\label{bgfromm}
ds^2  =  ds^2_{0123} + h_2 (ds^2_{45} + h_1 ds^2_{89}) + h_1 ds^2_{67}, \quad e^\phi  =  \sqrt{h_1h_2}, \quad
{\cal T}_{abi}^{(l)}  = c_l \epsilon_{abik} \partial_k h_{l},
\nd
with $c_l$ being a constant, ${\cal T}_{abi}^{(l)}$ being the torsion polynomials, and
($a, b$) being either ($4, 5$) or ($8, 9$).
Note that we have written the metric in a slightly suggestive way by combining the
$x^{4, 5}$ and $x^{8, 9}$ directions, along which we will be placing our K3 manifold.
Note also that if we make $h_1 = 1$ i.e remove all the small instantons and assume that
\bg\label{ask3}
ds^2_{45} ~ + ~ ds^2_{89} ~ \to ~ ds^2_{{\bf T}^4/{\mathbb Z}_2}~ \to ~ ds^2_{K3},
\nd
then we reproduce \eqref{nikara} provided $\Delta = \sqrt{h_2}$. However, for generic choice of $h_1$, the
metric in \eqref{bgfromm} tells us that the K3 is deformed to a non-K\"ahler manifold $\widetilde{\rm K3}$:
\bg\label{ask31}
ds^2_{45} ~ + ~ h_1 ds^2_{89} ~ \to ~ ds^2_{\bf P^1} ~ + ~ h_1 ds^2_{\bf P^1}~ \to ~  ds^2_{\widetilde{\rm K3}},
\nd
and so the heterotic solution is no longer a conformally Calabi-Yau two-fold. The fluxes and the dilaton
\eqref{bgfromm} are supported on this non-K\"ahler space. As far as we could see, this non-K\"ahler deformation
of the K3 manifold has not been studied before, so should be new. A simple way to analyze this manifold would be
to go to the orbifold limit of K3 where we can write the metric in terms of some blown-up 2-cycles and then
rescale the 2-cycles using $h_1$ in an unequal way as in \eqref{ask31}. This
procedure of course doesn't capture the full picture, but is nevertheless a good starting point. In the following subsection
we will elaborate more on this.


\subsection{Type IIB analysis at the orientifold point \label{model4subsecIIBO7}}

\begin{figure}[htb]
        \begin{center}
\includegraphics[height=4cm]{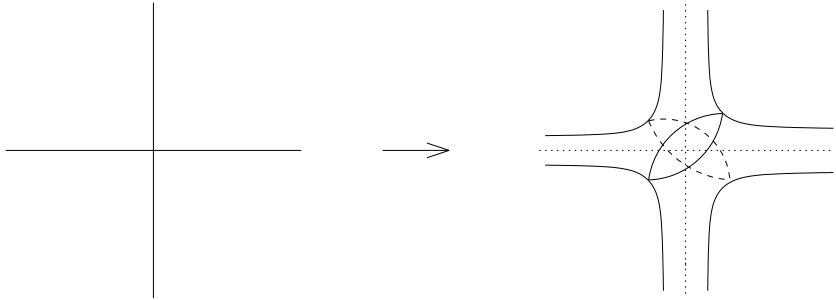}
        \end{center}
        \caption{The splitting of the O7-planes in the Gimon-Polchinski model to form a
hyperboloid like structure.} \label{gpmodel}
        \end{figure}

Let us now see how this
works for the explicit case of \eqref{bgfromm} assuming \eqref{ask31}.
Making an S-duality of the background given in \eqref{ask31} will give
us the corresponding type I background. The torsion three-form
will be replaced by the RR three-form of the type I theory. We will also
assume that the two warp factors are now given by\footnote{Clearly these warp factors are no longer
harmonic functions as ${\partial^2h_i\over (\partial x^8)^2} + {\partial^2h_i\over (\partial x^9)^2}$ do not
vanish for any non-trivial choice of the exponents.}:
\bg\label{2harmoni}
h_1 ~ \equiv ~ (1+x)^k, \qquad h_2 ~ = ~ (1+x)^\beta ~ = ~ h_1^{\beta/k}, \qquad k > 0
\nd
where $x$ is some inverse powers of $\sqrt{(x^8)^2 + (x^9)^2}$, and $\beta$ is the number of five-branes
stuck to the orbifold plane and orthogonal to the $k$ small instantons. Using these definitions,
one
can easily show that the corresponding type I background follows
from the following background in type IIB theory:
\bg\label{metiintwob}
ds^2 ~= ~ h_1^{-{k+\beta\over 2k}} ds^2_{0123} + h_1^{{\beta-k \over 2k}} ds^2_{\bf P^1} +
 h_1^{{\beta+k \over 2k}} ds^2_{\bf P^1} +  h_1^{{k-\beta \over 2k}} ds^2_{67},
\nd
where we have
explicitly shown the metric of the two non-compact $P^1 \equiv
{R^2\over {\cal I}_2}$. Observe that the type IIB manifold has
retained its basic form but the two $P^1$'s are now scaled
differently. The type IIB coupling constant, $g_B$, is no longer
a constant number as in \cite{DM1, DRS, becker1}, rather now it has
dependence on the warp factor. The dependence is easy to infer,
and is given by
\bg\label{iibcoup}
g_B ~=~ h_1^{-{k+3\beta\over 2k}},
\nd
where we are ignoring
constant factors. This therefore implies that the
Gimon-Polchinski model is at an orientifold limit with coupling
constant $\tau$ given by
\bg\label{gptau}
\tau \equiv C_0 + i~h_1^{{k+3\beta\over 2k}}
\nd
where $C_0$ is the
RR scalar (axion). Notice that ${\rm Im}~\tau > 0$ and therefore
this presumably incorporates possible non-perturbative
corrections to the D7-branes and O7-planes background. This
is because near an O7-plane the behaviour of
$\tau$ naively would be
\bg\label{taubeor}
\tau = -{2\over i\pi} {\rm ln}
\left(u - {\widetilde u}_i \right),
\nd
where ${\widetilde u}_i$ is the
position of one of the O7-plane on the ${\bf P^1}$ with
coordinates $u$ (similar behaviour is expected for the other
${\bf P^1}$). But this background of axio-dilaton should receive
correction because $\tau \to - i \infty$ as we approach the
orientifold plane. The correct behaviour therefore should be
\eqref{gptau}. Switching on blow-up models at the orientifold
singularities fuses two intersecting orientifold planes to a
complex hyperbola \cite{sengimon, sengimon2}. In {\bf figure \ref{gpmodel}} we have
represented a pair of O7-planes that become fused together.

Note that in the limit $k \to 0$ the above background should simplify. But the way we have expressed $h_1$ and
$h_2$ in \eqref{2harmoni} this limit is not easy to see. However one may easily convince oneself that when
$k\to 0$, the backgrounds \eqref{metiintwob} and \eqref{iibcoup} change to the following simpler form \cite{becker4}:
\bg\label{simbag}
ds^2 ~ = ~ \Delta^{-1} ds^2_{012367} + \Delta ds^2_{\bf P^1} + \Delta^{-1} ds^2_{\bf P^1}, \qquad g_B ~ = ~
{1\over \Delta^3}.
\nd
Returning back to \eqref{metiintwob},
the above background in type IIB theory can be
easily lifted to F-theory. In the absence of fluxes we know that
the F-theory background is a torus fibration over a ${\bf P}^1 \times
{\bf P}^1$ base. What happens now?  To answer this, note that
since both F-theory and M-theory
span the same moduli space\footnote{i.e F-theory on ${\mathbb X} \times {\bf S}^1$ is on the same moduli space as
M-theory on ${\mathbb X}$.},
this metric will also correspond to
the metric in M-theory. The gauge fluxes will correspond to
localized $G$-fluxes in M-theory. This then tells us that the metric in F/M-theory
will be:
\bg\label{backinF}
ds^2 = && ~h_1^{-{k+\beta\over 3k}} ds^2_{\rm spacetime} + h_1^{{2\beta-k\over 3k}} ds^2_{45}
+ h_1^{{2k-\beta\over 3k}}ds^2_{67}
+ h_1^{{2k+2\beta\over 3k}} ds^2_{89} \nonumber\\
&& +{1\over 4}\left(h_1^{-{k+\beta\over 3k}} - h_1^{{2k+2\beta\over 3k}}\right) (dz^2 +
d{\bar z}^2) + {1\over 2}\left(h_1^{-{k+\beta\over 3k}} + h_1^{{2k+2\beta\over 3k}}\right) \vert dz\vert^2 ,
\nd
where ($z, {\bar z}$) are
the coordinates of the fiber. In the absence of the small instantons, the F-theory background is simpler:
\bg\label{kof}
ds^2 ~ = ~ \Delta^{-{2\over 3}}ds^2_{\rm st} +  \Delta^{{4\over 3}} ds^2_{K3}
+  \Delta^{-{1\over 3}} ds^2_{67} +  {1\over 4} \left(\Delta^{-{2\over 3}} -  \Delta^{{4\over 3}}\right)
(dz^2 + d{\bar z}^2)
+ {1\over 2} \left(\Delta^{-{2\over 3}} +  \Delta^{{4\over 3}}\right)\vert dz\vert^2,\nonumber\\
\nd
where $ds^2_{\rm st}$ denotes the metric of the spacetime, as in \eqref{backinF}. The torsions in the heterotic
theory will become non-localized $G$-fluxes in F-theory, and the localized $G$-fluxes will become vector bundles as we
discussed above.

The issue of the vector bundle is now important. From the
F-theory side there are many enhanced global symmetry points. The one that we were
considering so far was $U(4)^4 \times U(4)^4$ which appeared from colliding $D_4$ singularities in the Tate's
algorithm \cite{tate} appropriately modified due to the two orientifold
actions\footnote{One might get worried by the fact that in F-theory colliding $D_4$ singularities may lead to
tensionless strings as shown in \cite{bikmsv}.
For our case this doesn't happen because the orbifold singularity
associated with the generator $gh$ in \eqref{gimgen} hides half a unit a
$B_{\rm NS}$ flux \cite{aspinwall} coming from the collapsed two-cycle.
This makes the tension of the string non-zero at the orbifold singularity.}.
To see how an $U(4)$ singularity
may arise from the $SO(8)$ singularity, consider the diagram: 
\begin{equation}
\xymatrix@1@C=25mm{
D_4~~
\ar[r]^{\left(\begin{smallmatrix}~{\cal M}&0\\~0 & -{\cal M}\end{smallmatrix}\right)}
\ar[r]&~~ A_3 ,
}\label{eq:freeX}
\end{equation}
where the matrix represents gauge transformation on the gauge group $SO(8)$ due to one of the orientifold projection
(say $h$) in \eqref{gimgen}. Such an action will break the $SO(8)$ gauge group to the one that commutes with the given
matrix which, in turn, is constructed out of a smaller $4\times 4$ matrix ${\cal M}$ given by:
\bg\label{matmat}
{\cal M} ~ \equiv ~ \left(\begin{matrix}~0&~0&~1&~0\\~0&~0&~0&~1\\-1&~0&~0&~0\\~0&-1&~0&~0\end{matrix}\right).
\nd
However,
the $U(4)^4 \times U(4)^4$ is not the simplest global symmetry. There exists a simpler one given by
the $SU(2)^8 \times SU(2)^8$ point
because one may think that at no point in the moduli space of the theory a single
brane can move freely. In order to move
 in this theory, branes need to be in a group of at least two, giving us the above mentioned
global symmetry group. The story is however not completely correct as the gauge group can be broken further.
For example there is a possibility that we can change the
Weierstrass equation \eqref{wseop} to:
\bg\label{wseopnow}
y^2 ~ = ~ x^3 + \left[f(u, v) + \delta f(u, v)\right]x + g(u, v) + \delta g(u, v)
\nd
with an appropriate choices for ($\delta f, \delta g$) so as to break the global symmetry group completely.
The discriminant
locus will change accordingly to reflect this. Such a deformation is reminiscent of the recent work on
${\cal N} = 1$ theory from ${\cal N} = 2$ theory in \cite{vafaT}. We will discuss more on this below as well as
in the discussion section.
Of course in the absence of ($\delta f, \delta g$) the global symmetry will
remain $SU(2)^8 \times SU(2)^8$ with pairs of seven-branes moving together.

All the above pictures should also be visible in the heterotic side. This becomes slighly involved because the
background torsion converts the usual Donaldson-Uhlenbeck-Yau equations to the following torsional equivalents:
\bg\label{vecbund}F_{ab} = F_{\bar a \bar b} = g^{a \bar
b}F_{a \bar b} = 0, \qquad {1\over 30}{\rm tr}~F \wedge F = {\rm tr}~R \wedge
R - {2\over \alpha'} d\ast d\phi .
\nd
In the limit where the axio-dilaton doesn't vanish globally,
which could happen generically for non-constant global and local
$G$-fluxes, we will no longer
be in the constant coupling point of the Gimon-Polchinski
model\footnote{Assuming $d{\cal T} = 0$,
then in the absence of the small instantons, the warp factor will satisfy $\square \Delta^2 = 0$.
Since we also require the warp factor to be small everywhere, we
can take it to be a linear function of the coordinates $z^i$,
where $z^i$ are defined on patches along $x^{4, 5, 8, 9}$ directions. This would imply that the torsion ${\cal T}$ will be a
constant form globally. The background therefore is given by:
\bg\label{consisol}
&&\Delta^2 = c_o + A~z^1 + B~z^2 + {\rm
c.c}, \qquad e^\phi = \Delta, \nonumber\\
&& {\cal T} =  A ~ dz^1 \wedge
dz^2 \wedge d{\bar z}^2 + B~dz^1 \wedge d{\bar z}^1 \wedge dz^2 +
{\rm c.c},
\nd
where $c_o$ is a constant that would determine the
size of the conformal K3 space. Putting $A = 0$ will reproduce
the model with non-trivial $\tau$. Now since $\square$ is
measured with respect to the unwarped metric, $\Delta^2$ will satisfy the
warp factor equation. Therefore the constant three-form would
follow from there. We also have to make sure that $\Delta^2$ is
small everywhere. This is possible because our space is compact
and therefore $z^1, z^2$ are measured only in a small patch.
Assuming $\vert z^1 \vert^2 + \vert z^2 \vert^2 \le \epsilon$, we
can have $\Delta^2$ small everywhere. The equation \eqref{consisol}
will therefore be the torsional background for the conformal K3
in the heterotic theory.}.
The background flux actually demands the $\tau$ to take an expression given in
\eqref{gptau}. This would determine the resulting positions of the
branes and planes in this scenario, and therefore the gauge
bundle.

In the limit where we are in the constant coupling scenario with four seven-branes (plus their images) on
top of O7-plane and parallel to the $u$-plane with similar configurations along the $v$-plane the gauge group is
what we studied before. In the heterotic side Wilson loops (associated with the distances between the
four D7/O7 system in the $u$- as well as the $v$-planes) break the heterotic gauge group to the required type IIB one.
If we demand the discriminant locus to take the form:
\bg\label{dislok}
\Delta ~ = ~ (u-u_1)^2 \widetilde{\Delta}(u, v),
\nd
then we have only a pair of seven-branes together, resulting in a classical $U(2)$ global symmetry \cite{gimpol}.
In the heterotic side
$U(2)$ gauge symmetry implies
\bg\label{trff}
{\rm tr}~ F\wedge F ~ = ~ 8\pi^2\left[c_2({\cal V}) - {1\over 2} c_1^2({\cal V})\right]
\nd
where $c_i({\cal V})$ are the Chern classes of the vector bundles ${\cal V}$. One can now use the first
equation in \eqref{vecbund} to
determine the primitive i.e (1, 1) part of the bundle using the inverse of the metric in \eqref{bgfromm}. Once this is
done, the torsion ${\cal T}$ in \eqref{bgfromm} may be used in \eqref{vecbund} to determine the local $U(2)$ gauge
symmetry in the heterotic side.

An interesting exercise is to get the $U(1)$ theory from the $U(2)$ theory. In the type IIB
side we can allow the full Weierstrass equation \eqref{wseopnow} so that all the seven-branes move
individually. A particular choice for ($\delta f, \delta g$) in \eqref{wseopnow} that can do this is the following
(see also \cite{sengimon}):
\bg\label{wseopnoww}
y^2 ~ = ~ x^3 + \left[f(u, v) + \sum_{m,n = 0}^1a_{mn} u^mv^n\right]x + g(u, v) + \sum_{m,n = 0}^2 b_{mn} u^m v^n
\nd
with an appropriate choices for ($a_{mn}, b_{mn}$) so as to break the global symmetry group completely. The discriminant
locus will change accordingly to reflect this.

In the following we will consider only the $SU(2)$ sector, or more appropriately the $U(2)$ sector, of the global symmetry
group broken to $U(1)$ in the heterotic side.
To analyse this, one may take the $F$ of $U(2)$ theory and write it as:
\bg\label{ftof}
F = \left(\begin{matrix}f+\overline{f}&0\\0&f+\overline{f}\end{matrix}\right)
\nd
where $\overline{f}$ is the
complex conjugate of $f$: this way $F$ is real. Next step is to demand primitivity and then determine ${\rm Re}~f$
from the torsional equation \eqref{vecbund} as:
\bg\label{fdeter}
{\rm Re}~f \wedge {\rm Re}~f = {15\over 4} \left[{\rm tr}~R \wedge R - {2 d\ast d\phi\over \alpha'}\right].
\nd
In an alternative scenario, the F-theory compactification on a
base ${\bf P}^1 \times {\bf P}^1$ is also connected to another Calabi-Yau with
Hodge numbers (51,3) (and probably also to (3, 51)). This connection was pointed out by
\cite{gopamu, dabholkarP, blumZ}. The difference between the two
compactifications: one with Hodge numbers $(3,243)$ and the other with
Hodge numbers $(51,3)$, is related to the number of tensor multiplets and charged hypermultiplets.
For the case studied, we saw that the heterotic dual has only one
tensor multiplet. To get more tensors in six dimensions, we have
to redefine the orientifold operation in the dual type I side. A
way to achieve this was shown in \cite{dabholkarP}. We can define the
orientifold operation in such a way that not only it reverses the
world sheet coordinate $\sigma$ to $\pi - \sigma$, but also flips
the sign of twist fields at all fixed points. In fact in the
closed string sector the two theories tally, but in the twisted
sector, we get 17 tensor multiplets instead of hypermultiplets.
Therefore even though the orientifolding action look similar in
both cases, the massless multiplets are quite different.

\begin{figure}
\setlength{\unitlength}{0.008in}%
$$\begin{picture}(445,266)(60,385)
\thinlines
\put(100,420){\circle*{6}}
\put(140,420){\circle*{6}}
\put(260,420){\circle*{6}}
\put(300,420){\circle*{6}}
\put(420,420){\circle*{6}}
\put(460,420){\circle*{6}}
\put(120,440){\circle*{6}}
\put(160,440){\circle*{6}}
\put(240,440){\circle*{6}}
\put(280,440){\circle*{6}}
\put(320,440){\circle*{6}}
\put(400,440){\circle*{6}}
\put(440,440){\circle*{6}}
\put(140,460){\circle*{6}}
\put(180,460){\circle*{6}}
\put(220,460){\circle*{6}}
\put(260,460){\circle*{6}}
\put(300,460){\circle*{6}}
\put(340,460){\circle*{6}}
\put(380,460){\circle*{6}}
\put(420,460){\circle*{6}}
\put(160,480){\circle*{6}}
\put(200,480){\circle*{6}}
\put(240,480){\circle*{6}}
\put(280,480){\circle*{6}}
\put(360,480){\circle*{6}}
\put(400,480){\circle*{6}}
\put(180,500){\circle*{6}}
\put(220,500){\circle*{6}}
\put(260,500){\circle*{6}}
\put(300,500){\circle*{6}}
\put(340,500){\circle*{6}}
\put(380,500){\circle*{6}}
\put(200,520){\circle*{6}}
\put(240,520){\circle*{6}}
\put(280,520){\circle*{6}}
\put(320,520){\circle*{6}}
\put(360,520){\circle*{6}}
\put(220,540){\circle*{6}}
\put(260,540){\circle*{6}}
\put(300,540){\circle*{6}}
\put(340,540){\circle*{6}}
\put(240,560){\circle*{6}}
\put(280,560){\circle*{6}}
\put(320,560){\circle*{6}}
\put(260,580){\circle*{6}}
\put(300,580){\circle*{6}}
\put(280,600){\circle*{6}}
\put(320,480){\circle*{6}}
\put(120,400){\circle{10}}
\put(280,400){\circle{10}}
\put(440,400){\circle{10}}
\put(200,440){\circle{10}}
\put(200,480){\circle{10}}
\put(360,440){\circle{10}}
\put(360,480){\circle{10}}
\put(280,480){\circle{10}}
\put(280,520){\circle{10}}
\put(280,560){\circle{10}}
\put(280,600){\circle{10}}
\put(280,440){\circle{10}}
\put(120,440){\circle{10}}
\put(440,440){\circle{10}}
\put( 80,400){\line( 1, 0){420}}
\put( 80,400){\line( 0, 1){240}}
\put(300,620){\circle*{6}}
\put(320,600){\circle*{6}}
\put(340,580){\circle*{6}}
\put(380,540){\circle*{6}}
\put(400,520){\circle*{6}}
\put(420,500){\circle*{6}}
\put(440,480){\circle*{6}}
\put(440,480){\circle{10}}
\put(460,460){\circle*{6}}
\put(480,440){\circle*{6}}
\put(360,520){\circle{10}}
\put(360,560){\circle*{6}}
\put( 75,379){\makebox(0,0)[lb]{\raisebox{0pt}[0pt][0pt]{0}}}
\put(175,379){\makebox(0,0)[lb]{\raisebox{0pt}[0pt][0pt]{5}}}
\put(270,379){\makebox(0,0)[lb]{\raisebox{0pt}[0pt][0pt]{10}}}
\put(370,379){\makebox(0,0)[lb]{\raisebox{0pt}[0pt][0pt]{15}}}
\put(470,379){\makebox(0,0)[lb]{\raisebox{0pt}[0pt][0pt]{20}}}
\put( 65,495){\makebox(0,0)[lb]{\raisebox{0pt}[0pt][0pt]{5}}}
\put( 55,595){\makebox(0,0)[lb]{\raisebox{0pt}[0pt][0pt]{10}}}
\put(505,395){\makebox(0,0)[lb]{\raisebox{0pt}[0pt][0pt]{$r$}}}
\put( 75,645){\makebox(0,0)[lb]{\raisebox{0pt}[0pt][0pt]{$a$}}}
\end{picture}$$
	\caption{The Voisin--Borcea examples. This figure is taken from \cite{MV2} where the readers may
find more details about the topological invariants of these manifolds. See also \cite{voisin, borcea}.}
	\label{figVB}
\end{figure}
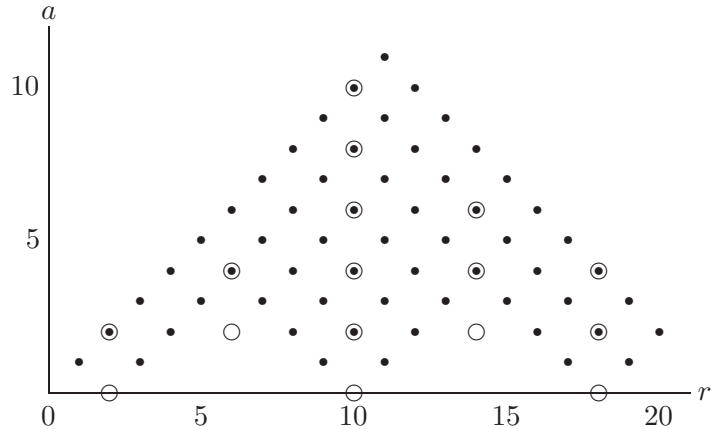

All these analysis were done without any probe D3-branes in type IIB theory.
We may now want to study D3-branes probing type IIB on ${\bf T}^4/\left({\mathbb Z}_2 \times {\mathbb Z}_2\right)$
with the ${\mathbb Z}_2$ actions defined non-trivially on ${\bf T}^4$.
The F-theory lift of these would be ${\bf T}^2$ fibrations over these
four-dimensional bases probed by multiple D3-branes. All
these examples are members of the Voisin-Borcea family of Calabi-Yau three-folds
\cite{voisin, borcea} as illustrated in {\bf figure \ref{figVB}}.
The simplest example that we studied here may now be generated by
choosing the following Nikulin involution \cite{nikulin}
\bg\label{nikubhai}
(r, a, \delta) ~ = ~ (18, 4, 0),
\nd
where ($r, a, \delta$) are defined in \cite{voisin, borcea, nikulin, MV2},
of the Voisin-Borcea three-fold and then {\it exchanging} the vector and the neutral hypermultiplets
of the model. Finally rearranging the seven-branes will give us the other gauge groups on the
heterotic side\footnote{Of course this is also the same play-ground where heterotic/heterotic duality of
\cite{dmwit} was presented.}.
Probing other Nikulin involuted K3 manifolds by multiple D3-branes will then give us other non-K\"ahler
deformations of the K3 manifolds in the heterotic side.
More details
on the various alternations of these models will be discussed in the
sequel.

\newpage

\section{Discussion and conclusions \label{conclusionsection}}

In this paper, we 
studied an exhaustive list of theories stemming from D3-branes probing
various orientifold (and also away from the orientifold) limits of F-theory. The simplest case, which we
called as Model 1, already gave us interesting physics related to $r+1$ points in moduli space where we have maximal Argyres-Douglas
points in the Coulomb branches of the theories governed by gauge groups $Sp(2r)$ where $r$ could be any integer.
We also argued the foremost question that would
naturally appear when we study $Sp(2r)$ gauge theories, namely, how many [$p, q$]
seven-branes do the classical orientifold plane split into.
Recall that for the $Sp(2)$ case the monopole/dyon
points in the Coulomb branch tell us that in the orientifold limit the quantum corrections will split the
orientifold into two non-local seven-branes. For the $Sp(2r)$ case there are numerous such points and additionally, the
Coulomb branch is multi-dimensional. So the natural question is how do, in F-theory, where we see only a one complex
dimensional Coulomb branch per probe D3-brane, the full class of solitonic states appear in this reduced moduli space? However from a
slightly different perspective, that we elucidated in the paper, there seems to be no apparent reason why the $r$
D3-brane probes should alter the physics of the splitting. Therefore we should, from this perspective only, conclude that
the number of splittings should be {\it two} irrespective of the number of probe D3-branes. Clearly a direct proof of
this statement is still lacking from the gauge theory side. Instead we started with the reasonable conjecture that the classical O7-plane will split into two points on the $u$-plane regardless of number of probe D3-branes, then we found various properties for (potentially 1-1) mapping between moduli space of the curve and that of brane dynamics. However, brane dynamics do not seem to cooperate very well with the combinatorics, and we have several puzzles which calls for further work. For example, it will be interesting to study more carefully the SW curve for $Sp(2r)$ with antisymmetric matter given in \cite{AMP}, and also compare their IIA-M theory picture with our F-theory picture.

Model 2 is related to the Gaiotto's model \cite{gaiotto}
where we have multiple D3-branes probing parallel seven-branes and orientifold
seven-planes wrapped on multi Taub-NUT spaces. We discussed various things in this model including connections to a class
of Gaiotto's model as well as T-dual maps to various brane networks of \cite{bbt}. However many things still remain. For
example all the conformal examples that we provided here are constructed out of seven-branes wrapped on
multi Taub-NUT spaces. These seven-branes are arranged in such a way that we can have F-theory at constant coupling. The
constant axio-dilaton that we studied here was calculated without taking the effects of the Taub-NUT geometries. This is
not {\it a priori} a problem as the seven-brane monodromies will not be affected by wrapping the seven-branes on the
Taub-NUT spaces. This should also be obvious from the dual brane network model where the seven-brane monodromies are
not affected by the presence of the [$p, q$] five-branes. 
On the other hand if there
there are background fluxes then the calculations need to be changed to incorporate various backreactions and anomaly
effects.

One of the major outcomes of \cite{gaiotto} is the so-called $T_N$ model
with only global symmetries
of $SU(N)^3$ and no gauge symmetries. 
For small $N$ we have studied a few examples in this paper. However the puzzle
is to deal with larger $N$. For models with $N > 8$, the number of seven-branes could become
larger than 24, which is an upper bound on the number of seven-branes in the F-theory set-up compactified on $S^2$. 

A way out of this is to allow for UV completions via a non-supersymmetric theory. For example one may allow
$24 + k$ seven-branes with $k >> 1$  and keep $k$ {\it anti} seven-branes at large $u$ in the $u$-plane. Since
large $u$ implies UV physics,
we are breaking the supersymmetry at high energy, but restoring the IR ${\cal N} = 2$
supersymmetry. The system would be stable, as all the tachyons are very massive here, and so integrating out these
massive modes we would have a well defined effective action at the IR. With additional seven-branes at hand, one might
be able to construct other $T_N$ models studied in \cite{gaiotto}. It would be interesting to see how arbitrary
global symmetries can now be accommodated from F-theory (see also \cite{kleban} for some recent discussions on
F-theory with more than 24 seven-branes).

In the presence of $m$ fractional D3-brane probes in the set-up that we used here, the situation is more interesting.
Now we can try to realize the other
models of Gaiotto with three {\it full punctures} and $m+1$ {\it simple punctures}. It will be interesting to see
how the ${\cal N} = 2$ dualities manifest here. Note that for this case there is no simple T-dual map to any brane
network models as the fractional branes are realized by ${\rm D5}$-$\overline{\rm D5}$ pairs.

The supergravity solutions that we studied here need bit more elaborations. Our way of approaching the sugra background
is very different from what was attempted before in Grana-Polchinski \cite{poln3} and more recently in
Gaiotto-Maldacena \cite{GM}. We took the backreactions of large number of ${\rm D5}$-$\overline{\rm D5}$ branes
in the presence of
\begin{itemize} \item World-volume electric and magnetic fluxes
\item Multi Taub-NUT spaces, and
\item Local and non-local seven-branes
\end{itemize}
to give us both the UV and the IR limits of the corresponding gauge theories. We find that the UV physics
is captured by a six-dimensional theory and the IR physics is captured by a four-dimensional theory. The precise
UV metric is given by \eqref{ltubaz}, which we copy below as:
\bg\label{ltubazgain}
ds^2_{\rm UV} = && {1\over \sqrt{V_2}}\Bigg[-f_1 V_1^{-1} dx_0^2 + f_3 dx_1^2 +
f_2 dx_2^2 + f_4 dx_3^2 + f_7 V_2dx_7^2
+ f_5 V_2 V_1^{-1}\left(dx^6 + \sum_\sigma F^\sigma_i dx^i\right)^2\Bigg] \nonumber\\
&& + \sqrt{V_2}\left[f_8 (dx_8^2
+ dx_9^2)
+ f_6 \tau_2(u) \Bigg\vert \eta^2(\tau(u))\prod_{i=1}^{24} {du\over (u - u_i)^{1/12}}\Bigg\vert^2 \right]
\nd
where the various coefficients are described in subsection \ref{subsecUVIRgrav}. Note that we have not been
able to determine the coefficients precisely there. These coeffcients are not difficult to determine, as there are
only finite number of sugra equations. However even in the absence of the precise warp factors the physics should be
clear. The backreactions of the seven-branes appear explicitly in the metric and contains all the informations that we
expect from F-theoretic embeddings, namely, the bound on the seven-branes and the associated singularity structures.
It will be interesting to compare the UV metric with the one in \cite{GM}.

On the other hand the physics probed by larger wavelengths i.e the IR physics is relatively simpler than
\eqref{ltubazgain}. This is given in \eqref{irgeometry}, which we copy below for the benefit of the readers:
\bg\label{irgeometrygain}
ds^2_{\rm IR} = && {\cal F}_1^{-1/2} ds^2_{0123} + {\cal F}_1^{1/2} ds^2_\perp =
{\cal F}_1^{-1/2} (-dx_0^2 + dx_1^2 + dx_2^2 + dx_3^2) + {\cal F}_2 \vert d\vec{w}\vert^2 \nonumber\\
&& + {\cal F}_3 \left(dx^6 + \sum_\sigma F^\sigma_i dx^i\right)^2 + {\cal F}_4
\tau_2(u) \Bigg\vert \eta^2(\tau(u))\prod_{i=1}^{24} {du\over (u - u_i)^{1/12}}\Bigg\vert^2.
\nd
As we clarified in the subsection \ref{subsecUVIRgrav}, the IR physics is captured by a four-dimensional theory, as
all the subtleties of the multi Taub-NUT space will be invisible at low energies. Again the warp factors still remain
to be determined, and it would be interesting to compare this to the analysis of \cite{GM}. We expect, among other
things, to complete these computations in the sequel.

Another thing that came out of our analysis with fractional D3-branes probing seven-branes wrapped on multi Taub-NUT
spaces is the ${\cal N} = 2$ cascade. In subsection \ref{subsecNonConf} we clarified how the ${\cal N} = 2$ cascade
may be related to the ${\cal N} = 1$ cascade. The idea is simple and follows a two-step procedure to go from
one cascade to another.

\begin{itemize}
\item Allow for a non-trivial fibration of the simplest Taub-NUT space over the compactified $u$-plane
as in \eqref{fibex}.
\item Blow-up the vanishing two-cycle in the resulting geometry {\it or} switch on a
$b \equiv B_{\rm NS}$ field
on the vanishing two-cycle. 
\end{itemize}
\noindent The second point is obviously the result of allowing a complexified K\"ahler parameter
${\cal K} = \omega + i b$ in the geometry. As is well known, for $b = 0$ we have Vafa's geometric transition
\cite{vafaGT} with ${\cal N} = 1$ cascade realised as an infinite sequence of flop transitions ending with a
conifold transition \cite{katzvafa}, at least in the {\it absence} of the seven-branes. Similarly for $\omega = 0$
i.e with a blown-down ${\bf P}^1$, we get the cascading dynamics of Klebanov-Strassler \cite{KSks}. In the presence of
seven-branes the picture is a little different. It is given by the Ouyang modification \cite{ouyang1, ouyang2} where
the final IR stage may even be conformal if one chooses the bi-fundamental matters more carefully. In any
case, the presence of the seven-branes typically tends to slow down the cascade. This can also be explained by the
brane construction of \cite{DOT, DOT2, DOT3}. The supergravity solution in the far IR (where we expect a four-dimensional
physics too) has been studied in great details in \cite{anke1, anke2, anke3, chen1}. The UV physics for the
${\cal N} = 1$ case will again be captured by a six-dimensional theory for the model of Vafa \cite{vafaGT}, whereas
it'll continue to remain four-dimensional for the Klebanov-Strassler model \cite{KSks}.

There is one more subtlety in the UV that we have been ignoring so far. In the presence of the seven-branes there is a
possibility that the UV physics may have Landau poles including UV divergences of the Wilson loops \cite{chuchai}.
This is a major problem for the ${\cal N} = 1$ cascading theories, and has recently been resolved by UV-completing
these theories by asymptotically conformal theories \cite{miamodel} at least for the Klebanov-Strassler case. For the
${\cal N} = 2$ case we might expect a somewhat similar situation. A way out of this would be to follow the same
procedure as outlined in \cite{miamodel}, namely, take the {\it far} UV to be given by an asymptotically conformal
theory with gauge group
\bg\label{ggac}
SU(k + M) ~ \times ~ SU(k+M)
\nd
and then at a certain scale $\Lambda = \Lambda_0$, high enough to be covering the
regime given by the metric \eqref{ltubazgain}, one of the gauge group in \eqref{ggac}
is Higgsed so that we can have the gauge group
\eqref{kebot}. Exactly at this scale, cascading dynamics can start leading finally to confinement at far IR. Before
Higgsing, both have the same Yang-Mills coupling, given by $g_{\rm YM}$. The UV beta function then will be
\bg\label{uvbeta}
\beta(g_{\rm YM}) ~ = ~ {g^3_{\rm YM} \over 16\pi} \sum_{n = 1}^\infty {{\cal D}_n \over \Lambda^n}, ~~~~~
\Lambda > \Lambda_0
\nd
with ${\cal D}_n$'s being constants. Thus at $\Lambda \to \infty$ we expect an exactly conformal theory. Below
$\Lambda_0$ the two gauge couplings wouldn't be equal and they will flow differently leading to our required
cascades.

Clearly a concrete construction of an UV complete picture is required at this stage. Another advantage of such a
completion is to 
get the equivalent story for ${\cal N} =1$ case. So far all sugra solutions were
restricted to far IR only, therefore
getting an UV complete background for models with geometric transitions \cite{vafaGT}
would be a welcome bonus.

Model 3 is an interesting extension to Model 2. If we {\it compactify} the Taub-NUT space to form
a K3 manifold, the resulting physics probed by D3-branes becomes very different. Firstly, due to
tadpole cancellations, we will not be able to put more than 24 D3-branes \cite{DRS}. One way to increase the
number of D3-branes, without breaking all the supersymmetries, is to switch on ${\overline{\rm D3}}$ branes with
world-volume electric and magnetic fluxes (on both kinds of branes) to kill the tachyons. Again the UV and IR
physics will be different, and the full four-dimensional physics can be recovered at high energies only.

Once we lift the picture to M-theory then, in the limit where the number of D3-${\overline{\rm D3}}$
in type IIB becomes very large, the resultant IR physics is captured by M(atrix) theory on ${\rm K3} \times {\rm K3}$. M(atrix)
theory on K3 has been studied before, but now due to tadpole cancellations we have to switch on $G$-fluxes. Thus the
physics will be captured by M(atrix) theory with background $G$-fluxes. In subsection \ref{MatrixTheory} we have
merely scratched the surface of the problem, as more detailed study is required to elucidate the physics there.

Another interesting speculation is the following. Imagine we replace the $G$-fluxes by M2-branes (satisfying the
tadpole cancellation). In type IIA we have a configuration of large number of D0-branes and 24 D2-branes. Imagine
we switch on $B_{\rm NS}$ fluxes on the D2-branes such that the world-volume theory becomes non-commutative and
can be represented by D0-branes \cite{swncg}. Then the whole system could be expressed by M(atrix) theory on
${\rm K3} \times {\rm K3} $ {\it without} any 
$G$-fluxes. It would be interesting to study this and compare with the one with
$G$-fluxes. This may shed some light into the issue of abelian instantons and fluxes that we discussed there.

Finally, Model 4 is a generalization of Model 1 with additional orthogonal set of O7-plane 
and D7-branes (classically).
One of the effects of these additional branes and planes is that the supersymmetry is reduced to ${\cal N} =1$, at least near the junction
of the O7-planes. 
This means that the $k$ probe D3-branes will see a ${\cal N} =1$  $Sp(2k) \times Sp(2k)$ gauge theory
near the junction. 

For this model we studied many things in the paper. We analyzed the heterotic dual and argued how we get a
non-K\"ahler K3 manifold from the wrapped small instantons in the heterotic side. Of course, as we mentioned in the
main text, we only studied the closed string sector of the model. Both the open string sector as well as the twisted
sector states are not addressed in much detail here. For example, we do not address at all the connection among various models with
Hodge numbers (3, 243), (3, 51) and (51, 3), that may be distinguished from their twisted
sector and open string states. All these examples fall in the class of Voisin-Borcea
three-folds \cite{voisin, borcea, MV1, MV2}, and it would be interesting to study them in detail.

Recently a way to study ${\cal N} =1$ theory in F-theory was done in \cite{vafaT} based on an earlier work of
\cite{AKS}. In the models of \cite{AKS, vafaT}, ${\cal N} = 1$ was generated by {\it tilting} the F-theory seven-branes
slightly by modifying the Weierstrass equation in the following way:
\bg\label{wetilt}
y^2 ~ = ~ x^3 + \left[f(u, v) + \delta f(u, v)\right]x + g(u, v) + \delta g(u, v),
\nd
such that this deformation may contribute to the superpotential that breaks the susy from ${\cal N} = 2$ to
${\cal N} = 1$. Our analysis of generation ${\cal N} = 1$, although uses similar form for the Weierstrass
equation as in \eqref{wseopnow} and \eqref{wseopnoww},
 is different from this as we take intersecting branes
that fall in the class of ${\cal N} =2$ theories broken to ${\cal N} = 1$ only at the intersection point of the
orientifolds. Tilting the two orthogonal seven-branes would presumably break all the supersymmetry at the
junction point of the orientifolds, but would allow ${\cal N} =1$ along the $u$- or the $v$-plane (where we had
${\cal N} = 2$ supersymmetry earlier). More details on this, and in particular the connection to integrable systems,
will be analysed in the sequel.

Finally, it would be interesting to extend the analysis of \eqref{dhop} to the intersecting seven-branes case. For
example if we consider F-theory on the four-fold of the form:
\bg\label{ffsua}
{\bf T}^2 \ltimes \left(K3 \ltimes {\bf P}^1\right),
\nd
which in turn would mean that type IIB is on the base $K3 \ltimes {\bf P}^1$ with the intersecting D7/O7 configurations.
Both the intersecting brane-plane configurations now wrap the ${\bf P}^1$. However in the presence of probe D3-branes,
Gauss' law would be an issue now, exactly as we faced earlier for model 3. The difference now is that the
orthogonal set of D7/O7's wrap respectively only orthogonal set of two cycles of the K3 manifold. Thus we can
invoke M(atrix) theory again by introducing a large number of D3-${\overline {\rm D3}}$ brane pairs to probe the system.

The situation could be generalised from \eqref{ffsua} to even more interesting direction by assuming that the type
IIB compactification is now on ${\bf P}^1$ fibered over del Pezzo surfaces\footnote{del Pezzo surfaces are manifolds
of complex dimension 2 with positive first Chern class. For us,
a simple defination would be to take ${\bf P}^1 \times
{\bf P}^1$ and blow up atmost {\it seven} generic points. Alternatively one may take ${\bf P}^2$ and blow-up
atmost {\it eight} generic points.}. This system will not have an heterotic dual, but could be analysed from
M(atrix) theory. It would be interesting to see if M(atrix) theory could shed some light on the so-called
{\it mysterious} duality \cite{vafaiqbal}.

\vskip.2in

\centerline{\bf Acknowledgements}

\vskip.1in

\noindent It is a great pleasure to thank Philip Argyres, Mike Douglas, Olivier DeWolfe, Mathias Gaberdiel,
Sergei Gukov, Jeff Harvey, Andy Neitzke, Al Shapere, Ashoke Sen, Yuji Tachikawa, Wati Taylor, Cumrun Vafa,
Johannes Walcher, and Brian Wecht for many
helpful correspondences and discussions. Special thanks go to Philip Argyres, Ashoke Sen and Yuji Tachikawa for insightful comments on
various parts of our draft.
Part of our work was completed during
PITP 2010, Chicago Great Lakes
string conference 2011, the String-Math 2011 conference at UPenn, and Strings 2011 at Uppsala. We thank the
organizers of these conferences for
providing a stimulating atmosphere for fruitful discussions. We also thank the JHEP referee for careful reading and many helpful suggestions and comments.
The work of K. D and J. S is supported in parts by NSERC grants and the work of A. W is supported in parts by
FQRNT scholarship and Schulich fellowship.


\newpage

\appendix

\section{How to obtain Seiberg-Witten curve and 1-form for pure $Sp(2r)$ theories
\label{sp2r01}}

Seiberg-Witten curve and Seiberg-Witten one-form for $Sp(2r)$ with $N_f=2r+2$ fundamental hypermultiplets is given in \cite{ArgyresShapere} as following:
\begin{eqnarray}
x y^{2}   & =&\left(  x\prod_{a=1}^{r}\left(  x-\phi_{a}^{2}\right)
+g \prod_{j=1}^{2r+2} m_j \right)  ^{2}-g^2 \prod_{j=1}^{2r+2} \left( x-m_j^2\right) , \label{AS12a}\\
\lambda & =&  \frac{\sqrt{x}}{2\pi i} d \log \left( \frac{   x\prod \left(  x-\phi_{a}^{2}\right) + g \prod  m_j -\sqrt{x} y   }{  x\prod \left(  x-\phi_{a}^{2}\right) + g \prod  m_j +\sqrt{x} y  } \right), \nonumber \\
 & =& a \frac{dx}{2\sqrt{x}}  \log \left( \frac{   x\prod \left(  x-\phi_{a}^{2}\right) + g \prod  m_j +\sqrt{x} y   }{  x\prod \left(  x-\phi_{a}^{2}\right) + g \prod  m_j -\sqrt{x} y  } \right), \label{AS12oneform}\\
\emph{g(}\tau)&=&\frac{\vartheta_{2}^{4}}{\vartheta_{3}^{4}+\vartheta_{4}^{4}} ,%
\label{AS12b}
\end{eqnarray}
in terms of the Jacobi theta functions given below:
\begin{equation}
\vartheta_{2}^{4}=16 q+ {\cal O}(q^3), \qquad \vartheta_{3}^{4}=1+8q+{\cal O}(q^2), \qquad \vartheta_{4}^{4}=1-8q+{\cal O}(q^2).
\label{AS15}
\end{equation}
Combining \eqref{AS12b} and \eqref{AS15} gives us the coupling constant $g(\tau)$ in terms of $q\equiv e^{i \pi \tau}$ as below:
\begin{equation}
g(\tau)=\frac{\vartheta_{2}^{4}}{\vartheta_{3}^{4}+\vartheta_{4}^{4}}%
=\frac{16q+{\cal O}(q^{3})}{1+8q+{\cal O}(q^{2})+1-8q+{\cal O}(q^{2})}=8q+{\cal O}(q^{3}).
\end{equation}

We want to obtain Seiberg-Witten curve for pure (no flavor, $N_{f}=0$) case, which is achieved by taking all the flavors to be infinitely massive $m_{j}\sim M\rightarrow\infty$, while keeping \begin{equation} 8\Lambda^{2r+2}=8qM^{2r+2}=gM^{2r+2} \label{noflavor} \end{equation} finite, as argued in \cite{ArgyresShapere}.
  Inside \eqref{noflavor} allows, $g\prod_{j=1}^{2r+2}m_{j}=gM^{2r+2}$ to be replaced with
$8\Lambda^{2r+2}$ and $g^{2}\prod_{j=1}^{2r+2}(x-m_{j}^{2})=\left(  gM^{2r+2}\right)
^{2}$ with $\left(  8\Lambda^{2r+2}\right)^{2}$ in \eqref{AS12a} and \eqref{AS12oneform}. The former, the Seiberg-Witten curve now becomes
\begin{align}
x y^{2}  &  =\left(  x\prod_{a=1}^{r}\left(  x-\phi_{a}^{2}\right)
+8\Lambda^{2r+2}\right)  ^{2}-\left(  8\Lambda^{2r+2}\right)  ^{2} \nonumber \\
&  =\left(  A+B\right)  ^{2}-\left(  B\right)  ^{2}=A(A+2B)\nonumber \\
&  =\left(  x\prod_{a=1}^{r}\left(  x-\phi_{a}^{2}\right)  \right)  \left(
x\prod_{a=1}^{r}\left(  x-\phi_{a}^{2}\right)  +16\Lambda^{2r+2}\right). \label{ASnoFlavorCurve1}
\end{align}
Dividing with $x$ on both sides of \eqref{ASnoFlavorCurve1}, we finally get:%
\begin{equation}
y^{2}=\left(  \ \prod_{a=1}^{r}\left(  x-\phi_{a}^{2}\right)  \right)
\left(  x\prod_{a=1}^{r}\left(  x-\phi_{a}^{2}\right)  +16\Lambda
^{2r+2}\right). \label{ASnoFlavorCurve}
\end{equation} The Seiberg-Witten one-form becomes
\begin{eqnarray}
\lambda & =& a \frac{dx}{2\sqrt{x}}  \log \left( \frac{   x\prod \left(  x-\phi_{a}^{2}\right) + g \prod  m_j +\sqrt{x} y   }{  x\prod \left(  x-\phi_{a}^{2}\right) + g \prod  m_j -\sqrt{x} y  } \right),\nonumber \\
& =& a \frac{dx}{2\sqrt{x}}  \log \left( \frac{   x\prod \left(  x-\phi_{a}^{2}\right) + 8\Lambda^{2r+2} +\sqrt{x} y   }{  x\prod \left(  x-\phi_{a}^{2}\right) + 8\Lambda^{2r+2} -\sqrt{x} y  } \right). \label{ASnoFlavor1Form}
\end{eqnarray}
The final result in \eqref{ASnoFlavorCurve} and \eqref{ASnoFlavor1Form} is used as the Seiberg-Witten curve and one-form for pure $Sp(2r)$ gauge theory throughout this paper, for example in \eqref{curve3} and \eqref{OneForm}.

\newpage


\section{Speculative viewpoints of Model 1 related to one-on-one mapping}

In this section we will discuss two speculative viewpoints of model 1 that will show us how the 1-1 mapping could in principle be highly 
non-trivial. We begin with a naive 1-1 mapping. We call it a naive mapping, because in subsubsection \ref{towardMap}, we pointed out physical
inconsistencies in the mapping. However, it is instructive to go through this type of reasoning because it will help us
to identify important physical ingredients, hitherto missing from our discussions, that will be absolutely essential to
find the correct 1-1 mapping. Interestingly, the mapping discussed in the following subsection has an apparently clean
combinatorics, that again will help us to identify the essential ingredients towards the correct 1-1 mapping.

\subsection{A naive 1-1 mapping between moduli spaces of brane and curve \label{naiveMap}}

To argue that the system at hand is more subtle than what one would have guessed,
we will first discuss a naive 1-1 mapping between moduli space of D3-branes and that of Seiberg Witten curve keeping the anti-symmetric matter very heavy
so as to  
continue the discussion for single D3-brane case given in {\bf table \ref{sp2table}}.
The biggest motivation for this naive mapping is that we somehow want to have 1-1 mapping between the moduli spaces. There are $r+1$ maximal Argyres-Douglas points in $  {\cal M}_{f_{Sp(2r)}}$, and on the brane picture we see only 2 monopole dyon points on the $u$-plane. This naive mapping has a way to 1-1 map these loci.
Assuming Bose-Einstein statistics among D3-brane probes, there are $r+1$ ways to put $r$ D3-branes at two special locations. For example, we can put $i$ D3-branes at the first location, and the rest $r-i$ D3-branes at the second location, while $i$ can take $r+1$ different values $i=0, 1, \cdots, r$ \footnote{In general, putting $r$ identical objects at $s$ different locations (or putting $r$ objects in $s$ different groups where the group size is nonnegative), there are
$ \left(
\begin{array}
[c]{c}%
r+s-1\\
s-1
\end{array}
\right) $
ways.}. Therefore putting $r$ D3-branes at either monopole or dyon 7-brane will have $r+1$ different ways, and there is no free parameter left. This way we singled out $r+1$ points in ${\cal M}_{r {\rm D3}}$. The naive map in this subsubsection proposes that these $r+1$ points are images of $r+1$ maximal Argyres-Douglas points in $  {\cal M}_{f_{Sp(2r)}}$\footnote{As we saw in subsubsection \ref{towardMap} when we will check this against physics we learn that only 2 out of these $r+1$ points in ${\cal M}_{{\rm D3}}$ are actually maximal Argyres-Douglas points, namely, when we put all D3-branes all together at monopole or dyon.}.

\begin{table}[h!]
 \begin{center}
\begin{tabular}{|c|c|c|c|c|}\hline \multicolumn{4}{|c|}{${\cal M}_{2{\rm D3}}$} & Expected ~{${\cal M}_{f_{Sp(4)}}$} \\ \cline{1-4}
monopole & dyon & $\infty$ & elsewhere & \\ \hline
2&0&0 &0&2 mutually non-local massless BPS \\
0&2&0&0& at 3 maximal Argyres Douglas points\\
1&1&0&0& of $\Sigma^Q_i \cap \Sigma^Q_j$\\ \hline
1&0&1&0& 2 mutually local massless BPS\\
0&1&1&0&  at $\Sigma^Q_i \cap \Sigma^C_j$\\ \hline
1&0&0&1& $\Delta_x f_C \ne 0$ but $\Delta_x f_Q = 0$ \\
0&1&0&1& one massless dyon\\ \hline
0&0&2 &0& Two $\Sigma^C$ vanish (related to $v=0$ loci) \\ \hline
0&0&1 &1& Some sort of $M_\infty$ ? \\  \hline
0&0&0 &$\{1,1\}$ & generic location with $\Delta_x f \ne 0$ \\ \hline
0&0&0 &$\{2\}$&  $\Delta_x f_C = 0$ but $\Delta_x f_Q \ne 0$ \\ \hline
  \end{tabular}
\end{center}
\caption{(A naive 1-1 map) Two D3-branes probing the quantum corrected orientifold background. The number of {\emph {leftover}} D3-branes (put elsewhere) corresponds to the number of degrees of freedom.}
\label{sp4tableNaive}
\end{table}

In {\bf tables \ref{sp2table}, \ref{sp4tableNaive}}, and {\bf \ref{sp2rtableNaive}}, we discuss the naive mappings for $r=1,2$ and higher respectively. Each row denotes how a locus in $ {\cal M}_{\mathrm D3}$ (left) maps to a locus in
${\cal M}_{f_{Sp(2r)}}$ (right), by the 1-1 map ${\cal C}$.
In {\bf tables \ref{sp4tableNaive}}, and {\bf \ref{sp2rtableNaive}} we denoted ${\cal M}_{f_{Sp(2r)}}$ as ``expected
${\cal M}_{f_{Sp(2r)}}$'' to justify what we {\it expect} and not what we {\it get}.
First four columns denote location of D3-branes (different locus in $ {\cal M}_{\mathrm D3}$).
First three columns are counting D3-branes located at three special points on $u$-plane, as introduced in \eqref{notationM}. The fourth column shows all the {\it leftover} D3-branes which are elsewhere. Some of them could be on top of each other, so we introduce $\{ ~ \}$ in order to denote their groupings. Number of elements inside $\{ ~ \}$ are also the number of free parameters (number of freedom to choose locations for each D3-brane group) of that loci in the moduli space.

Let us look at the rank 2 case more closely in this light. The cleanest mapping
should be between point-like singularities given in the first three lines in {\bf table \ref{sp4tableNaive}}. Point-like singular loci in the moduli space ${\cal M}_{2 {\mathrm D3}}$ are given by [2, 0], [1, 1], and [0, 2], with no D3-branes elsewhere. They reflect codimension-2 singularity of $f_C$, therefore we expect them to map to the
three maximal Argyres-Douglas points, given by a cusp-like point in {\bf figure \ref{sp4jihye1}} and its two other $Z_3$ cousins at $|v| = 3$\footnote{We do not know how to make this 3-3 map into 1-1, but we will follow-up on this later (it might be some trivial symmetry breaking).}.

The case [0,0,2] of {\bf table \ref{sp4tableNaive}} can be divided into two sub-cases $[0, 0, \{1, 1\}]$ and $[0, 0, \{2\}]$, where inside $\{ ~ \}$ we denote how D3-branes are grouped.	$[0, 0, \{1, 1\}]$	corresponds to	2 D3	 branes each at	generic location	on $u$-plane - a generic point at ${\cal M}_{2 {\mathrm D3}}$ with total 2 degrees of freedom. In moduli space of the pure Sp(4) curve $  {\cal M}_{f_{Sp(4)}}$,
we expect that it maps to a region with $\Delta_x f_Q \ne 0$. An example is all the region in the figure in the right of {\bf figure \ref{sp4jihye1}} excluding the red lines $\Sigma^Q_i$ .
On the other hand, $[0, 0, \{2\}]$ is when two D3-branes are on top of each other, the degree of freedom is 1,
and so we expect that it satisfies  $\Delta_x f_Q \ne 0 =\Delta_x f_ C$. Now it should develop a singularity for $f_C$ seen as $\Sigma^C_i$ blue lines in {\bf figure \ref{sp4jihye1}}.

Next is the codimension 1 loci in the moduli space, for example given by $[1, 0, 1]$ and $[0, 1, 1]$ for D3 language in  $ {\cal M}_{\mathrm D3}$. We expect that these two should somehow correspond to three $\Sigma^Q_i$'s in {\bf figure \ref{sp4jihye1}}. It might be that some of $\Sigma^Q_i$ are connected so that we have 2-2 mapping instead of 2-3 mapping. For example, in {\bf figure \ref{sp4jihye1}} $\Sigma^Q_0$ and $\Sigma^Q_2$ intersect. In fact, all the $\Sigma^Q_i$ loci is spanned by 1-parameter as given in \eqref{SigmaQofSp4}.

\begin{table}[h!]
 \begin{center}
\begin{tabular}{|c|c|c|c|c|}\hline \multicolumn{4}{|c|}{${\cal M}_{r {\rm D3}}$} & Expected ~ {${\cal M}_{f_{Sp(2r)}}$} \\ \cline{1-4}
monopole & dyon & $\infty$ & elsewhere & \\ \hline
$r$&0&0 &0& $r+1$ maximal Argyres Douglas points \\
$r-1$&1&0&0& with $r$ mutually non-local massless BPS \\
$\cdots$ & $\cdots$ &  0 & 0 & \\
0&$r$&0&0&  \\ \hline
 1&0&0&$\{ 1^{r-1} \}$& $\Delta_x f_C \ne 0$ but $\Delta_x f_Q = 0$ \\
0&1&0&$\{ 1^{r-1} \}$& $\Sigma^Q_i$ loci with one massless dyon\\ \hline
0&0&$r_\infty$ &$\{r_1,r_2,\cdots\}$& $\Delta_x f_Q \ne 0$, gauge enhancement from colliding D3's   \\ \hline  \end{tabular}
\end{center}
  \caption{(A naive 1-1 map) Multiple D3-branes probing the quantum corrected
orientifold background. (A partial table) The number groupings of D3-branes put elsewhere corresponds to the number of degrees of freedom.}
  \label{sp2rtableNaive}
\end{table}

 Similarly, we can discuss the mapping between moduli spaces for rank $r$ case. In the last column of the
{\bf table \ref{sp2rtableNaive}},
D3-brane elsewhere, we could again group them in $\{ ~ \}$ and reduce degrees of freedom.
For example $\{ 1^{r-1} \}$ means $\{ 1, 1, \cdots, 1  \}$, and $(r-1)$ leftover D3-branes are moving on $u$-plane all independently. Their expected mappings are shown in {\bf table \ref{sp2rtableNaive}}. Similarly, following our nose
we would map the
maximal Argyres-Douglas points to the point-like loci in ${\cal M}_{r {\mathrm D3}}$ given as the first group in
{\bf table \ref{sp2rtableNaive}}.


\subsection{A fantasy on how to save the naive 1-1 mapping between moduli spaces of Seiberg-Witten curve and D3-brane picture for pure $Sp(2r)$ theory \label{fantasy}} In subsubsection \ref{towardMap}, we discussed physical flaws in the naive map given above in subsubsection \ref{naiveMap}. However, given mathematical beauty of the 1-1 map (though physically naive) and out of curiosity, we will fantasize some conditions under which the naive mapping would have survived physical tests\footnote{Note that we are still ignoring the dynamics of the anti-symmetric matter as we have assumed it to be very heavy.}. 

We begin by noticing that the state denoted by this
\begin{equation}
[ r_m, r_d ; r_{\infty}, \{ r_1, r_2, \cdots \}  ] =[ 1, 1 ; 0, \{ 0 \}  ]
\end{equation}
might be Argyres-Douglas within our fantasy - if 7-brane charges play a certain role.
Here we have put a D3 on monopole 7 brane, and another D3 on dyon 7 brane. We have massless strings in $3_1$-$7_m$ and $3_2$-$7_d$ sectors.
Using notation of {\bf table \ref{massless}}, they will be charged as in {\bf table \ref{local}}. These charges have zero intersection number, and they are mutually local to each other.
 \begin{table}[htb]
 \begin{center}
\begin{tabular}{|c|cccc|}\hline
Sector &    $p_1$ & $p_2$   &  $q_1$ & $q_2$ \\ \hline
$3_1$-$7_m$ & $\big( (0\big.$, & 0),  &   $  ( 1$ , & $\big.0) \big)$  \\
$3_2$-$7_d$ & $\big( (0\big.$, & 1),   &  (0,&$-1  ) \big)$ \\  \hline
  \end{tabular}
\end{center}
\caption{$3_1$-$7_m$ $\cap$ $3_2$-$7_d$ = 0}
\label{local}
\end{table}

Here we try to cook up a scenario where these two would have been mutually non-local, with charge vectors intersecting with non-zero intersection number. For this, we will introduce extra elements in the charge vectors.
In {\bf table \ref{massless}}, we denoted charges with respect to $U(1)$'s of D3-branes, ignoring $U(1)$'s related to seven-branes.
If we consider seven-branes' $U(1)$'s as well, we will get a following charge configuration as in {\bf table \ref{fantasy1}}.
Note that we introduced new elements in extra paranthesis, to denote electric and magnetic charges with respect to $U(1)$'s of seven-branes. $U(1)$'s of 3- and 7-branes give gauge and global symmetry respectively, therefore we want to put them on different footings. Note that we put extra coefficient $c$ in front of charges with respect to $U(1)$'s of 7-brane. We recover {\bf table \ref{massless}} by $c \rightarrow 0$, where these extra elements were suppressed.
 \begin{table}[h!]
 \begin{center}
\begin{tabular}{|c|cccccccc|}\hline
Sector &    $p_1$ & $p_2$ & $p_m$ & $p_d$ &  $q_1$ & $q_2$ & $q_m$ & $q_d$\\ \hline
$3_1$-$7_m$ & $\big( (0\big.$, & 0), &$c$(0,& $0)$, & $(1$ ,&0), & $ c(-1$ , & $\big.0) \big)$  \\
$3_2$-$7_d$ & $\big( (0\big.$, & 1), &$c$(0,& $ -1)$, & (0, & $-1$), & $c$(0,&$1  ) \big)$ \\  \hline
  \end{tabular}
\end{center}
\caption{$3_1$-$7_m$ $\cap$ $3_2$-$7_d$ = 0}
\label{fantasy1}
\end{table}


We can further imagine a situation where seven-branes share a same $U(1)$.
By some non-trivial interactions, if the two 7-branes know about each other, and
by the fact that they come from same O7 ancestor, we might hope that they are actually charged with respect to the same $U(1)$. Also they have only codimension real 2 with respect to the spacetime directions. It might be that the flux has nowhere to hide, and all the 7-brane actually know about each other\footnote{One could also speculate a similar scenario of putting D3-brane along the $u$-plane in absence of 7-branes, and puting two D3's on $u$-plane at monopole and dyon points. Then the charge configuration would be as below, almost same as the {\bf table \ref{fantasy2}} except for changing $c=1$. We have Argyres-Douglas theories.
 \begin{center}
\begin{tabular}{|c|cccccc|}\hline
Sector &    $p_1$ & $p_2$ & $p_u$ &  $q_1$ & $q_2$ & $q_u$\\ \hline
$3_1$-$3_u$ & $\big( (0\big.$, & 0, &$0)$, & $(1$ ,&0, & $ -1\big. ) \big)$  \\
$3_2$-$3_u$ & $\big( (0\big.$, & 1, &$-1)$, & (0, & $-1$, & $1 ) \big)$ \\  \hline
  \end{tabular}
 \end{center}

However, not only the states given above, but also all possible $(p,q)$ strings will become massless in 3-3 sectors as below.
  \begin{center}
\begin{tabular}{|c|cccccc|}\hline
Sector &    $p_1$ & $p_2$ & $p_u$ &  $q_1$ & $q_2$ & $q_u$\\ \hline
$3_1$-$3_u$ & $\big( (p_{1u}\big.$, & 0, &$p_{1u})$, & $(q_{1u}$ ,&0, & $ -q_{1u}\big. ) \big)$  \\
$3_2$-$3_u$ & $\big( (0\big.$, & $p_{2u}$, &$-p_{2u})$, & (0, & $-q_{2u}$, & $q_{2u} ) \big)$ \\  \hline
  \end{tabular}  \end{center}
 }. Then the extra elements given in {\bf table \ref{fantasy1}} as ${(p,q)}_{m,d}$ will be contained in just one extra column ${(p,q)}_{7}$ as below in {\bf table \ref{fantasy2}}. Now we see that these two states are mutually non-local.
 \begin{table}[h!]
 \begin{center}
\begin{tabular}{|c|cccccc|}\hline
Sector &    $p_1$ & $p_2$ & $p_7$ &  $q_1$ & $q_2$ & $q_7$\\ \hline
$3_1$-$7_m$ & $\big( (0\big.$, & 0), &$c(0)$, & $(1$, &0), & $ c(-1\big. ) \big)$  \\
$3_2$-$7_d$ & $\big( (0\big.$, & 1), &$c(-1)$, & (0, & $-1$), & $c(1 ) \big)$ \\  \hline
  \end{tabular}
\end{center}
\caption{$3_1$-$7_m$ $\cap$ $3_2$-$7_d$ $ = -c^2$. Mutually non-local if $c\ne 0$.}
\label{fantasy2}
\end{table}

For rank 2 case, what this means is that $[1,1,0]$ will be Argyres-Douglas point just like $[2,0,0]$ and $[0,2,0]$. This way, we recover 3 maximal Argyres-Douglas points in ${\cal M}_{2 {\rm D3}}$ given {\bf table \ref{sp4tableNaive}}, instead of 2 points given in {\bf table \ref{sp4table}}. Similar argument for higher rank will recover all $r+1$ maximal Argyres-Douglas points in ${\cal M}_{r {\rm D3}}$ of {\bf table \ref{sp2rtableNaive}}, instead of 2 points given in {\bf table \ref{sp2rtable}}, and it may be helpful to do similar counting of (pairs of) sectors as given below {\bf table \ref{massless}}. Before developing this direction any further, most importantly, we need to verify the role of $U(1)$ charges of seven-branes, which will be discussed in future work.


\newpage

\section{Higgsing and Dualities for $Sp(2r)$ \label{higgsing01}}

In this appendix, we will review some properties of the $Sp(2r)$ quantum corrected curve as well as possible higgsing scenarios worked out by Argyres and Shapere.
Consider the curve and the meromorphic\footnote{The $\vartheta_2(\tau)$ are defined in the following
way:
\begin{eqnarray}&&\vartheta_2(\tau)=\sum_{n\in\mathbb Z}{q^{(n+1/2)^2}}\rightarrow \vartheta_2^4=16q+O(q^3),\\
&& \vartheta_3(\tau)=\sum_{n\in\mathbb Z}{q^{n^2}}\rightarrow \vartheta_3^4=1+8q+O(q^2),\\
&&\vartheta_4(\tau)=\sum_{n\in\mathbb Z}{(-1)^n\ q^{n^2}}\rightarrow \vartheta_4^4=1-8q+O(q^2),\\
&&\vartheta_2^4-\vartheta^4_3+\vartheta^4_4=0, ~~~~q\equiv e^{i\pi\tau}\equiv e^{i\vartheta}e^{-8\pi^2/g^2}.\end{eqnarray}}
1-form for $Sp(2r)$ with $N_f=2r+2$ fundamental hypermultiplets i.e conformal case \cite{ArgyresShapere}:

\begin{eqnarray}\label{curve}
xy^2&=&\left(x\prod_{a=1}^r{(x-\phi_a^2)}+g\prod_{j=1}^{2r+2}{m_j}\right)^2-g^2\prod_{j=1}^{2r+2}{(x-m_j^2)},\\
\lambda&=&\frac{\sqrt{x}}{2\pi i}d\left(\frac{x\prod{(x-\phi_a^2)}+g\prod m_j-\sqrt{x} y}{x\prod{(x-\phi_a^2)}+g\prod m_j+\sqrt{x} y}\right),\\
g(\tau)&=&\frac{\vartheta^4_2}{\vartheta^4_3+\vartheta^4_4}.
\end{eqnarray}

The masses transform in the adjoint representation of $SO(2N_f)$ flavor group.

\begin{itemize}
\item For $r=1$, this curve reduces to the $SU(2)$ curve of \cite{sw2} with 4-flavors which has $PSL(2,\mathbb Z)$ duality group acting on the coupling constant and transforming the masses by the $S_3$ outer automorphisms of the $SO(8)$ flavor symmetry.
\item For $r>1$, the curve is invariant under $\Gamma_0(2)\subset PSL(2,\mathbb Z)$ duality group generated by $T:\  \tau\to \tau+1,\ \ \prod m_j\to-\prod m_j$, and
$ST^2S:\tau\to\tau/(1-2\tau)$.\\ $\Gamma_0(2)$ can be characterized as the set of $SL(2,\mathbb Z)$ matrices whose lower off-diagonal element is even.
\item The asymptotic free theories with $N_f<2r+2$ flavors are obtained by taking $2r+2-N_f$ masses $\sim M\to\infty$ , while keeping $\Lambda^{2r+2-N_f}=qM^{2r+2-N_f}$ finite.
\end{itemize}
Recall that $N=2\ SU(2)$ gauge theory with $4$-flavors enjoys $S$-duality which acts through $SL(2,\mathbb Z)$ and where the parameter space is given by $H/SL(2,Z)$ with $H$ beeing the upper half plane.  On the other hand, $SU(N)$ gauge theory with $N_f=2N_c$ is also superconformal and $S$-duality acts here as a proper subgroup $\Gamma_0(2)$ of $SL(2,Z)$ \cite{plesser,gaiotto}.  Thus, as expected, both $Sp(2r)$ and $SU(N)$ gauge theories share the same $S$-duality group.

\subsection{Higgsing of $Sp(2r)$}
On the Coulomb branch, the adjoint chiral superfield $\Phi$ has expectation values that can be diagonalized as

\begin{equation}
\langle \phi \rangle=diag(\phi_1,\dots,\phi_r,-\phi_1,\dots,-\phi_r).
\end{equation}
Before we dive into the higgsing mechanism at finite coupling, we start by analyzing the classical case by taking the $\Lambda\to0$ limit, $\tau\to+i\infty$ in the scale invariant case. In this limit of
\begin{equation}
q\equiv e^{i\pi \tau}\to0, \qquad g(\tau)=\frac{\vartheta_2^4}{\vartheta_3^4+\vartheta^4_4}=\frac{16q+{\cal O}(q^3)}{2+{\cal O}(q^2)}\to 0+{\cal O}(1/q),
\end{equation}
 the curve (\ref{curve}) becomes:
\begin{equation}\label{curve1}
y^2=\prod_{a=1}^{r}{(x-\phi_a^2)^2}.
\end{equation}
How can we read from the curve that the gauge group is $Sp(2r)$?   Recall that the gauge theory is described by the configuration of the D3-branes  and that the positions of these ${\mathrm D3}_a$'s are given by the $\phi_a$'s. From the curve (\ref{curve1}), we see that for every $\phi_a$ there exists $-\phi_a$ with $a=1,\dots r$.  Thus, there are $2r$ values of $\phi_a$, representing the position of $r$ D3-branes and their images.  Next, remember from \cite{sw1,sw2} that enhance gauge symmetry appears when $\langle \phi \rangle=0$ i.e when $\phi_a=0\ \forall a$.  In the type $IIB$/orientifold language, this translates as having a stack of $r$ D3-branes on top of the orientifold point ($\phi_a$ can be viewed as the distance from the O7 to the ${\mathrm D3}_a$).  Thus, when $\phi_a=0\ \forall a$ the gauge group is $Sp(2r)$. For arbitrary values of $\phi_a$'s where $\phi_a\neq\phi_b$  and $\phi_a\neq0$ the gauge group is $U(1)^r$ where $r$ is the rank of the gauge group.  In other words, we get $U(1)^r$ gauge group when we have $r$ D3-branes spread out on the $u$-plane away from the O7.

In the weak coupling limit $\Lambda\to0$, the branch point of the curve (\ref{curve}) are at the zeros of (\ref{curve1}).  When two (or more) of these coincide one or more cycles of the Riemann surface degenerate, corresponding to some charged state becoming massless.  Such massless states appear at weak coupling whenever two of the $\phi_a$ coincide (corresponding to an unbroken $SU(2)$ gauge subgroup of $Sp(2r)$) or one of the $\phi_a$ vanishes (corresponding to an unbroken $Sp(2)$ gauge group) \cite{ArgyresShapere}.  In the type $IIB$/orientifold language, we understand this picture as follows: when $\phi_a=\phi_b$ two D3-branes are on top of each other, away from the O7 and generating an unbroken $SU(2)$ gauge group.  When $\phi_a=0$ the ${\mathrm D3}_a$ is on top of the orientifold plane, and generates an unbroken $Sp(2)$ gauge group.

At zero coupling, one should get the breaking $Sp(2r)\to Sp(2r-2k)\times SU(k)\times U(1)$ through an expectation value $\langle \phi \rangle=diag(M,\dots, M,0,\dots 0)$ with $r-k$ 0's and $k$ M's.  Thus, this breaking is realized by putting $r-k$ D3-branes on top of the O7 and having a stack of $k$ D3-branes away from the O7.

Non-perturbative corrections are captured in the full curve  (\ref{curve}) with finite values of the gauge coupling $g(\tau)$.  The general (maximal) adjoint breakings of the $Sp(2r)$ gauge theories with non-perturbative corrections are given in \cite{ArgyresShapere}.  In the following, we review these breaking patterns and write the associated curves.  (The semiclassical scale invariant $Sp(2r)$ factor will have finite coupling $\tau$ if we send $\Lambda\to\infty$ keeping $g\equiv\left(\frac{\Lambda}{M}\right)^{2k}$ constant).
Considering the full curve, the breaking of $Sp(2r)$ with $N_f=2r+2$ down to $Sp(2r-2)\times U(1)$ with $2r$ light hypermultiplets is achieved by \cite{ArgyresShapere}:

\begin{eqnarray}
Sp(2r)&\to& Sp(2r-2)\times U(1),\\
\phi_a&=&\left\{{\begin{array}{cc}
\phi'_a&\ \ \ \ \ \ \ \ \ \ \ \ \ a=1,\dots r-1,\\
M&\ a=r.
\end{array}}\right .\\
m_j&=&\left\{{\begin{array}{cc}
h(q)m'_j&j=1,\dots 2r,\\
k(q)M&\ \ \ \ \ \ \ \ j=2r+1,2r+2.
\end{array}}\right.
\end{eqnarray}
Here $h(q),k(q)=1+{\cal O}(q)$.  The limit $M\to\infty$ keeping $\phi'_a$ and $m'_j$ fixed achieves the desired breaking.  The function $h(q)$ and $k(q)=1$ can be chosen to equal to 1.  One then finds that the curve (\ref{curve}) reduces to a curve of the same form with $r\to r-1$ and $g_{r-1}{(q_{r-1})}=g_{r}(q_r)$.  At weak coupling and nonperturbatively, we choose $\tau_{r-1}=\tau_r$ \cite{ArgyresShapere}.
The proposed curve, after the breaking, is

\begin{eqnarray}
xy^2=\left(x\prod_{a=1}^{r-1}(x-{\phi'_a}^2)+g\prod^{2r}_{j=1}{m'_j}\right)^2-g^2\prod^{2r}_{j=1}(x-{m'_j}^2).
\end{eqnarray}
Another adjoint breaking is given by
\begin{equation}
Sp(2r)\to SU(r)\times U(1).
\end{equation}
The semiclassical breaking of $Sp(2r)$ with $N_f=2r+2$ down to $SU(r)\times U(1)$ with $2r$ light hypermultiplets is achieved by tuning
\begin{eqnarray}\label{break2}
\phi_a&=&M+\tilde\phi_a,\ \ \ \ \ \ \ \ \ \ \ \ \ \ \ \ \ \ \ \ \ \ \  \ \ \ \ \sum_{a=1}^{r}{\tilde\phi_a}=0,\\\label{break3}
m_j&=&\left\{{\begin{array}{cc}
M+\tilde m_j+2h(q)\tilde u& \ \ \ \ \ \ \ \ \tilde u\equiv \frac{1}{2r}\sum^{2r}_{j=1}{\tilde m_j},\\
0& \ \ \ \ \ \ \ \ \ j=2r+1,2r+2.
\end{array}}\right.
\end{eqnarray}
Here again, we take the limit $M\to\infty$, while keeping $\tilde\phi_a$ and $\tilde m_j$ fixed and where $h(q)\sim O(q)$.  Substituting (\ref{break2}), (\ref{break3}) into (\ref{curve}) shifting $x\to M^2+2M\tilde x,\ y\to M^{r+1}\tilde y$ expanding around $|\tilde x|<<|M|$ and tuning $h(q)$ appropriately, we recover the $SU(r)$ curve \cite{ArgyresShapere}:
\begin{eqnarray}\label{curveB2}
y^2&=&\prod^{r}_{a=1}{(x-\phi_a)^2}+4h(h+1)\prod^{2r+1}_{j=1}{(x-m_j-2h\mu)},\ \ \ \ \ \sum{\phi_a}=0,\\
\lambda&=&\frac{x-2h\mu}{2\pi i}d\log\left(\frac{\prod(x-\phi_a)-y}{\prod(x-\phi_a)+y}\right),\\
h(\tau)&=&\frac{\vartheta_2^4}{\vartheta_4^4-\vartheta^4_2}.
\end{eqnarray}
\newpage

\section{Modular parameter of zero and 1-instanton for $Sp(2r)$ \label{modular01}}

In this section, we will digress a little bit by introducing $N_f$ flavors in the gauge theory side.
These $N_f$ hypermultiplets, transform under the defining representation of the gauge group, of dimension $N_c$, and with bare masses $m_j,j=1,\dots,N_f$. There are arbitrary number of hypermultiplets, as long as the theorie remains asymp. free.

The classical moduli space is $r$-dimensional and is parametrized by the independent eigenvalues (order parameters) $\bar{a}_k$,  $k=1,\dots,r$ of $\phi$, the complex scalar field of the $N=2$ chiral multiplet in the adjoint representation of the gauge group
\begin{eqnarray}
Sp(2r):\quad \phi=diag(\bar{a}_1,-\bar{a}_1,\dots,\bar{a}_r,-\bar{a}_r).
\end{eqnarray}
The renormalized order parameters $a_k$'s of the theory, their duals $a_{D,k}$'s and the prepotential $\mathcal{F}$ are given by
\begin{equation}
2\pi ia_k=\oint_{A_k}{d\lambda},\ \ \ \ \ 2\pi ia_{D,k}=\oint_{B_k}{d\lambda},\ \ \ \ \ a_{D,k}=\frac{\partial\mathcal{F}}{\partial a_k}.
\end{equation}
One way of determining the prepotential of classical groups is to constraint the curve, prepotential, and order parameters - both classical and quantum-  of $SU(N)$ theory \cite{dhoker}.

The hyperelliptic curve\footnote{Note: different hyperelliptic curves have been proposed for the same gauge groups with the same hypermultiplet content. As shown explicitly in \cite{dhoker2} the corresponding effective prepotentials are the same for each of these different  models of curves because the effective prepotential is unchanged under analytic reparametrizations of the classical order parameters.  Recall that the relation between the classical and quantum order parameters is given in \cite{dhoker}.} associated to $Sp(2r)$ and the associated meromorphic 1-form have been proposed as follows
\begin{eqnarray}
y^2&=&A^2(x)-B(x),\\
d\lambda&=&\frac{xdx}{y}\left(A'-\frac{1}{2}A\frac{B'}{B}\right),
\end{eqnarray}
with
\begin{eqnarray}
A(x)&=&x^2\prod^{r}_{k=1}{(x^2-{\bar{a}_k}^2)}+A_0,\\
B(x)&=&\Lambda^{4r-2N_f+4}\prod^{N_f}_{j=1}{(x^2-{m_j}^2)},
\end{eqnarray}
where
$A_0=\Lambda^{2r-N_f+2}\prod^{N_f}_{j=1}{m_j}$.

As shown in \cite{dhoker}, the method of finding the relevant physical quantities of $Sp(2r)$ from constraining those of $Sp(2r)$ works for $A_0=0$, thus, for at least one hypermultiplet of exactly zero mass.  In fact, they show that for exactly two massless hypermultiplets the rescaled curve for $Sp(2r)$ gauge group admits an even simpler form, which they denote as $Sp(2r)''$:
\begin{eqnarray}
Sp(2r)'':\quad&&A(x)=\prod^r_{k=1}{(x^2-\bar{a}_k^2)},\\
(m_{N_{f-1}}=m_{N_f}=0)\quad &&B(x)=\Lambda^{4r-2N_f+4}\prod^{N_f-2}_{j=1}{(x^2-m_j^2)}.
\end{eqnarray}

We will now compute explicitely the modular parameter $\tau^{ij}\equiv\frac{\partial^2\mathcal{F}(a)}{\partial a_i\partial a_i}$ for $Sp(2r)''$ where
\begin{equation}
\mathcal{F}_{G;N_f}(a_1,\dots, a_{N_c};m_1,\dots,m_{N_f};\Lambda)=\mathcal{F}^{(0)}_{G;N_f}+\sum^{\infty}_{d=1}{\mathcal{F}^{(d)}_{G;N_f}}(a_1,\dots, a_{N_c};m_1,\dots,m_{N_f};\Lambda),
\end{equation}
where $\mathcal{F}^{(0)}_{G;N_f}=\mathcal{F}^{(0)}_{G;N_f}(a_1,\dots, a_{N_c};m_1,\dots,m_{N_f};\Lambda)
$ represents the zero-instanton contribution which corresponds to the classical plus perturbative corrections and is given by \cite{dhoker}
\begin{eqnarray}
&&\frac{4\pi}{i}\mathcal{F}^{(0)}_{G;N_f}(a_1,\dots, a_{N_c};m_1,\dots,m_{N_f}) \nonumber \\
&=&\sum^{r}_{k\neq l}\sum_{\epsilon=\pm 1}(a_k+\epsilon a_l)^2\log\frac{(a_k+\epsilon a_l)^2}{\Lambda^2}
+4\sum^{r}_{k=1}a_k^2\log\frac{a_k^2}{\Lambda^2}-\sum^{r}_{k=1}\sum^{N_f}_{j=1}\sum_{\epsilon=\pm1}(a_k+\epsilon m_j)^2\log\frac{(a_k+\epsilon m_j)^2}{\Lambda^2}. \nonumber \\
\end{eqnarray}
The 1-instanton contribution is given by
\begin{equation}
\mathcal{F}^{(1)}_{G;N_f}=\frac{1}{4\pi i}\bar{\Lambda}^2\sum_{k=1}^{r}{\Sigma_k(a_k)},
\end{equation}
with \begin{equation}
Sp(2r)'':\quad {\bar\Lambda}=\Lambda^{2r+2-N_f},\quad {\Sigma_k(x)}=(x+a_k)^{-2}\prod^{N_f}_{j=1}{(x^2-m_j^2)}\prod_{l\neq k}{(x^2-a_l^2)}^{-2}.
\end{equation}
The $2-$instanton contribution is
\begin{equation}
\mathcal{F}^{(2)}_{G;N_f}=\frac{1}{16\pi i}\bar{\Lambda}^4\left[\sum^r_{k\neq l}\sum_{\epsilon\neq1}{\frac{\Sigma_k(a_k)\Sigma_l(a_l)}{(a_k+\epsilon a_l)^2}}+\frac{1}{4}\sum^r_{k=1}{\Sigma_k(a_k)\left. \frac{\partial^2\Sigma_k(x)}{\partial x^2} \right|_{x=a_k}}\right].
\end{equation}
We compute the modular parameter  for the zero and $1$-instanton contribution
\begin{equation}
\tau^{ij}=\frac{\partial^2\mathcal{F}_{G;N_f}}{\partial a_i\partial a_j}=\frac{\partial^2\sum^\infty_{d=0}{\mathcal{F}^{(d)}}}{\partial a_i\partial a_j},
\end{equation}
where the contribution from the classical and perturbative corrections to $\tau^{ij}$ are given by
\begin{eqnarray}
\frac{\pi}{i}\tau^{(0)ij}\equiv\frac{\pi}{i}\frac{\partial^2\mathcal{F}^{(0)}}{\partial a_i\partial a_j}&=&\sum^r_{i\neq l}\sum_{\epsilon \neq 1}\left[\delta^j_i\left(\log\frac{(a_i+\epsilon a_l)^2}{\Lambda^2}+3\right)+\epsilon(1-\delta^j_i)\left(\log\frac{(a_i+\epsilon a_l)^2}{\Lambda^2}+3\right)\right]\nonumber \\
&+&2\delta^j_i\left(\log\frac{a_i^2}{\Lambda^2}+3\right)-\frac{1}{2}\sum_{n=1}^{N_f}\sum_{\epsilon\neq 1}\left[\delta^j_i\left(\log\frac{(a_i+\epsilon m_n)^2}{\Lambda^2}+3\right)\right].
\end{eqnarray}
After simplification, the 1-instanton contribution is:
\begin{eqnarray}
\frac{4\pi i}{\bar\Lambda^2}\tau^{(1)ij}&=&\delta _{ij}\left(
\begin{array}{c}
\frac{6}{a_{i}^{2}}+\sum_{J}\frac{-6}{a_{i}^{2}-m_{J}^{2}}+4a_{i}^{2}\
\sum_{I\neq J}\frac{1}{a_{i}^{2}-m_{I}^{2}}\frac{\ 1}{a_{i}^{2}-m_{J}^{2}}\
\nonumber \\
+\sum_{l\neq i}\frac{12}{\left( a_{i}^{2}-a_{l}^{2}\right) }+\sum_{l\neq i}%
\frac{8a_{i}^{2}}{\left( a_{i}^{2}-a_{l}^{2}\right) ^{2}} \nonumber \\
+16a_i^2\left( \ \sum_{l\neq i}\frac{1}{\left( a_{i}^{2}-a_{l}^{2}\right) }%
\right) ^{2} \nonumber \\
-16a_{i}^{2}\left( \ \sum_{l\neq i}\frac{1}{\left(
a_{i}^{2}-a_{l}^{2}\right) }\right) \left( \ \sum_{J}\frac{1\ }{%
a_{j}^{2}-m_{J}^{2}}\ \right)
\end{array}%
\right) \Sigma _{i}  \label{1tau} \nonumber \\
&&+\sum_{k\neq i}\delta _{ij}\frac{4\left( a_{k}^{2}+a_{i}^{2}\right) }{%
\left( a_{k}^{2}-a_{i}^{2}\right) ^{2}}\ \Sigma _{k}+\sum_{k\neq i,j}\frac{%
16a_{i}a_{j}}{\left( a_{k}^{2}-a_{i}^{2}\right) \left(
a_{k}^{2}-a_{j}^{2}\right) }\Sigma _{k} \nonumber \\
&&-\ \left( 1-\delta _{ij}\right) \frac{8a_{i}a_{j}}{\left(
a_{i}^{2}-a_{j}^{2}\right) ^{2}}\left( \Sigma _{i}+\Sigma _{j}\right) \nonumber \\
&+&\Sigma _{i}\ \left( 1-\delta _{ij}\right) \frac{8a_{i}a_{j}}{\left(
a_{i}^{2}-a_{j}^{2}\right) }\left( \frac{-1\ }{a_{i}^{2}}+\sum_{I}\frac{1\ }{%
a_{i}^{2}-m_{I}^{2}}+\sum_{l\neq i}\frac{-2\ }{\left(
a_{i}^{2}-a_{l}^{2}\right) }\right) \nonumber  \\
&+&\Sigma _{j}\left( 1-\delta _{ij}\right) \frac{8a_{i}a_{j}}{\left(
a_{j}^{2}-a_{i}^{2}\right) }\left( \frac{-1\ }{a_{j}^{2}}+\sum_{I}\frac{1\ }{%
a_{j}^{2}-m_{I}^{2}}+\sum_{l\neq j}\frac{-2\ }{\left(
a_{j}^{2}-a_{l}^{2}\right) }\right).
\end{eqnarray}

\newpage

\section{Tachyons in the brane anti-brane theory in the presence of fluxes \label{tachyonappendix}}

In this appendix we will discuss how the tachyons behave in the system where we
have wrapped ${\rm D5}$-$\overline{\rm D5}$ on a 2-cycle of a Taub-NUT space.
Unfortunately this exercise is rather non-trivial to perform because of the background
curvature. Therefore we will simplify the system by replacing the 2-cycle by a
toroidal 2-cycle with fluxes and study the behavior of the tachyons\footnote{A somewhat similar
behavior of tachyons were studied in a different context for a system
of a D3 parallel to a D7 in the presence of non-primitive fluxes in \cite{dhhk}. See also \cite{marina} for yet
another application.}.
There is also a non trivial $B$-flux on the 2-cycle, and since the
2-cycle is of vanishing size the $B$-field is actually infinite. Similarly
there is also a gauge flux on the brane.

The quantization of open strings connecting a brane to an antibrane is
well-known.
In the Neveu-Schwarz-Ramond (NSR) formalism, it is identical to the usual
quantization of open strings except that the Gliozzi-Scherk-Olive (GSO) projection is
opposite to the usual one. Hence we keep the ``anti-GSO''
states. These include, at the lowest levels, a tachyon with negative mass square $M^2 = -\half$.
In addition we find a set of massless fermions which are obtained by
dimensionally reducing a single 10-dimensional Majorana-Weyl fermion,
of opposite chirality to the usual GSO-projected one, down to $p+1$
dimensions.

In the case of interest here, this quantization is modified for two
reasons. First, there is a constant B-field, experienced by both the
brane and the antibrane, and also a world-volume field strength $F$ on
one of the pair. Second, the brane and antibrane are both wrapped
around a 2-cycle of vanishing size.

Let us work out the quantization of open strings joining a D5-brane to
a ${\overline{\rm D5}}$-brane in the presence of fluxes. Let $b_1 = (F_1 - B)$ and
$b_2 = (F_2 - B)$, where $F_i$ are the worldvolume gauge fields on the
$i$th brane and $B$ is the constant spacetime $B$-field. Also, let $z
= x^4 + i x^5$. We have chosen to allow nonzero $F$ and $B$ values only
along the two directions $x^{4,5}$.

The boundary conditions are:
\bg\label{boundcon}
&&\left(\del_\sigma z + b_1\, \del_t z\right)_{\sigma = 0} = 0, \nonumber\\
&&\left(\del_\sigma z + b_2\, \del_t z\right)_{\sigma = \pi}= 0,
\nd
and a similar condition for $\bar z$ (with $b_i \leftrightarrow
-b_i$).

Let us now write the mode expansion as:
\bg\label{modeexp}
z = \sum_n A_{n + \nu}\, e^{(n + \nu)(t + i\sigma)}
+ \sum_n B_{n + \nu}\, e^{(n + \nu)(t - i\sigma)}.
\nd
The first boundary condition yields
\bg\label{firstbound}
A_{n + \nu} = B_{n + \nu}\,{1+ib_1\over 1-ib_1},
\nd
while the second one gives
\bg\label{nueqn}
e^{2\pi i\nu} = {(1-ib_1)(1+ib_2)\over
(1+ib_1)(1-ib_2)}.
\nd
Now recall that the 2-cycle on which $b_i$ are valued is of zero size,
which is the same as saying that the value of the field $b_i$ is
infinite, to keep constant flux. Thus we really need the above formula
for infinite $b_1,b_2$, except that they can each be separately
positive or negative infinity. Solving the above, one finds:
\bg\label{nusolv}
\nu ={1\over 2\pi}\left( - {\rm tan}^{-1} {2b_1\over 1 - b_1^2}
 + {\rm tan}^{-1} {2b_2\over 1 - b_2^2} \right).
\nd
Now the relevant values are:
\bg\label{relevvalues}
 \lim_{b\rightarrow -\infty}~
{\rm tan}^{-1} {2b\over 1 - b^2} &=& 0, \nonumber\\
 \lim_{b\rightarrow 0}~ {\rm tan}^{-1} {2b\over 1 - b^2} &=& \pi , \nonumber\\
 \lim_{b\rightarrow \infty}~ {\rm tan}^{-1} {2b\over 1 -
b^2}  &=& 2\pi.
\nd
We can use \eqref{relevvalues} to
evaluate it for the relevant possibilities
($\vert b_i\vert =0, \infty$ for each
$i$). The result is easily seen to be
\bg\label{nuresult}
\nu = \half\Big( {\rm sign}(b_2) - {\rm sign}(b_1) \Big),
\nd
where ${\rm sign}(b_i) = 0, \pm 1$.

Note that for our purposes, $b_1 = F_1 - B$, $b_2 = F_2 - B$. Hence
the case which we expect to be BPS comes about when $F_1=0$, so that
${\rm sign}(b_1) = -1$ while ${\rm sign}(b_2) = 1$, and $\nu=1$. On the
other hand, with $F_1=F_2=0$ we would find $\nu=0$ and this is the
case where we do expect a tachyon.

It only remains to find out the zero-point energy as a function of
$\nu$, in the NS sector (which is where the tachyon appeared, in the
absence of flux). We use:
\bg\label{zeroptsum}
\sum_{n\geq 0}~ (n+\nu) = -{1\over 12}(6\nu^2
-6\nu +1).
\nd
The bosons along direction $x^{4,5}$ are quantised with mode numbers
$n+\nu$ and the fermions have mode numbers
$n \pm \vert \nu - {1\over 2}
\vert$. Thus the zero point energy of the system will be
\bg\label{zeroptexp} E = 2E(0) + E(\nu) + 2E(0) -E(0) -3E({1/2}) -
E(\vert \nu - 1/2\vert).
\nd
The first term comes from $x^{0,1,2,3}$, the second from $x^{4,5}$,
the third from $x^{6,7,8,9}$ and the fourth from the bosonic ghosts. The
remaining terms are fermionic contributions. Adding up all the
contributions, we get:
\bg\label{eground} E = -{1\over 2}\left(\vert \nu - {1\over 2} \vert +
{1\over 2}\right).
\nd
The case of no fluxes is $\nu = 0$ while the case of fluxes relevant to
fractional branes, as we argued above, is $\nu = 1$. From the above formula
we seem to find that both $\nu = 0$ and $\nu = 1$,
the
ground-state energy is $E=-\half$ and hence there is a tachyon.

However the actual result is more subtle because of the GSO
projection.
At zero flux, along with the open string tachyon there is
always a massless state created by a world-sheet fermion (in the NS
sector) $\psi_{-\half}$.  This is in fact a spacetime scalar or vector
(depending on whether the fermion mode has an index transverse to the
brane or along the brane). Now when there is flux, this mode (for the
directions along which the flux is present) becomes
$\psi_{-\vert \nu-\half\vert}$. Thus the corresponding state has energy
\bg\label{london}
E=
-\half\left(\vert \nu-\half\vert + \half
\right) + \vert \nu-\half\vert =
\half\left(\vert \nu-\half \vert  -\half\right).
\nd
Thus altogether we have a pair of low-lying states, one of energy
$-\half(\vert \nu-\half\vert  + \half)$ and the other of energy
$\half(\vert \nu-\half\vert  -\half)$. At
$\nu=0$ these states have energies
$-\half,0$ respectively, and at $\nu=1$ they also have energies
$-\half,0$. But if we tune $\nu$ continuously from 0 to 1, we find
that at $\nu=\half$ the states become degenerate in energy, with both
having $E=-{1\over 4}$. It turns out that at this point the two states
cross each other.

To see this more explicitly, observe that the energies of the pair of
states can equivalently be written $-\half\nu$ and $\half(1-\nu)$ for
all $\nu$, without any mod sign. In this way of writing it, the
energies vary smoothly with $\nu$. These expressions, and not the
earlier ones involving modulus signs, are the correct ones if we want
to follow the evolution of a given state (with a given sign under GSO
projection) as $\nu$ varies. Now we see that the tachyon at $\nu=0$
becomes massless at $\nu=1$. On the other hand the massless state at
$\nu=0$ becomes tachyonic at $\nu=1$. But since we are in a sector
with anti-GSO projection, the latter state is projected out for any
$\nu$! The physical (anti-GSO) state, which is tachyonic at $\nu=0$
really does become massless at $\nu=1$. Thus we have shown that the
tachyon disappears in the presence of flux, as desired. In the {\bf figure \ref{jolanda}}
 below we sketch the behavior of the tachyon for this system:
\begin{figure}[htb]
        \begin{center}
\includegraphics[height=6cm]{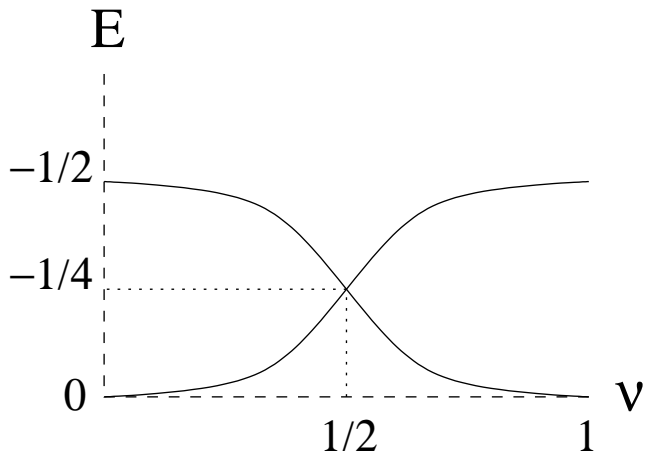}
        \end{center}
        \caption{The behavior of the tachyon in the brane anti-brane system.} \label{jolanda}
        \end{figure}
\noindent As we mentioned earlier, there are two subtleties in the problem
which we have ignored. First,
the $B$ and the $F$ fields on the 2-cycle are not constant so the situation is little subtle. We have discussed the
non-constant ($B, F$) in \cite{marina}.
The second subtlety has to do with the spherical cycle. We have taken the
brane and the antibrane to be wrapped on a toroidal cycle. But we think that the physics may not be that
different for the spherical case. 

\newpage

\section{Connection to integrable systems \label{integrab}}

Now that we have developed a somewhat clearer picture of Model 4, it is time to see whether there could be some
connection to integrable systems of the kind studied in \cite{dw}. To analyze the story here,
we will begin this section by reviewing the connection between the curves describing Riemann surfaces of $\mathcal{N}=2$ supersymmetric gauge theories and algebraically complete integrable Hamiltonian systems\footnote{We would like to thank Sergei Gukov and Johannes Walcher for valuable discussions.}.

Families of elliptic and hyperelliptic curves have been introduce to solve $\mathcal{N}=2$ supersymmetric gauge theories with different gauge groups.
For a gauge group $G$ of rank $r$, these curves parametrize a family of fiber bundles $Y\hookrightarrow U$ where $X$ is a complex manifold of complex dimension $r$ and $U$ is the complex $u$-plane parametrizing the Coulomb branch whose complex dimension is equal to $r$.  The fibers of the map $Y\hookrightarrow U$ are Riemann surfaces of genus equal to the rank $r$ of the gauge group $G$.  These Riemann surface degenerate at finitely many singularities located on the $u$-plane.  We denote the fiber corresponding to $u$ as $Y_u$.
Taking the Jacobian of these Riemann surfaces generates complex $r-$dimensional tori that are now the fibers of $X\hookrightarrow U$ where $X$ is an algebraically complete integrable system.  Note that the rank 1 case doesn't see the difference between the Riemann surface and its Jacobian.

For $r=1$, a family of elliptic curves was introduced in \cite{sw1,sw2} to solve $\mathcal{N}=2$ $SU(2)$ gauge theories with and without  matter.  The solution required the introduction of a meromorphic differential one-form $\lambda$ on $X_u$ which varies holomorphically with $u=$Tr$\phi^2$.  Stable states with magnetic and electric charge $(n_m,n_e)$ have masses given by the BPS formula
\begin{equation}
M^2=2|Z|^2=2|n_ea(u)+n_ma_D(u)|^2,
\end{equation}
where $\vec{a}=(a_D,a)$ are the fundamental period of $\lambda$ on the curve describing the space of vacua
and are given by
\begin{eqnarray}
a_D^i=\int_{\alpha_i}{\lambda},\ \ \ \ \ \ \ \ \ \   a^i=\int_{\beta_i}{\lambda},
\end{eqnarray}
where $\alpha_i,\beta_i$ form a basis of homology cycles on the curve.  One can build a two-form holomorphic object $\omega=d\lambda$ where the holomorphy 
condition insures positivity of the metric on $U$.  In this new coordinate, one gets
\begin{equation}
\vec{a}=\int_{\vec\gamma}\lambda,\ \ \ \ \ \ \ \ d\vec a/du=\int_{\vec\gamma}{d\lambda/du},\ \ \ \ \ \ \ \ d\vec a=\int_{\vec\gamma}{\omega},
\end{equation}
where $\vec\gamma$ is a set of fundamental one-cycles on the fiber and $d$ is the exterior derivative on $U$.

For a gauge group $G$ of rank $r>1$, the base $U$ is a complex manifold of dimension $r$ and the physics is described by a family of $r-$dimensional complex tori $X_u$.   Once again a meromorphic one-form $\lambda$ is required to solve part of the theory and one can circumvent all the ambiguities associated to this quantity by introducing a closed holomorphic 2-form $\omega=d\lambda$ with
\begin{equation}
\omega=\sum_i{du^i\frac{d\lambda}{du^i}},
\end{equation}
where $u_i$=Tr$\langle\phi^i\rangle$ are gauge-invariant order parameters parametrizing the Coulomb branch.  The metric will be positive definite if and only if $\omega$ is non degenerate, that is, the matrix of components of $\omega$ (in any local coordinate system on $X$) is invertible.  In particular, $\omega$ can
be written locally as
\begin{equation}\label{omega}
\omega=\sum_i{dx_i\wedge du^i},
\end{equation}
where $u^i,i=1\dots r$ are coordinates on $U$ and $x_i,i=1\dots r$ are along the fibers.  $\omega$ is non-degenerate at least where the physics is non-singular.  Thus, the geometry, away from singular fibers,  is that of a family of abelian varieties $X\hookrightarrow U$ endowed with a complex symplectic structure $\omega$ on $X$ that takes the form (\ref{omega}).  One can then define a Poisson bracket of local holomorphic functions $f,g$ as
\begin{equation}\label{Poisson}
\{f,g\}:=\frac{rdf\wedge dg\wedge \omega^{r-1}}{\omega^r}.
\end{equation}
The condition $d\omega=0$ would imply that this Poisson bracket obeys the Jacobi identity.  Moreover, using equations (\ref{omega}) and (\ref{Poisson}), the Poisson bracket of the order parameters on the $u$-plane
$\{u^i,u^j\}$ vanishes
\begin{equation}
\{u^i,u^j\}=\frac{rdu^i\wedge du^j\wedge\left(\sum_{k}{dx_k\wedge du^k}\right)^{r-1}}{\left(\sum_k{dx_k\wedge du^k}\right)^{r}}=0,
\end{equation}
indicating that the $u^i$ form a maximal set of commuting Hamiltonians \cite{dw,markman}.
The object $X$ with the above properties is called an algebraic completely integrable Hamiltionian system.
Conversely, given an algebraic completely Hamiltonian integrable system $X$, on can define $U$ to be the space spanned by commuting Hamiltonians and create a map $X\hookrightarrow U$.  If this map if proper, one can prove that the fibers are complex tori and $\omega$ has can be written as (\ref{omega})  \cite{dw,markman}.

\subsection{$\mathcal{N}=2$ $SU(n)$ gauge theories as algebraically complete integrable systems}

Donagi-Witten \cite{dw} have identified the Coulomb branch of $d=4$ $\mathcal{N}=2$ $SU(n)$ supersymmetric gauge theories with a hypermultiplet in the adjoint representation as described by an integrable Hamiltonian system.  In particular, they have shown that an integrable system for any gauge group can be described by a spectral curve parametrizing a Riemann surface of genus $g$ equal to the rank of the gauge group.   This curve is the $n$-fold cover of the ``bare'' spectral curve which parametrizes a genus 1 Riemann surface.  Moreover, it was shown in \cite{wittenM} that this spectral curve has a M-theory brane interpretation.  As we will see in the sections bellow, some of our F-theory constructions can also be understood as integrable systems.   Under a reparametrization worked out in \cite{dw}, the spectral curve for pure $\mathcal{N}=2$ $SU(n)$ gauge theory matches the $SU(n)$ hyperelliptic curve of \cite{ArgyresFaraggiSU, KLYTsimpleADE}.  Indeed, this latter curve is a function of $u$-plane coordinates $u_k=$Tr$\langle\phi^k\rangle$ where the $u$'s form a set of maximally commuting Hamiltonians, that is, they have vanishing Poisson brackets $\{u^i,u^j\}$.  As explained above, this is often considered as the basis for asserting that we are dealing with a complex, or even algebraic, analogue of a complete integrable Hamiltonian system \cite{dw}.

\subsection{Model 1 and 2 as algebraically complete integrable systems}
An integrable system for $d=4$ $\mathcal{N}=2$ $Sp(2r)$ supersymmetric gauge theory with vanishing bare mass of the antisymmetric tensor multiplet and massive fundamental matter was constructed in \cite{dougschwa}.  In the present paper, we have studied the embedding of the hyperelliptic curves for $Sp(2r)$ with and without matter \cite{ArgyresShapere} into a consistent F-theory brane picture. Models 1 and 2 studied in the present paper have the necessary and sufficient ingredients to be  algebraically complete integrable Hamiltonian systems.  In particular, they have the right elliptically fibered geometries, admit a symplectic structure that takes the form of (\ref{omega}), and have vanishing poisson bracket of their order parameters on the Coulomb branch.  Recall that Model 1 consists of F-theory on an elliptically fibered $\rm K3$ where the full geometry is given by ${\rm K3}\times {\mathbb R}^4\times \mathbb R_{0123}$ which is type IIB on $T^2/{\mathbb Z}_2\times {\mathbb R}^4\times \mathbb R_{0123}$ and Model 2 is F-theory on an elliptically fibered $\rm K3$ ${\rm K3}\times {\mathbb R}^4/{\mathbb Z}_2\times \mathbb R_{0123}$ which is type IIB on $T^2/{\mathbb Z}_2\times {\mathbb R}^4/{\mathbb Z}_2\times \mathbb R_{0123}$ .  The coulomb branch of both these models is on the $\mathbb CP^1$ of $\rm K3$.  Also,  one can naively understand the $Sp(2r)$ gauge theory as being an algebraically complete integrable system from the fact that the order parameters of $Sp(2r)$ are nothing but a naturally constrained versions of those of $SU(2r)$  \cite{dhoker}.

\subsection{Model 3 as an algebraically complete integrable system}
Model 3 can be understood as simply compactifying the running directions of Model 2 into a $\rm K3$.  One then obtains the following geometry: $\mathcal{N}=2$ F-theory on ${\rm K3} \times {\bf T}^4/{\mathbb Z}_2\times \mathbb R_{0123}$ or  simply put ${\rm K3} \times {\rm K3}$ which is type IIB on ${\bf T}^2/{\mathbb Z}_2\times {\bf T}^4/{\mathbb Z}_2\times \mathbb R_{0123}$ .  The Coulomb branch of Model 2 was left intact here.  Thus, Model 3 in an algebraically complete integrable system with gauge group $SU(r)$ or $SU(r|r)$ in the presence of D3 and anti-D3 branes.  As discussed in the beginning of chapter 5, this geometry can be brought to a F-theory $\mathcal{N}=1$ system of elliptically fibered Calabi-Yau four-fold.
Here, the underlying $\mathcal{N}=1$ theory doesn't have a integrable system description as far as we know.

\subsection{Model 4 as an algebraically complete integrable system?}
\begin{figure}[htb]
        \begin{center}
\includegraphics[height=6cm]{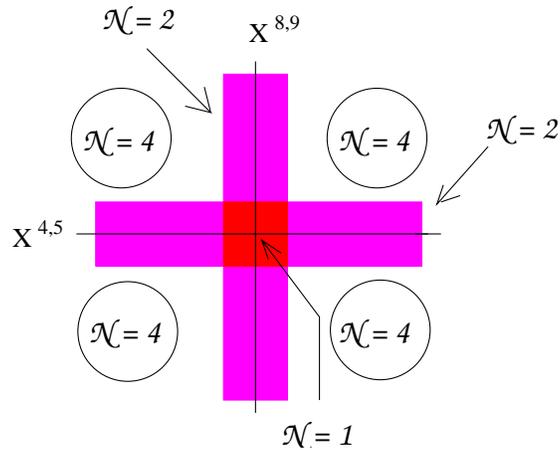}   \end{center}
        \caption{The various manifestaions of supersymmetries in the intersecting seven-brane model. Away from one set of
seven-branes we expect ${\cal N} = 2$ supersymmetry. Right at the intersection the supersymmetry is reduced to
${\cal N} = 1$. Far away from both set of seven-branes, the supersymmetry is maximal, i.e ${\cal N} = 4$. Note that the
above figure has to be translated along $x^{6, 7}$ directions to get the full scenario.}
        \label{susygp}
        \end{figure}

Model 4 being an $\mathcal{N}=1$ theory of multiple D3-branes probing intersecting O7-planes with gauge group $Sp(2r)\times Sp(2r)$ on a T-dual version of type IIB on ${\bf T}^4/({\mathbb Z}_2\times {\mathbb Z}_2)$ \cite{gimpol,sengimon,vafaT}, we don't seem to have the required geometry and form of the symplectic structure to obtain an algebraic integrable system.  However, it is interesting to point out the different regimes of this theory that do admit an integrable description of the system.  First, when the D3-branes are far away from the system of intersecting O7-planes, the theory regains its full supersymmetry and we obtain the celebrated ${\cal N}=4$ theory which is a well known integrable system.  Secondly, in the limit away from the intersecting point, the $u$-branch and $v$-branch each become $\mathcal{N}=2$ IIB theories on $T^2/{\mathbb Z}_2$  which, as argued above, are algebraically integrable systems.
This is shown in {\bf figure \ref{susygp}}.
Here the $u$ and $v$ coordinates refer to \cite{sengimon} with $u^2=w$, $v^2=z$ and $u=x^4+ix^5$, $v=x^8+ix^9$.  We believe this theory is an interesting playground for better understanding integrable systems for $\mathcal N=1$ theories.

\newpage
\bibliographystyle{JHEP}
\bibliography{SWcurve}

\end{document}